\documentclass[
aps,
onecolumn,
12pt,
tightenlines,
nofootinbib,
superscriptaddress,
floatfix,
prd
]{revtex4-2}
\usepackage{subcaption}
\usepackage{amsmath} 
\usepackage[utf8]{inputenc}
\usepackage[OT1]{fontenc}
\usepackage{siunitx}
\usepackage{siunitx}
\usepackage{physics}
\usepackage{braket}
\usepackage{amsmath}	
\usepackage{amsthm}		
\usepackage{amssymb}	

\usepackage{mathrsfs, mathtools}

\usepackage{letltxmacro} 

\usepackage{color}
\usepackage{bm}

\usepackage{tgpagella}

\usepackage{graphicx}

\usepackage{fancyhdr}

\usepackage{lastpage}

\usepackage{titlesec}

\usepackage{rotating}
\usepackage{setspace}

\usepackage[printonlyused]{acronym}
\usepackage{aas_macros}
\usepackage{xspace}

\usepackage{datetime}

\usepackage{lettrine}

\usepackage{anyfontsize}

\usepackage[colorlinks=true
,urlcolor=DARKBLUE
,anchorcolor=DARKBLUE
,citecolor=DARKBLUE
,filecolor=DARKBLUE
,linkcolor=DARKBLUE
,menucolor=DARKBLUE
]{hyperref}

\titleformat{\section}
       {\centering\normalfont\fontsize{14}{17}\bfseries}{\thesection}{1em}{}

\numberwithin{equation}{section}

\makeatletter 
    \renewcommand{\thefigure}{{\bf\@arabic\c@figure}}
\makeatother

\usepackage{tcolorbox}

\definecolor{grey}{rgb}{0.4,0.4,0.4}
\definecolor{dullmagenta}{rgb}{0.4,0,0.4}
\definecolor{darkblue}{rgb}{0,0,0.4}
\definecolor{midblue}{rgb}{0,0,0.7}
\definecolor{midred}{rgb}{0.5,0,0}
\definecolor{orange}{rgb}{1,0.5,0}
\definecolor{lightbrown}{rgb}{0.75,0.5,0.25}
\definecolor{tan}{cmyk}{0.14,0.42,0.56,0}
\definecolor{djunglegreen}{cmyk}{0.99,0,0.52,0}
\definecolor{lightgreen}{rgb}{0,1,0}
\definecolor{olivegreen}{cmyk}{0.64,0,0.95,0.40}
\definecolor{midgreen}{rgb}{0.0,0.675,0.0}
\definecolor{darkgreen}{rgb}{0,0.5,0}
\definecolor{pink}{rgb}{1,0.078,0.57}

\definecolor{MONZA}{HTML}{CF000F}
\definecolor{DARKBLUE}{HTML}{00008b}
\definecolor{DARKMAGENTA}{HTML}{8b008b}

\newcommand{\vs}{\vspace}

\makeatletter
\let\oldr@@t\r@@t
\def\r@@t#1#2{%
\setbox0=\hbox{$\oldr@@t#1{#2\,}$}\dimen0=\ht0
\advance\dimen0-0.2\ht0
\setbox2=\hbox{\vrule height\ht0 depth -\dimen0}%
{\box0\lower0.4pt\box2}}
\LetLtxMacro{\oldsqrt}{\sqrt}
\renewcommand*{\sqrt}[2][\ ]{\oldsqrt[#1]{#2}}
\makeatother

\newcommand{\la}{\ensuremath{\leftarrow}}

\renewcommand{\u}{\underline}

\newcommand{\be}{\begin{equation}}
\newcommand{\ee}{\end{equation}}
\newcommand{\ba}{\begin{eqnarray}}
\newcommand{\ea}{\end{eqnarray}}

\smallskipamount
\settimeformat{ampmtime}

\def\ga{\mathrel{\raise.3ex\hbox{$>$\kern-.75em\lower1ex\hbox{$\sim$}}}}
\def\la{\mathrel{\raise.3ex\hbox{$<$\kern-.75em\lower1ex\hbox{$\sim$}}}}

\newcommand{\Msun}{\ensuremath{\,\rm{M}_{\odot}}\xspace}
\newcommand{\Ms}{\Msun}
\newcommand{\msun}{\Msun}

\newcommand{\kms}{{\rm {km\, s^{-1}}}}

\acrodef{BH}{black hole}
\acrodefplural{BH}[BHs]{black holes}

\acrodef{BBH}{Binary black hole}
\acrodefplural{BBH}[BBHs]{binary black holes}

\acrodef{IMBH}{intermediate-mass black hole}
\acrodefplural{IMBH}[IMBHs]{intermediate-mass black holes}

\acrodef{SMBH}{supermassive black hole}
\acrodefplural{SMBH}[SMBHs]{supermassive black holes}

\acrodef{PBH}{primordial black hole}
\acrodefplural{PBH}[PBHs]{primordial black holes}

\acrodef{BHXB}{black hole X-ray binary}
\acrodefplural{BHXB}{black hole X-ray binaries}

\acrodef{GC}{globular cluster}
\acrodefplural{GC}[GCs]{globular clusters}

\acrodef{NC}{nuclear cluster}
\acrodefplural{NC}[NCs]{nuclear clusters}

\acrodef{OC}{open cluster}
\acrodefplural{OC}[OCs]{open clusters}

\acrodef{YMC}{young massive cluster}
\acrodefplural{YMC}[YMCs]{young massive clusters}

\acrodef{MLT}{mixing-length theory}
\acrodef{HR}{Hertzsprung-Russell}
\acrodef{MS}{main sequence}
\acrodef{ZAMS}{zero-age MS}
\acrodef{CHeB}{core helium burning}
\acrodef{RGB}{red giant branch}
\acrodef{RG}{red giant}
\acrodef{RSG}{red super-giant}
\acrodef{HB}{horizontal branch}
\acrodef{AGB}{asymptotic giant branch}
\acrodef{TP-AGB}{thermally-pulsating asymptotic giant branch}
\acrodef{LBV}{luminous blue variable}
\acrodef{WR}{Wolf-Rayet}

\acrodef{WD}{white dwarf}
\acrodefplural{WDs}{white dwarfs}
\acrodef{NS}{neutron star}
\acrodefplural{NSs}{neutron stars}
\acrodef{CO}{compact object}
\acrodefplural{COs}{compact objects}

\acrodef{CC}{core collapse}
\acrodef{ECSN}{electron-capture supernova}
\acrodef{PI}{pair instability}
\acrodef{PISN}{pair instability supernova}
\acrodef{PPISN}{pulsational pair instability supernova}
\acrodef{compas}{\texttt{COMPAS}}
\acrodef{mobse}{\texttt{MOBSE}}
\acrodef{cBH}{Black hole}
\acrodefplural{cBH}[BHs]{Black holes}

\acrodef{sBH}{stellar-mass black hole}
\acrodefplural{sBH}[sBHs]{Stellar-mass black holes}

\acrodef{IMBH}{intermediate-mass black hole}
\acrodefplural{IMBH}[IMBHs]{intermediate-mass black holes}

\acrodef{SMBH}{supermassive black hole}
\acrodefplural{SMBH}[SMBHs]{Supermassive black holes}

\acrodef{IMRI}{intermediate-mass ratio inspiral}
\acrodefplural{IMRI}[IMRIs]{intermediate-mass ratio inspirals}

\acrodef{BBH}{binary black hole}
\acrodefplural{BBH}{binary black holes}

\acrodef{GC}{globular cluster}
\acrodefplural{GC}[GCs]{Globular clusters}

\acrodef{NSC}{nuclear star cluster}
\acrodefplural{NSC}[NSCs]{Nuclear star clusters}

\acrodef{OC}{open cluster}
\acrodefplural{OC}[OCs]{open clusters}

\acrodef{YMC}{young massive cluster}
\acrodefplural{YMC}[YMCs]{Young massive clusters}

\acrodef{AGN}[AGN]{active galactic nucleus}
\acrodefplural{AGN}[AGNs]{active galactic nuclei}

\acrodef{IMF}{initial mass function}

\acrodef{GW}{gravitational wave}
\acrodefplural{GW}{gravitational waves}
\acrodef{LVK}{LIGO-Virgo-KAGRA collaboration}
\acrodef{LISA}{Laser Interferometer Space Antenna}

\acrodef{VMS}{very massive star}
\acrodefplural{VMSs}[VMS]{very massive stars}

\acrodef{ZAMS}{zero-age main sequence}
\acrodef{MS}{main sequence}
\acrodef{WD}{white dwarf}
\acrodefplural{WDs}{white dwarfs}
\acrodef{NS}{neutron star}
\acrodefplural{NSs}{neutron stars}
\acrodef{CO}{compact object}
\acrodefplural{COs}{compact objects}
\acrodef{DCO}{double compact object}
\acrodefplural{DCOs}{double compact objects}
\acrodef{RLOF}{Roche Lobe Overflow}

\acrodef{CE}{Common Envelope}

\acrodef{MW}{Milky Way}

\acrodef{SSE}{\texttt{SSE}}
\acrodef{BSE}{\texttt{BSE}}
\acrodef{MOCCA}{\texttt{MOCCA}}

\headwidth=\textwidth

\let\originalparagraph\paragraph
\renewcommand{\paragraph}[2][:]{\originalparagraph{\textbf{#2#1}}}

\usepackage{titlesec}
\usepackage{titletoc}

\titleformat{\section}[block]{\large\centering\bfseries}{\thesection.}{0.5em}{}
\titleformat{\subsection}[block]{\centering\bfseries}{\thesubsection.}{0.5em}{}
\titleformat{\subsubsection}[block]{\centering\bfseries}{\thesubsubsection.}{0.5em}{}

\titlecontents{section}
  [-2pt] 
  {\addvspace{0.25em}} 
  {\bfseries\contentslabel{0.1em}\hspace{2.3em}} 
  {\titlerule*[0.5pc]{.}\contentspage} 
  {\bfseries\titlerule*[0.5pc]{.}\contentspage} 

\begin{document}

\title{\Large {\huge I}ntermediate-{\huge M}ass {\huge B}lack {\huge H}oles \\ in {\huge S}tar {\huge C}lusters and {\huge D}warf {\huge G}alaxies}


\author{Abbas Askar}
\email{askar@camk.edu.pl}
\affiliation{Nicolaus Copernicus Astronomical Center,
    Polish Academy of Sciences,
    ul. Bartycka 18,
    00-716 Warsaw,
    Poland
    }
\author{Vivienne F. Baldassare}
\email{vivienne.baldassare@wsu.edu}
\affiliation{Department of Physics and Astronomy, Washington State University, Pullman, WA 99163, USA }

\author{Mar Mezcua}
\email{marmezcua.astro@gmail.com}
\affiliation{Institute of Space Sciences (ICE, CSIC), Campus UAB, Carrer de Magrans, 08193 Barcelona, Spain}
\affiliation{Institut d'Estudis Espacials de Catalunya (IEEC), Carrer Gran Capit\`a, 08034 Barcelona, Spain}

\date{\formatdate{\day}{\month}{\year}, \currenttime}

\begin{abstract}
\vs{1mm}
\begin{tcolorbox}
Black holes (BHs) with masses between 100 to 100,000 times the mass of the Sun ($\msun$) are classified as intermediate-mass black holes (IMBHs), potentially representing a crucial link between stellar-mass and supermassive BHs. Stellar-mass BHs are endpoints of the evolution of stars initially more massive than roughly 20 $\msun$ and generally weigh about 10 to 100 $\msun$. Supermassive BHs are found in the centre of many galaxies and weigh between $10^{6}$ to $10^{10} \ \msun$. The origin of supermassive BHs remains an unresolved problem in astrophysics, with many viable pathways suggesting that they undergo an intermediate-mass phase. Whether IMBHs really stand as an independent category of BHs or rather they represent the heaviest stellar mass and the lightest supermassive BHs is still unclear, mostly owing to the lack of an observational smoking gun. The first part of this chapter discusses proposed formation channels of IMBHs and focuses on their formation and growth in dense stellar environments like globular and nuclear star clusters. It also highlights how the growth of IMBHs through mergers with other BHs is important from the point of view of gravitational waves and seeding of supermassive BHs in our Universe. The second part of the chapter focuses on the multi-wavelength observational constraints on IMBHs in dense star clusters and dwarf galactic nuclei. It also examines the potential insights that future gravitational wave detectors could offer in unraveling the mystery surrounding IMBHs.

\end{tcolorbox}
\end{abstract}

\begin{center}
    \phantom{\fontsize{50}{50}\selectfont I}\\
    {
    \fontsize{30}{10}\selectfont 
    {\fontsize{30}{20}\selectfont I}ntermediate-{\fontsize{30}{20}\selectfont M}ass 
    {\fontsize{30}{20}\selectfont B}lack
    {\fontsize{30}{20}\selectfont H}oles \\
    [5mm]
    in \\
    [5mm]
    {\fontsize{30}{20}\selectfont S}tar 
    {\fontsize{30}{20}\selectfont C}lusters \\
    [5mm]
    and \\
    [5mm]
    {\fontsize{30}{20}\selectfont D}warf
    {\fontsize{30}{20}\selectfont G}alaxies}
    \\[90mm]
    {\noindent\makebox[\linewidth]{\resizebox{0.3333\linewidth}{1pt}{$\bullet$}}\bigskip}\\[2mm]
    {\fontsize{18}{5}\selectfont Abbas Askar$^*$}\\[6mm]
    {\fontsize{18}{5}\selectfont Vivienne F. Baldassare$^*$}\\[6mm]
    {\fontsize{18}{5}\selectfont Mar Mezcua$^*$}\\[6mm]
    {\fontsize{12}{5}\selectfont $*$All authors equally contributed to the chapter}
\end{center}
\newpage
\maketitle

\newpage
\tableofcontents
\addtocontents{toc}{\vspace*{-1.2cm}}
\newpage


\part{Intermediate-Mass Black Holes in Star Clusters and Dwarf Galaxies: Formation and Growth 
\\ {\large Author: Abbas Askar}}
\label{Sec21}
\section{Introduction}
\label{sec:Introduction}

\subsection{Context and definition}\label{sec:1.1}

As the name suggests, \ac{IMBH} characterizes \acp{BH} that have masses which are intermediate between two distinct mass classifications of \acp{BH} that have been observed in our Universe. These two mass categories of \acp{BH} are summarized below. 

\begin{enumerate}
    \item \textbf{\acp{sBH}} have masses that are roughly between 10 to 100 times the mass of our Sun ($\msun$). These \acp{BH} are born when massive stars ($\rm M_{birth} \gtrsim 20 \ \msun$) exhaust their nuclear fuel and their remnant core collapses due to gravity (See {\color{blue} \textbf{Chapter 1}}). Given a typical \ac{IMF} \citep[e.g.,][]{2001MNRAS.322..231K,2003PASP..115..763C}, about one or two in every 1000 stars would end its life as an \ac{sBH}. Observational evidence for the existence of \ac{sBH}s began accumulating in 1960s and 1970s with the advent of X-ray astronomy. Multiple luminous X-ray sources were detected by early space-based missions \citep[e.g.][see {\color{blue} \textbf{Chapter 1.V}} for further details]{1965Sci...147..394B,1971ApJ...165L..27G,1982ApJ...253..485P}. By the 1980s it had became clear that many of these high-energy electromagnetic radiation sources are close binary systems in which an \ac{sBH} ($\sim 5$ to $20 \ \msun$) is accreting material from a companion star \citep{1972Natur.235...37W,1973A&A....24..337S,1975ARA&A..13..381E,1984SSRv...38..353L,1992ARA&A..30..287C}\footnote{catalog of known sBH in Galactic X-ray binaries is available on the following link: \url{https://stellarcollapse.org/sites/default/files/table.pdf} \citep{2014bsee.confE..37W}}.
    This accreted material feeds into the \ac{BH} and as it does it can get heated to millions of degrees resulting in the production of X-rays.
    In 2016, the first direct detection of \acp{GW} from the merger of two \acp{sBH} was announced by the \ac{LVK} \citep{2016PhRvL.116f1102A}. This discovery confirmed that \acp{BBH} exist and can merge to form a more massive \ac{BH}. Since 2016, more than 80 merging \acp{BBH} have been detected by the \ac{LVK} \cite{2021arXiv211103606T}. Each detected merger event allows the inference of the mass of the two merging \acp{BH} and the mass of the merged \ac{BH}. Therefore, these detections are providing new insights into the demographics of \acp{BH} in our Universe. \ac{sBH} masses inferred from the \ac{LVK} observations range between $\sim 5$ to $180 \ \msun$ \footnote{GWTC (Gravitational Wave Transient Catalog) is a catalog of all \ac{GW} merger events that have been observed by the \ac{LVK}. Thanks to the `The Gravitational Wave Open Science Center (GWOSC)' of the \ac{LVK}, it can be viewed online on the following link: \url{https://www.gw-openscience.org/eventapi/html/GWTC/}}.

    The final mass of an \ac{sBH} that forms through the evolution of an isolated massive star strongly depends on its birth mass (\ac{ZAMS} mass) and metallicity. These parameters govern how quickly the progenitor star will evolve, how much of its birth mass will be lost through stellar winds and whether it would undergo a supernova (SN) explosion in the final stages of its evolution. Many of these important phases of the evolution of massive stars are poorly understood and this contributes to the uncertainty in the mass function and mass limits of \acp{sBH}. Observations and theoretical studies suggest that masses of \acp{sBH} forming via the isolated evolution of a single massive star range between $\sim 3 - 5 \ \msun$ \citep{2010ApJ...725.1918O,2011ApJ...741..103F,2022ApJ...937...73Y} to up $\sim 50 - 70 \ \msun$ \citep[e.g.,][]{2016A&A...594A..97B,2017MNRAS.470.4739S,2020ApJ...902L..36F,2022ApJ...937..112F}. More massive \acp{sBH} could be produced through the complex evolution of massive stars born in close binary or multiple stellar systems \citep{2019MNRAS.485..889S,2020ApJ...897..100V,2020A&A...638A..94O,2022ApJ...924...56S}. 
   
    \item \textbf{\acp{SMBH}} are found in the nuclei of many galaxies including the \ac{MW}. Their masses range from about $10^{6}$ to $10^{10} \ \msun$, though the lower boundary is not well constrained. Observational evidence for \acp{SMBH} emerged  with the discovery of extremely bright and relatively compact extragalactic sources that came to be known as \ac{AGN}. These were postulated to be accreting \acp{SMBH} located in the nucleus of their host galaxy. Similar to \acp{sBH} accreting material from companion stars in close binary systems, an \ac{AGN} shines brightly as gas feeding into an \ac{SMBH} heats up in an accretion disk \citep[see][and references therein]{1984ARA&A..22..471R}.
    Additionally, long-term monitoring of the motion of stars in the Galactic center through infrared observations show that these stars are orbiting a $\sim 4\times10^{6} \ \msun$ \ac{SMBH} that is located in the very center of our Galaxy \citep{2019A&A...625L..10G,2008ApJ...689.1044G,2002Natur.419..694S}. The observed masses of \acp{SMBH} tightly correlate with the velocity dispersion ($\sigma$) of stars in the bulge of their host galaxies \citep[known as the M-$\sigma$ relation;][]{2000ApJ...539L...9F,2000ApJ...539L..13G,2009ApJ...698..198G}. The exact formation and growth mechanism of these \acp{SMBH} remains an unresolved problem in astrophysics (for further details, see {\color{blue} \textbf{Chapter 3}}).
\end{enumerate}

\begin{figure}[hbt!]
\centering
\includegraphics[width = 0.75\textwidth]{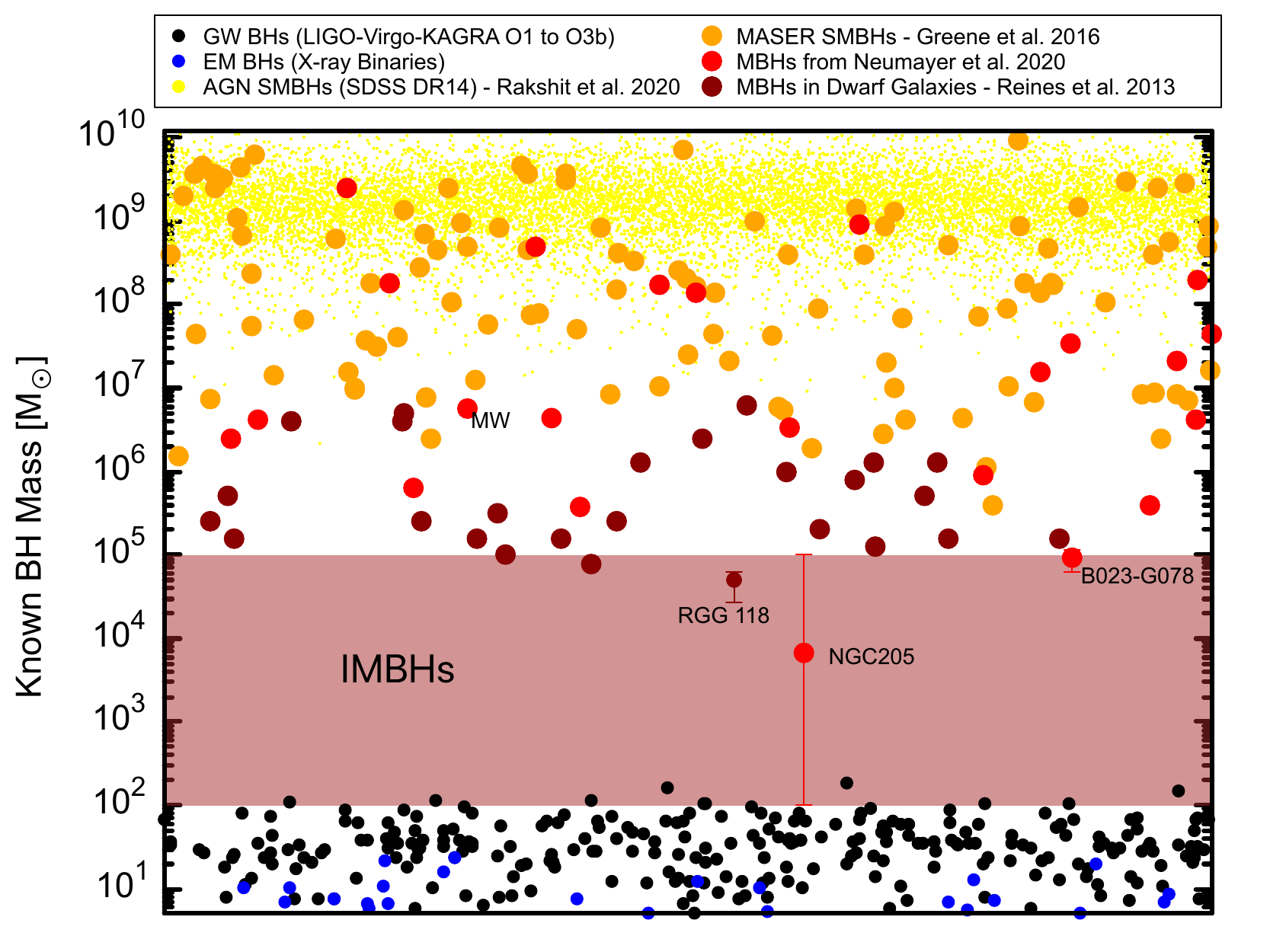}
\caption{The coloured circular points show masses (y-axis) of known \acp{BH} discovered through different methods. The x-axis has been randomly spread in order for the points to become viewable and has no significance. \acp{BH} with masses between $5$ to $180 \ \msun$ that maybe classified as \acp{sBH} are shown with small black circles \citep[masses inferred through GW observations by the \ac{LVK};][]{2021arXiv211103606T} and blue circles \citep[masses of \acp{BH} in Galactic X-ray binaries that are sources of electromagnetic (EM) radiation\textsuperscript{1};][]{2014bsee.confE..37W}. The small yellow circles show \ac{SMBH} masses inferred using emission line data for more than 10,000 \acp{AGN} in the Sloan Digital Sky Survey (SDSS) Data Release 14 \citep{2020ApJS..249...17R} with a signal-to-noise ratio larger than 10 and \ac{BH} mass errors less than 0.03 dex. These masses range between $10^{8}$ to $10^{10}$ $\msun$. More known \ac{SMBH} masses are shown with larger orange \citep[precision mass measurements of \acp{SMBH} from water megamaser disks;][]{2016ApJ...826L..32G}, brown \citep[\ac{SMBH} masses measured in low-luminosity AGNs of dwarf galaxies;][]{2013ApJ...775..116R,2015ApJ...809L..14B} and red circles \citep[masses for SMBHs taken from Table 3 in][and references therein]{2020A&ARv..28....4N}. It can be seen that known \ac{SMBH} masses mostly range between $10^{5}$ to $10^{10} \ \msun$. The red shaded region shows the \ac{IMBH} mass range between $\sim 10^{2}$ to $10^{5} \ \msun$. As examples, masses of two \ac{IMBH} candidates that were estimated by fitting dynamical models to observed kinematic data are labelled in the figure. This includes an IMBH candidate at the center of NGC 205 \citep{2019ApJ...872..104N} and in the center of a dense stellar system located within the M31 galaxy \citep{2022ApJ...924...48P}. Additionally, the mass of an \ac{IMBH} candidate discovered in the nucleus of the active dwarf galaxy RGG 118 \citep{2015ApJ...809L..14B} has also been labelled.}
\vspace*{-1mm}
\label{fig:known-bh-masses}
\end{figure}

\acp{IMBH} are categorized as \acp{BH} that are more massive than \acp{sBH} and less massive than \acp{SMBH}. This means that they are defined as \acp{BH} having masses that are between about $\sim 10^{2}$ to $\sim 10^{5} \ \msun$. Fig.\,\ref{fig:known-bh-masses} shows masses of observed \acp{sBH} and \acp{SMBH} that have been measured using different approaches. Unlike \acp{sBH} and \acp{SMBH}, \acp{IMBH} have remained observationally elusive. However, in recent years there has been growing observational evidence for the existence of \acp{IMBH} with masses bordering close to the lower and upper-mass limits of $10^{2} \ \msun$ and $10^{5} \ \msun$ (see {\color{blue}\textbf{Chapter 2.II}}). This includes the \ac{GW} merger event GW190521 that was discovered by the \ac{LVK} in 2020 \citep{2020ApJ...900L..13A}. In this merger event, a $147.4_{-16.0}^{+40.0}$ $\msun$ \ac{IMBH} formed from the merger of $105.5_{-24.1}^{+45.3}$ $\msun$ and $57.2_{-30.1}^{+27.1}$ $\msun$ BHs \citep{2020ApJ...900L..13A}. This discovery provides the first concrete evidence for the presence of an \ac{IMBH} larger than a $100$ $\msun$. Few more such merger events have been observed by the \ac{LVK} \citep[e.g., GW190426\_190642, GW200220\_061928][]{2021arXiv211103606T}. With the emergence of \ac{GW} astronomy, we now know that low-mass \acp{IMBH} can form and grow through mergers with other \acp{BH}\footnote{The term \acp{IMBH} first began to appear in published papers from the late 1990s that were predicting the type of merging \acp{BH} that could potentially be observed with future \ac{GW} missions \citep[e.g.,][]{1997CQGra..14.1425S,1998PhRvD..57.4535F}.} 

Several promising \ac{IMBH} candidates with masses ranging between $10^3$ to $10^{5}$ $\msun$ have also been identified. For instance, \citet{2022ApJ...924...48P} carried out high-resolution kinematic observations of an extragalactic star cluster in the Andromeda (M31) galaxy. Dynamical modelling of the observed data strongly suggests that there is a $10^{5}$ $\msun$ \ac{IMBH} in the center of this cluster. Other observational properties of this star cluster strongly suggest that it is a tidally stripped nucleus of a once more massive galaxy \citep{2022ApJ...924...48P}.  Recently, another \ac{IMBH} candidate was indirectly identified through a luminous X-ray burst caused by the tidal disruption of a star by a 20,000 $\msun$ \ac{IMBH} in an extragalactic star cluster \citep{2018NatAs...2..656L,2021ApJ...918...46W}. More details on observational methods to identify IMBHs can be found in {\color{blue}\textbf{Chapter 2b}}.

\begin{center}\fbox{%
    \parbox{0.95\textwidth}{
    {An \textbf{Intermediate-mass black hole (IMBH)}} is defined as having a mass in the range of $\sim 100-10^5 \mathrm{M}_{\odot}$}}\end{center}

\subsection{Relevance}\label{sec:1.2}

Whether \acp{IMBH} exist and how they may form and grow is particularly relevant for understanding the origin of \acp{SMBH} in our Universe (this is discussed in more detail in {\color{blue} \textbf{Chapter 3}}). If \acp{SMBH} formed from the growth of lower mass \acp{BH} then that means at some stage their masses were in the \ac{IMBH} mass range \citep{2012RPPh...75l4901V,2020ARA&A..58..257G}. Therefore, understanding potential pathways for forming and growing \acp{IMBH} can shed light on the origin of \acp{SMBH}. More broadly, understanding how \acp{BH} grow is relevant for a diverse range of theoretical and observational research areas that include \ac{GW} astronomy, high-energy electromagnetic astrophysics, and the formation and evolution of galaxies and their stellar structures. For this reason, \acp{IMBH} are an active topic of research and several comprehensive review papers have focused on them from a theoretical and observational perspective \citep[e.g.,][]{2004IJMPD..13....1M,2004cbhg.symp...37V,2016IAUS..312..181L,2017mbhe.confE..51K,2017IJMPD..2630021M,2020ARA&A..58..257G}.

This chapter describes proposed theoretical formation and growth channels (see Section \ref{sec:Channels}) for \acp{IMBH} and their relevance for \ac{GW} astronomy (see Section \ref{sec:gws}). In particular, it focuses on how \acp{BH} can potentially form and grow within dynamically active environments like star clusters (see Section \ref{subsec:clusters}). \acp{BH} can become more massive by merging with other \acp{BH}. Such mergers produce \acp{GW} and in the context of \acp{IMBH}, they may play a role in potentially seeding or contributing to the growth of \acp{SMBH}. Binary \ac{IMBH} formation and evolution (see Section \ref{subsec:binary-imbh}) and \ac{IMBH} mergers with \acp{SMBH} in galactic nuclei (see Section \ref{subsec:heavy-imri}) are also discussed. Observing these mergers using \acp{GW} and their detectability with future \acp{GW} detectors like \ac{LISA} is also discussed (see Section \ref{sec:gws}).

\section{Formation pathways of IMBHs}
\label{sec:Channels}

In this subsection, various proposed theoretical channels for \ac{IMBH} formation are reviewed. In subsection \ref{sec:popIII}), the formation of \ac{IMBH} through the stellar evolution of massive metal-poor stars in the very early Universe is very briefly discussed. In subsection \ref{subsec:clusters}, the different pathways by which \acp{IMBH} can form and grow within dense star clusters are explained. In subsection \ref{subsec:agn-disks}, \ac{IMBH} formation within accretion disks around \acp{SMBH} is very briefly discussed.




\subsection{Evolutionary remnants of massive metal-free and metal-poor stars}\label{sec:popIII}

It has been hypothesized that low-mass \acp{IMBH} of the order of $\sim 100$ to $1000 \ \msun$ could form through the stellar evolution of metal-free (population III) and metal-poor (population II) \acp{VMS} with \ac{ZAMS} masses $\gtrsim 200 \ \msun$. Very massive population III (pop III) stars are postulated to be born in the early Universe (at redshift $z \gtrsim 10$) \citep{2001ApJ...551L..27M,2013RPPh...76k2901B}. These first stars are expected to be nearly devoid of elements heavier than hydrogen and helium \citep{2004ARA&A..42...79B}. Absence of heavier elements in their protoclouds makes cooling inefficient and as a result pop III star can have larger birth masses compared to stars that form later in the Universe \citep{2001ApJ...548...19N}. Compared to massive relatively metal-rich stars, pop III stars are not expected to lose a significant amount of their initial mass through stellar winds during their evolution \citep{2007ApJ...654...66O,2020MNRAS.495.4170T}. Therefore, as they approach the end of their lives (after a few million years of evolution), they can be sufficiently massive to directly collapse to form \acp{BH} that have masses $\gtrsim 200 \msun$ \citep{2001ApJ...551L..27M,2003A&A...399..617M,2021MNRAS.505L..69H}. This pathway could result in the formation of low-mass \acp{IMBH} ($\lesssim 10^3 \ \msun$) in the early Universe. However, due to a dearth of observations and limitations of numerical simulations there are many uncertainties in formation, birth environment, mass function, multiplicity and evolution of pop III stars \citep{2015ComAC...2....3G,2017MNRAS.469..448B,2020MNRAS.495.4170T}. For instance, numerical simulations investigating pop III star formation predict a wide range of birth masses and maximum initial stellar mass that range from a few to up to thousand solar masses \citep[e.g.,][]{2002Sci...295...93A,2002ApJ...567..532H,2011ApJ...727..110C,2014ApJ...781...60H,2015MNRAS.448..568H,2020OJAp....3E..15R,2022ApJ...925...28L}. These results pertaining to the massive end of the pop III \ac{IMF} are sensitive to the treatment of key physical processes which govern star formation that include radiative feedback, gas accretion, cooling and fragmentation \citep{2012MNRAS.422..290S,2014ApJ...792...32S,2016MNRAS.462.1307S,2023MNRAS.518.1601T}. Additionally, pop III stars may also form in multiple stellar systems or star clusters. The evolution of binary systems comprising pop III stars may also lead to the formation of an IMBH \citep{2020ApJ...903L..40L,2021MNRAS.505.2170T,2021MNRAS.505L..69H}. Dynamics within pop III star clusters may also contribute to the formation and growth of an \ac{IMBH} \citep{2017MNRAS.472.1677S,2021MNRAS.506.5451L,2022MNRAS.515.5106W,2023arXiv231105393L}. The pathways that can lead to \ac{IMBH} formation within dense star clusters are covered in Section \ref{subsec:clusters}. 


Theoretical and numerical studies of the evolution of metal-poor ($\rm Z \lesssim 10^{-3}$ or $\lesssim 5\%$ of solar metallicity ($\rm Z_{\odot}$)) \ac{VMS}s ($\rm M_{\mathrm{ZAMS}} \gtrsim 250 \mathrm{M}_{\odot}$) population II (pop II) stars also suggest that they may evolve and directly collapse at the end of their lives to form an \ac{IMBH} in excess of $200 \ \msun$ \citep{2017MNRAS.470.4739S}. These pop II stars are slightly more metal-rich than pop III stars, the ones with metallicities $\lesssim 5\%$ of $Z_{\odot}$ also do not lose a significant fraction of their initial mass due to weaker stellar winds during the different phases of their evolution \citep{2001A&A...369..574V,2005A&A...442..587V,2021MNRAS.504.2051V}. Therefore, as these massive metal-poor stars approach the end of their lives, they can avoid exploding as supernova (SN) and can directly collapse to form low-mass \acp{IMBH} \citep{2017MNRAS.470.4739S}.

While this is the simplest way to form an \ac{IMBH}, there are many uncertainties surrounding the key processes involved in the evolution of these massive stars (see Section \ref{subsec:bh-formation-kicks-retention} for details). Among others, this includes the exact dependence of the strength of the stellar winds on the mass and chemical composition of the star, whether or not a SN can occur and what are the final BH masses produced through the evolution of \ac{VMS} \citep{2012ApJ...749...91F,2017ApJ...836..244W,2021hgwa.bookE..16M}. Furthermore, there is no concrete observational evidence to suggest that the star formation process can produce single \ac{VMS} with $\rm M_{\mathrm{ZAMS}} \gtrsim 250$. Observations of young star clusters and star forming regions, such as the Arches Cluster in our Galaxy and R136 star cluster in the Large Magellanic Cloud suggest that the upper mass limit of newly born stars ranges between $\sim 150$ to $\sim 250 \ \msun$ \citep{2005Natur.434..192F,2022ApJ...935..162K,2022A&A...663A..36B}. This suggests that metal-poor \ac{VMS} that could be potential progenitors of low-mass \acp{IMBH} could be very rare in our Universe. However, there is a possibility that such stars could form through multiple stellar collisions in dense stellar environments like cores of globular and young massive clusters (see Section \ref{subsec:fast-runway} for details). Additionally, evolution of two or more massive stars in close binary or multiple stellar systems could also lead to the formation of a \ac{VMS} that could be an \ac{IMBH} progenitor (see Section \ref{subsec:mergers-in-binaries} for details).
 
\paragraph{Accretion driven growth of stellar-mass BHs that formed from  pop III and metal-poor stars} \label{subsec:popIII}

\acp{sBH} or low-mass \acp{IMBH} that form via the evolution of massive pop III or metal-poor stars could significantly grow by accreting surrounding gas \citep{2001ApJ...551L..27M,2012NewAR..56...93A,2014Sci...345.1330A,2014ApJ...784L..38M}.
This growth is governed by how much gas is available and how efficiently these \acp{BH} can accrete surrounding gas \citep{2005ApJ...633..624V,2006MNRAS.373L..90K}. Seed \acp{BH} of around $100 \ \msun$ that may form via the evolution of pop III stars might be able to grow at Eddington or super-Eddington rates within gas-rich environments. More details on how pop III stars can contribute to the SMBH seeding and growth are discussed in {\color{blue} \textbf{Chapter 3.IV}} \citep[see also][]{2020ARA&A..58...27I,2023MNRAS.518.4672S}. For a \ac{BH} accreting at the Eddington limit, it's e-folding time is about 43 Myr \citep{2014ApJ...784L..38M}. Therefore, it might be possible to grow seed \acp{BH} into \acp{IMBH} larger than $10^{3} \ \msun$ within few hundred million years. Results from hydrodynamical simulations of \acp{BH} embedded in dense metal-poor gas clouds demonstrate that hyper-Eddington gas accretion on to the \ac{BH} may result in its rapid growth \citep{2016MNRAS.459.3738I,2016MNRAS.460.4122R,2023MNRAS.518.3606S}. Gas accretion may also be a viable way of growing \acp{sBH} within dense stellar and gas-rich environments \citep[e.g.,][]{2010ApJ...713L..41V,2013MNRAS.429.2997L,2021MNRAS.501.1413N}. This is covered in more detail in Section \ref{subsec:gas-accretion}. Furthermore, within dense environments, low-mass seed \acp{IMBH} may also grow by merging with other \acp{BH} \citep{2003ApJ...582..559V,2022MNRAS.511.2631A}. This is covered in more detail in Section \ref{subsec:hierarchial-mergers}.






\subsection{IMBH formation in dense stellar environments} \label{subsec:clusters}

This section provides a comprehensive overview of the proposed pathways by which an \ac{IMBH} can form within dense stellar environments.
Most stars, including massive ones that are progenitors of \acp{sBH} are observed to be born in stellar clusters and associations \citep[e.g.,][and references therein]{2003ARA&A..41...57L,2004MNRAS.349..735B,2005ASSL..324...87T}. Star formation occurs within giant molecular clouds (GMCs) that can have typical masses ranging from $10^{4}$ to $\gtrsim 10^{7} \ \msun$ \citep{2007ARA&A..45..565M,2012MNRAS.427..688L}. As gas cools within these clouds, the gas pressure acting against gravity decreases and the cloud becomes unstable to gravitational collapse. Local regions of high density within these molecular clouds fragment and eventually form multiple stars in gravitationally bound stellar clusters. In the cores of some of these star clusters (e.g., globular clusters or nuclear star clusters), the density of stars can be up to a million times higher than the density of stars in the solar neighbourhood. Due to the high stellar densities in these environments, stars can frequently interact with each other due to gravity. These interactions between stars drive the dynamical evolution of most star clusters, making them ideal laboratories for understanding how self-gravitating systems evolve. The fate of these star clusters depends on a variety of processes occurring within them as well as external factors (see Section \ref{subsec:dynamical-evolution-clusters}). The interplay between dynamical interactions and stellar/binary evolution makes dense star clusters unique sites for the production of various exotic stellar objects \citep[][see also {\color{blue} \textbf{Chapter 1.IV}}]{2013LRR....16....4B,2013pss5.book..879D}. Furthermore, these close dynamical encounters between stars can also facilitate the formation and growth of \acp{IMBH}.

Star clusters come in different varieties and are classified according to their ages, shape/size and location. Below we discuss three classifications of star clusters that are particularly relevant from the point of view of \acp{IMBH} formation and describe their key characteristics. 

\begin{itemize}
    \item  {\textbf{\acp{GC}}} are dense and spherical star clusters with ages that range from a few to up to 13 Gyr. \acp{GC} orbit their host galaxies and have been observed around extragalactic galaxies of all morphological types \citep{2006ARA&A..44..193B}. The specific frequency of \acp{GC} correlates with the mass of their host galaxy \citep{2013ApJ...772...82H}. Roughly 150 \acp{GC} have been observed in our Galaxy \citep[][updated 2010]{1996AJ....112.1487H}. Most of these have present-day masses of $\gtrsim 10^{4} \ \msun$ to up to $\sim 10^{6} \ \msun$. \acp{GC} have a characteristic core-halo structure with a dense and bright central core surrounded by a less dense outer halo. The central densities ($\rho_c$) of \acp{GC} can be as high as $\sim10^5 \ \msun \ \mathrm{pc}^{-3}$. Galactic GCs come in different sizes and have a median core radius of about 1 pc and diameters that range from a few to several tens of pc \citep{2013LRR....16....4B}. Based on the present-day properties of Galactic GCs, their central escape velocities range from few tens to up to a hundred $\kms$ \citep{2002ApJ...568L..23G,2018MNRAS.478.1520B,2020PASA...37...46B}. While in the local Universe GCs only make up a fraction of a galaxy's total stellar mass \citep[$\lesssim 1\%$][]{2013ApJ...772...82H}, it has been suggested that this fraction may have been as high as $10\%$ at redshift ($z$) $\gtrsim 3$ when the Universe was about 2 billion years old \citep{2010ApJ...718.1266M}. A substantial number of these GCs may have dissolved due to internal dynamical processes and the influence of the galactic tidal field \citep{1990ApJ...351..121C,1997ApJ...474..223G}.

    \item  {\textbf{\acp{NSC}}} are massive ($\rm \sim 10^{5}-10^{9}\ M_{\odot}$) and dense star clusters ($\rm \sim 10^{6}-10^{7}\ M_{\odot} \rm \ pc^{-3}$) located in the nuclei of most observed galaxies. Up to about 80\% of galaxies with stellar masses between $10^{8}$ to $10^{11} \ \msun$ have an \ac{NSC} in their centers and these are the densest star clusters in the local Universe \cite[see][and references therein]{2020A&ARv..28....4N}. Often an \ac{NSC} co-exists with an \ac{SMBH}. For instance, the \ac{NSC} in our Galaxy has an estimated mass of about $2 \times 10^{7} \ \msun$ \citep{2017MNRAS.466.4040F} and at its center is the \ac{SMBH}, Sagittarius A* which has a mass of about $4 \times 10^{6} \ \msun$ \citep{2019A&A...625L..10G,2019Sci...365..664D}. There are mainly two proposed scenarios for the formation and growth of NSCs: (1) through the migration and merger of star clusters in the galactic nucleus \citep[e.g.,][]{1975ApJ...196..407T,1993ApJ...415..616C,2012ApJ...750..111A}, and (2) through the in-situ star formation from high density gas in the galactic center \citep[e.g.,][]{1982A&A...105..342L,2004ApJ...605L..13M}. Observations of \acp{NSC} in galaxies with stellar masses in excess of $10^{9} \ \msun$ (e.g., the Milky Way) exhibit an age and metallicity spread of stars which suggests that the growth of the \acp{NSC} occurred through multiple episodes of star formation. It is likely that both in-situ star formation and star cluster mergers contribute to the growth of the \ac{NSC} \citep{2020A&ARv..28....4N}. Given their very high central densities, central escape velocities from \ac{NSC} may range from a few hundred to several hundreds of $\kms$ \citep{2019PhRvD.100d1301G,2019MNRAS.486.5008A,2020MNRAS.498.4591F}. 
    
    \item {\textbf{\acp{YMC}}} are young (ages that are typically $\lesssim 100 \ \mathrm{Myr}$) and massive star clusters ($\gtrsim 10^{4} \ \msun$). A significant fraction of stars are born in young clusters within star forming regions \citep{2003ARA&A..41...57L,2009A&A...498L..37P}. Most of these are open clusters (OCs) with masses ($\lesssim 10^{3} \ \msun$). Many of these young clusters can become unbound in relatively short timescales (few to tens of Myr) due to a number of different mechanisms which include residual gas expulsion \citep{2007MNRAS.380.1589B}, two-body relaxation, external tidal fields and shocks from disk or bulge passages or passing giant molecular clouds \citep{1997ApJ...474..223G,2011MNRAS.413.2509G,2019ARA&A..57..227K}. Dissolution of these clusters results in stars being expelled into the field. YMCs are massive and denser ($\rm \gtrsim 10^{3} \ M_{\odot} \rm \ pc^{-3}$) compared to OCs and may survive for longer times. 
\end{itemize}

\paragraph{Dynamical evolution of star clusters}\label{subsec:dynamical-evolution-clusters}

\begin{figure}[!htb]
\centering
\includegraphics[scale=0.2]{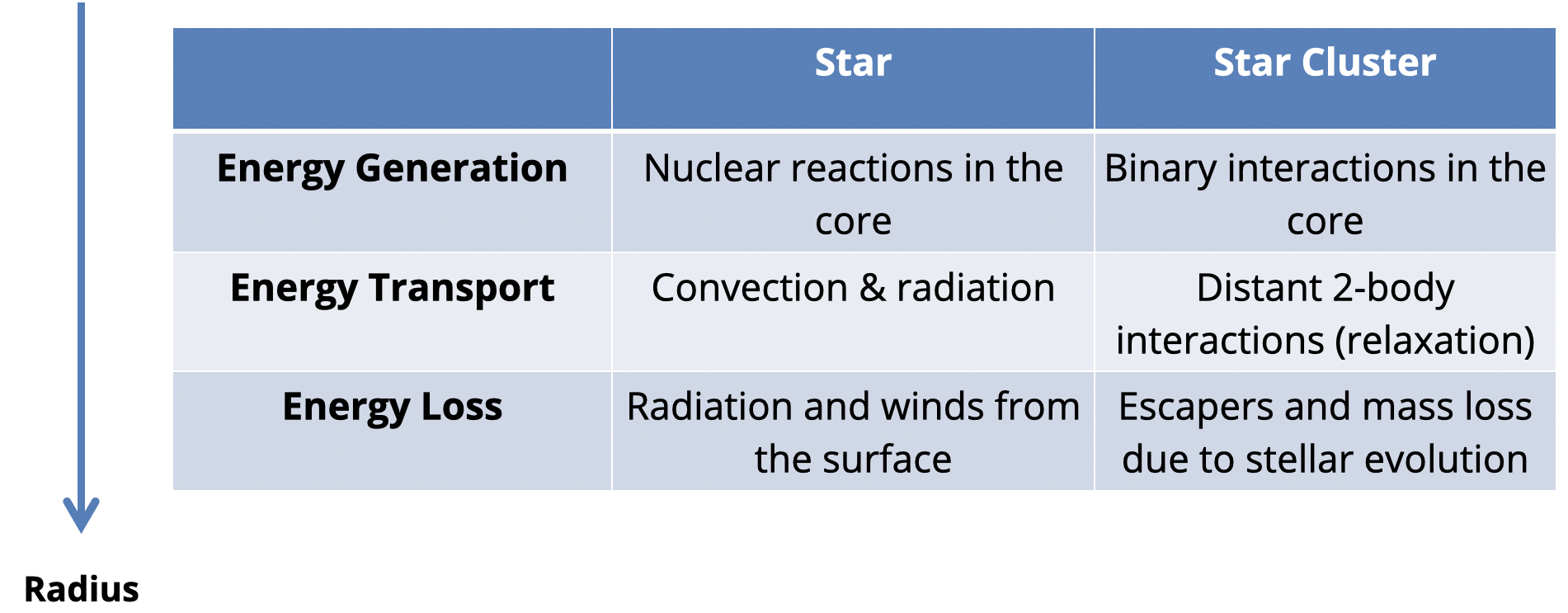}
\caption{Analogous physical processes between the evolution of a star cluster and the evolution of stars is shown. In order to support themselves against gravitational collapse, both stars and star clusters need to generate energy. Stars generate energy through nuclear reactions in their core whereas star clusters generate energy via strong interactions between binaries and other stars in the core. Stars transport energy via convection and radiation while in stellar clusters the main
mechanism for energy transport is distant two-body encounters. Stars lose energy through stellar winds and radiation from their surface. Star clusters also lose energy through stars that gain enough kinetic energy to escape the cluster and also through the evolution of stars that are within the cluster (Fig. idea credit: \citet{2016EAS....80...39B})}
\label{analogy-fig}
\end{figure}

This section briefly describes how dense star clusters (particularly GCs) evolve and what are the important dynamical processes that take place within them. Stellar clusters are in many ways similar to stars as both are self-gravitating systems that contain interacting particles. Fig. \ref{analogy-fig} shows how stars and star clusters have analogous physical processes for energy generation, transport and energy loss. For spherical self-gravitating systems like GCs to be stable, the inward pull of gravity has to be supported by the outward pressure that comes from the velocity dispersion of stars on random orbits (similar to how gas pressure supports stars on the main sequence from gravitational collapse). Such self-gravitating systems obey the virial theorem:

\begin{equation}\label{virial-eq}
2T+W=0
\end{equation}

where T is the total kinetic energy of the system and W is the total gravitational potential energy. 

Equation \ref{virial-eq} states that for a star cluster to be in virial equilibrium, the total kinetic energy ($T$) of the system should be equal to minus half times the total gravitational potential energy ($W$). Strong and weak gravitational encounters play a significant role in the evolution of star clusters and its structural properties. There are three characteristic radii which define the structure of a typical star cluster: the core radius ($r_{c}$), half-mass radius ($r_{h}$) and tidal radius ($r_{t}$). The half-mass radius is the radius within which half of the cluster mass is enclosed and the tidal radius defines the distance where the gravitational influence of the host galaxy balances the gravitational force of the cluster itself. Beyond this radius the gravitational field of the galaxy becomes more important than that of the star cluster. The core radius of a star cluster is typically defined as the radius at which the central surface density or surface brightness of the GC becomes half of its central value. 

As described above, two-body encounters between stars drives the dynamical evolution of a star cluster. The timescale over which these two-body encounters become significant is known as the relaxation time ($t_{\rm relax}$). The relaxation timescale of a spherical self-gravitating system is proportional to the velocity dispersion of the stars and is inversely proportional to its mass and density. A stellar system is described as being collisional (evolution driven by two-body relaxation) if its relaxation time is comparable or less than its age. The core of a typical GC has a relaxation time that is of the order of a few hundred million years and its relaxation time is about several hundred million to a billion years. Therefore, relaxation is the most important process that governs the long-term evolution of star clusters like GCs that are nearly as old as the universe. The half-mass relaxation time of a star cluster is given in Equation \ref{relax-eq} \citep{1987degc.book.....S}:
\begin{equation}\label{relax-eq}
{t_{{\rm{rh}}}} \sim \frac{0.138{N^{1/2}}{r_{h}^{3/2}}}{{{\langle m_{\star}\rangle}^{1/2}}{G^{1/2}}\ln (\gamma N)}
\end{equation}
where $N$ is the total number of stars inside the half-mass radius ($r_{h}$) and $m_{\star}$ is their average mass. The $\ln(\gamma N)$ is known as the Coulomb logarithm and it is a factor which describes the relative effectiveness of small and large angle encounters for a star cluster with given density and velocity distributions \citep{2013LRR....16....4B}. The relaxation time of a GC is significantly longer than its crossing time ($t_{c}$). As the name suggests, crossing time is the time needed for a star to cross a given characteristic cluster radii which would be equal to that radius divided by the velocity of the star. 

The crossing time can be related to cluster radius $R$ and the velocity dispersion of stars $\sigma$ \citep[e.g.,][]{2008gady.book.....B,2013LRR....16....4B} via 
\begin{equation}\label{eq:cross}
    t_{\rm{cross}} = \frac{R}{\sigma}.
\end{equation}
This crossing time relates to the cluster relaxation time through \citep[see Eq 1.38 and Eq 7.1 in][]{2008gady.book.....B}:  
    \begin{equation}\label{eq:relaxation-time}
    t_{\rm{rlx}} \simeq n_{\text {cross }} t_{\text {cross }} \simeq \frac{0.1 N}{\ln N} t_{\mathrm{cross}}
\end{equation}
where $N$ denotes the number of objects in the cluster. The relaxation time of a cluster can also be estimated using the following expression from \citet{2010ARA&A..48..431P}:

\begin{equation}\label{eq:relaxation2}
    t_{\rm{rlx}} \sim 15 \operatorname{Myr}\left(\frac{M}{10^4 M_{\odot}}\right)^{1 / 2}\left(\frac{R}{1 \mathrm{pc}}\right)^{3 / 2}\left(\frac{1 M_{\odot}}{m}\right)
\end{equation}
Where $M$ is the total cluster mass, $R$ is cluster radius and $m$ is the average mass of stars in the cluster.
Most GCs have velocity dispersion of of the order of $10 \ \kms$ with half-mass radii ranging from a few to up 10 pc.
The ratio between the relaxation time and the crossing time is proportional to the number of stars in the system.

Two-body relaxation is responsible for transporting energy within the cluster. Through relaxation, energy from stars in the cluster core is dissipated to stars that are in the halo. There are three very important consequences of two-body relaxation in the context of GC evolution.

\begin{itemize}

\item Evaporation: As kinetic energy is exchanged between stars due to two-body relaxation, it can lead to stars evaporating from the cluster. If a star gains enough kinetic energy, it can become unbound and escape from the cluster. For typical GCs, the time to evaporate completely is of the order of several tens to a hundred relaxation times. However, evaporation can be accelerated due to a strong tidal field or tidal shocks experienced by the GC due to passages through a galactic disk. 

\item Mass Segregation: Another physical process that plays an important role in the dynamical evolution of a GC is mass segregation. This process is a consequence of energy equipartition that results from two-body relaxation. Energy equipartition requires that stars in GCs would evolve to have roughly the same kinetic energy throughout the cluster. This means that massive stars would have lower average velocities than lower mass stars that would have higher velocities. As more massive stars will slow down effectively due to two-body relaxation, they will move slower and will sink deeper in the cluster potential while most low mass stars will move faster. Due to this process, as GCs evolve, the massive stars will quickly segregate to the cluster center while the lowest mass stars will occupy the halo.  

\item Core Collapse: Self-gravitating systems have a negative heat capacity, so as the stars in the core of the cluster dissipate kinetic energy by moving from the core to the halo, the cluster core begins to contract. As the core of the cluster will contract, the velocities of the stars in the core will increase to maintain the kinetic energy required for the system to stay in virial equilibrium. This will lead to faster dissipation of energy and as a result the cluster core will keep on contracting while its density will become extremely high. This process is known as core collapse and this continues until it is counteracted by the formation of hard binaries (can be dynamically formed in three-body encounters or can also be initial binaries\footnote{This refers to primordial binaries present in the cluster at birth.}) at the center of GCs. Binary systems are classified as hard or soft depending on their binding energy. Typically, hard binaries are those in which the component stars have orbital velocities larger than the velocity dispersion of stars in the GC. Therefore, the binding energy of hard binaries (left hand side of Eq. \ref{eq:hard-binary}) is larger than the mean kinetic energy of cluster stars (right hand side of Eq. \ref{eq:hard-binary}):
\begin{equation}\label{eq:hard-binary}
    \frac{G m_{1} m_{2}}{2 a}>\frac{1}{2}\langle m\rangle \sigma^{2}
\end{equation}
where $m_{1}$ and $m_{2}$ are masses of the two binary components, $a$ is the semi-major axis of the binary, $\langle m\rangle$ is the average mass of stars and $\sigma$ is the velocity dispersion.
These hard binaries prevent further cluster collapse and act as an energy source by exchanging their binding energy with surrounding stars in three-body encounters. The timescale for core collapse depends on the half-mass relaxation time of the cluster. Those with shorter half-mass relaxation time will evolve relatively quickly towards core collapse compared to ones with longer half-mass relaxation time. 

\end{itemize}

Strong dynamical interactions between multiple stars can also play an important role in the GC evolution. As densities become large during core collapse of GC, strong interactions between three stars can lead to the formation of binary systems. While there are many possible outcomes of strong dynamical interactions between binary-single stars and binary-binary stars, on average hard binaries in GCs are more likely to get harder due to strong interactions and soft binaries (binary systems in which stars have orbital velocities lower than the velocity dispersion of stars in the GC) tend to get softer. This is known as the Heggie-Hills law \citep{1975MNRAS.173..729H,1975AJ.....80..809H}. As a consequence of this law, many strong dynamical interactions can lead to the mergers of stars in a hard binary system and the disruption of soft binaries.





There are external factors which are also important for the dynamical evolution of GCs. This includes the tidal evaporation of the GC due to the galactic tidal field. As discussed above, GCs closer to the galactic disk, bar or bulge can lose a significant amount of mass due to tidal evaporation. Massive GCs also sink towards the center of their host galaxy due to dynamical friction. It is possible that massive GCs that may have formed close to the center of their galaxy may have merged with the galactic center within a few billion years due to mass segregation.

Also, the dynamical evolution of a GC is dependent on the stellar and binary evolution of stars within the GC. During the very early evolution of a GC, massive stars will quickly lose mass due to stellar evolution which will result in significant mass loss from the cluster. From the discussion above, we know that binaries play a pivotal role in the evolution of a GC, so the evolution of binaries as well as the population of initial binaries in GC can have a significant impact on the dynamical evolution of the clusters.

Uncertainties concerning the exact state of GCs at their formation (e.g. the distribution of position and velocities of stars within the cluster, their structural parameters, primordial binary fraction etc) adds to the complexity of understanding and properly modelling the evolution of GCs with time. Inclusion of external factors that can also affect dynamical evolution of a cluster and the uncertainties in those factors (e.g., an evolving tidal field of a galaxy) also further complicates the problem. 

\paragraph{Modelling the evolution of star clusters}\label{subsec:model}

As discussed in the previous section, the evolution of star clusters depends on a number of physical processes and external factors which makes it challenging to realistically model their evolution. There are a variety of ongoing evolutionary processes in the cluster that can have very different and discrepant times and length scales \citep{2003gmbp.book.....H,2023LRCA....9....3S}. For instance, it is essential to accurately model close encounters between three or more stars as this process results in the exchange of energy and angular momentum. Accurately modelling this process which occurs at much smaller time-length scales than the time-length for two-body relaxation is necessary for properly modelling the energy flow inside a star cluster. Stars of different masses are also evolving at different timescales during different stages of their stellar evolution. Physical processes in dense and massive star clusters can vary from time scales of minutes, hours and years to millions and billions of years. Realistic modelling of massive star clusters, like GCs, requires proper treatment of all these processes which are essential in driving their dynamical evolution.

\begin{figure}[hbt!]
\centering
\includegraphics[scale=0.3]{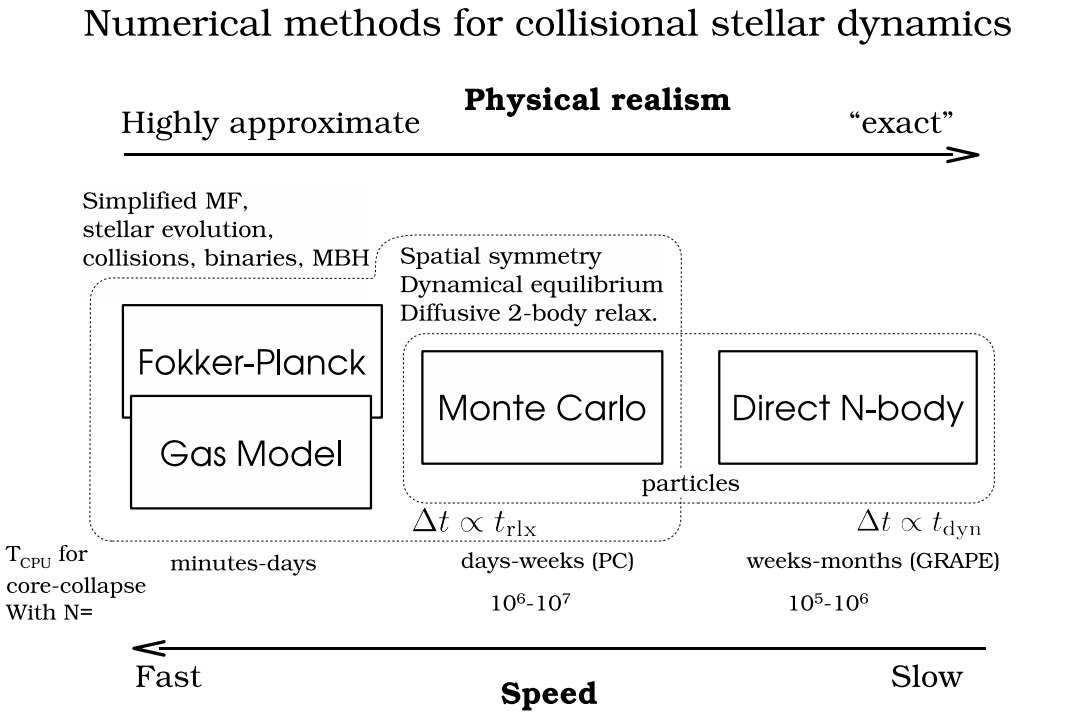}
\caption{Figure (taken from Fig. 3 in \citet{2007CQGra..24R.113A}) illustrates the various numerical methods that are used to model the evolution of dense stellar systems like GCs. Scales for physical realism and computational speed for each method are also shown. }
\label{comparison-fig}
\end{figure}

With advancements in computational technology over the last few decades, extensive work has been done in modelling the dynamical evolution of star clusters using different numerical algorithms in extensive simulation codes, particularly direct \textit{N}-body and Monte Carlo codes. These codes need to couple algorithms for computing the dynamical evolution of stars with prescriptions for stellar, binary evolution and the influence of a tidal field. For simply evolving the dynamics of star clusters, there are essentially four different methods \citep{2013pss5.book..879D,2015ASSL..413..225M, 2016EAS....80...39B}: direct \textit{N}-body calculations, Fokker-Planck codes, gas models and Monte Carlo \textit{N}-body codes. Each approach differs in terms of computational speed and physical realism that it provides, as is shown in Fig. \ref{comparison-fig} \citep{2007CQGra..24R.113A}. All approaches start with an initial distribution for positions and velocities of stars given according to an initial model such as an isothermal sphere, the \citet{1911MNRAS..71..460P} model or the \citet{1966AJ.....71...64K} model. This distribution of stars is then evolved in time.

In this chapter, the direct \textit{N}-body and Monte Carlo methods are described in more detail. These two approaches are the most prevalent and well-suited for realistically modelling the evolution of a dense star cluster that are most likely to form an \ac{IMBH}. Many of the results presented in the Sections within this Chapter were obtained from star cluster models simulated using these methods.

{\textbf{Direct \textit{N}-body Codes}}:

In the direct \textit{N}-body approach, the gravitational force between each star is directly calculated by numerically solving the Newtonian equations of motion. By doing this, the acceleration on each star resulting from the gravitational force of all others can be computed. The position and velocity of the star can be advanced forward in time and the process is repeated to simulate the motion of stars over time. For a star in a cluster, $i$, the cumulative acceleration from all other stars (represented by the index ($j$) in the cluster is given by
\begin{equation}
\ddot{\mathbf{r}}_i=-\sum_{j \neq i} \frac{G m_j\left(\mathbf{r}_i-\mathbf{r}_j\right)}{\left|\mathbf{r}_i-\mathbf{r}_j\right|^3}
\end{equation}
where $m_{j}$ is the mass of star $j$ and $r_j - r_i$ is the distance between stars $i$ and $j$. 

This is  the most accurate and conceptually straightforward approach to properly treat the dynamical evolution of a GC. However, this is also the most computationally expensive approach. Computational time for \textit{N}-body simulations scales with $N^{3-4}$ \citep{2003gmbp.book.....H,2016EAS....80...39B,2023LRCA....9....3S} where $N$ is the number of stars in the GC. Even with parallelization and GPU acceleration, simulating the evolution of a realistic globular cluster with several hundred thousands to millions of stars up to a Hubble time can take from several months up to a year \citep{2016MNRAS.458.1450W}. Computational time further increases if the simulated GC contains a large number of initial binaries or has initially high central densities. The most commonly used \textit{N}-body code for computing the evolution of collisional systems are the NBODYX series of codes developed primarily by Aarseth over the last 50 years \citep{2003gnbs.book.....A,2008LNP...760.....A,2023LRCA....9....3S}. The latest in the NBODYX series of codes allows one to setup and simulate the evolution of a GC. The code has hardwired routines for carrying out stellar and binary evolution of stars. Significant efforts have been made to parallelize NBODY codes and this has led to the development of NBODY6++GPU code that makes use of accelerated hardware to improve computational time \citep{2015MNRAS.450.4070W,2022MNRAS.509.2075A}.

While the \textit{N}-body approach is conceptually straightforward, there are many complications and caveats involved in its actual implementation -- from using accurate integrators to carefully managing timesteps and computational techniques for different processes. Due to these reasons, \textit{N}-body codes that include all the processes that are important in driving the evolution of a GC are quite complicated. Simulating a large number of stars with the \textit{N}-body code remains computationally challenging and scaling results from simulations of low $N$ to large $N$ comes at the cost of physical accuracy especially when other physical processes, such as stellar and binary evolution are included \citep{2013LRR....16....4B}.   

{\textbf{Monte Carlo \textit{N}-body Codes}}:

Direct \textit{N}-body codes are considered to be very reliable and can handle the important physical processes needed to follow the long-term evolution of a star cluster. However, computationally they are too expensive to simulate initially massive and dense star clusters in a reasonable amount of time. Another approach that can be used to effectively model the evolution of large GCs is the Monte Carlo (MC) method for stellar dynamics. This method, which was first developed by H\'{e}non in late 1960s and early 1970s \citep{1971Ap&SS..13..284H,1971Ap&SS..14..151H} combines the statistical approach of the Fokker-Planck methods for treating two-body relaxation with the particle based approach of \textit{N}-body simulations. Although a number of limiting assumptions go into the MC method, its particle based approach allows for the implementation of additional real physical processes and effects which are not possible with methods that focus purely on evolving distribution functions.

In H\'{e}non's MC algorithm, at each time step the total potential of the system is calculated using the mass and radial position for each star. Using this global potential, an orbit-averaged approximation is used to generate plane-rosette orbits that are characterized by energy ($E$) and angular momentum ($J$). Each star is perturbed by a weak encounter and the parameters for this encounter are calculated between two neighbouring stars. Taking this encounter to be statistically representative of the encounters that the star will undergo over a timestep, the result of this one perturbation can be used to compute the total perturbation the star would receive over the timestep \citep{2013LRR....16....4B,2013pss5.book..879D}. Through this algorithm, energy and angular momentum of the orbit can be altered to mimic the effects of gravitational encounters using the theory of relaxation as a diffusive process. This means that the timestep of a MC code should be a fraction of the total relaxation time of the GC. It also requires that the timestep is longer than the crossing time (it was discussed in Section \ref{subsec:dynamical-evolution-clusters} that the relaxation time of a realistic GC is much longer than its crossing time). Once new orbits and the new positions and velocities of stars have been determined, the new potential is calculated at the next timestep. The main advantage of using the MC method is that in terms of computation it is significantly faster than the \textit{N}-body code, the computation time in the MC algorithm scales with the number of stars as $N \ln N$. This means that realistic GCs comprising $10^{6}$ stars can be simulated within a few days to a week. It also allows implementation of additional physical processes that are important in realistically simulating a GC with relative ease. However, there are also certain parameters in MC approach that need to be calibrated with the results of direct \textit{N}-body simulations. This highlights the importance of comparing MC results with the results from direct \textit{N}-body simulations \citep{2013MNRAS.431.2184G,2016MNRAS.463.2109R,2017MNRAS.470.1729M}.

Although the MC method is very fast, there are limitations and assumptions under which it can be used. MC method can only be applied to spherically symmetric non-rotating clusters. The method cannot be used reliably for low $N$ ($N \lesssim 10000$) systems. MC also assumes that the cluster is in dynamical equilibrium as discussed above. This allows one to take timesteps that are much larger than orbital timescales and are a fraction of the cluster's relaxation time. This assumption also implies that MC method is not suitable for computing the short-term evolution of star clusters or evolution involving violent relaxation. These are the underlying assumptions that need to be fulfilled in order to use the MC method. GCs are spherically symmetric systems (and do not show evidence for significant rotation) that contain a large number of stars and are long living. These properties of GCs make the MC method highly suitable for modelling their long-term evolution.

H\'{e}non's MC scheme was significantly improved in the early 1980s by Stod\'{o}{\l}kiewicz, who introduced proper treatment for binary stars within the MC scheme and also added prescriptions for stellar evolution and other physical processes to his MC code to compute the evolution of a realistic star clusters \citep{1982AcA....32...63S,1986AcA....36...19S}. The MC scheme was further improved and revived in the late 1990s by \citet{1998MNRAS.298.1239G,2001MNRAS.324..218G,2003MNRAS.343..781G}. Other efforts were also made in developing MC codes around the same time by \citet{2000ApJ...540..969J,2001ApJ...550..691J} and \citet{2001A&A...375..711F}. Currently, there are two MC codes that are being actively developed and are frequently used for simulating massive globular clusters: the MOCCA code \citep{2013MNRAS.429.1221H,2013MNRAS.431.2184G,2015MNRAS.454.3150G} and the CMC code \citep{2013ApJS..204...15P,2022ApJS..258...22R}. 

As the MC method is significantly faster than direct \textit{N}-body codes, it can be used to simulate the evolution of dense GCs containing millions of stars up to a Hubble time on the computational timescale of weeks. This opens up the possibility of simulating a large suite of star cluster models in which the initial parameters (e.g., the initial structure, central density, relaxation time, primordial binary fraction, tidal and galactocentric radii and BH natal kicks) can be probed to see how changing these parameters can influence the subsequent evolution of a massive star cluster and how they affect the formation and properties of stellar exotica, \ac{sBH} binaries and \acp{IMBH}. Examples of large suites of star cluster models simulated with MC codes include the MOCCA-Survey Database I \citep{2017MNRAS.464.3090A} and II \citep{2022MNRAS.514.5879M}, the CMC cluster catalog \citep{2020ApJS..247...48K} and CMC-FIRE-2 star cluster models \citep{2022arXiv220316547R}.

\paragraph{Formation and retention of stellar-mass BHs in star clusters: natal kicks and dynamics}\label{subsec:bh-formation-kicks-retention}

Before discussing the different formation channels of \acp{IMBH} in star clusters, we will briefly review what happens to \acp{sBH} and their progenitors within star clusters. In Section \ref{sec:1.1}, it was
explained that \acp{sBH} form when massive stars (with $\rm M_{\rm ZAMS} \gtrsim 15-20$) end their lives. According to a typical IMF \citep[e.g.,][]{1955ApJ...121..161S,2001MNRAS.322..231K,2003PASP..115..763C}, 1 to 2 in every 1000 stars would be massive enough to evolve into an \ac{sBH}. This means that a massive and dense star cluster which initially has about $10^{6}$ stars
must have at least a thousand sBH progenitors. The timescales over which these massive stars evolve into sBHs depend on their initial mass. Stars with $\rm M_{\rm ZAMS} \gtrsim 50 \ \msun$ are expected to evolve into BHs within a few million years, while those with  $\rm M_{\rm ZAMS} \sim 20 \ \msun$ evolve within few tens of millions of years (Myr). Within 30 Myr, all \ac{sBH} progenitors within the cluster are expected to have evolved and formed \ac{sBH}. 

As highlighted earlier, stellar evolution of \ac{sBH} progenitors is not fully understood. In addition to uncertain dependencies on properties of the star (e.g., mass, metallicity, rotation, surface magnetic fields) there are also very few constraints on parameters that govern the physical processes associated with the different phases of the evolution of single massive massive star (see {\color{blue}\textbf{Chapter 1.I}}). This includes nuclear reaction rates, convective overshooting, electron-scattering Eddington factor, mass loss due to stellar winds, and SN mechanisms. All these processes can have an important influence on the final \ac{sBH} mass produced through stellar evolution. The situation becomes even more complicated for massive stars in close binary systems (see {\color{blue}\textbf{Chapter 1.II}} and {\color{blue}\textbf{Chapter 1.III}}). Depending on initial separation and orbital eccentricity, physical processes in binary evolution like tidal dissipation, mass transfer, and common-envelope evolution can significantly affect the final mass of both binary and single \ac{sBH}. 

One important aspect of \ac{sBH} formation that is important with regards to their retention in star clusters is whether a newly formed \ac{sBH} receives a significant natal kick. If this natal kick exceeds the escape velocity of the star cluster then the newly formed \ac{sBH} would be ejected out of its birth cluster. Evidence from observations of proper motions of pulsars suggests that neutron stars that form via core-collapse SN receive a substantial natal kick following the SN explosion \citep{2005MNRAS.360..974H,2017A&A...608A..57V,2020MNRAS.494.3663I}. The distribution of the natal kick velocities inferred from these observations follow a bimodal Maxwellian distribution with mean kick velocities as high as high as few hundred $\kms$. The exact mechanism for these natal kicks is not known but theoretical studies and simulations hypothesize that they depend on the processes involved in the SN mechanism \citep[e.g., asymmetric mass ejection in the very early phases of SN ejection][]{2017ApJ...837...84J,2018MNRAS.480.5657B,2019MNRAS.484.3307M}. For \acp{sBH}, there has been debate as to whether they receive the same natal kick as neutron stars \citep[e.g.,][]{2012MNRAS.425.2799R,2013MNRAS.434.1355J,2017MNRAS.467..298R,2019MNRAS.489.3116A,2021ApJ...920..157C,2022ApJ...930..159A}. 

If the origin of the natal kick is connected with asymmetries in mass ejection during the supernova explosion then conservation of linear momentum would require that natal kicks of \acp{sBH} are scaled according to the \ac{sBH} mass\footnote{Which can be a few to up to 50 times more massive than a typical neutron star.} Numerical codes used to carry out stellar and binary evolution population synthesis, and to evolve stars within stellar cluster simulations often draw natal kicks randomly from the same distribution as neutron stars and then they are reduced accounting for the final mass of the \ac{sBH} \citep{2020ApJ...891..141G,2020MNRAS.499.3214M,2020ApJ...898...71B}. Additionally, several studies have also suggested that relatively massive \acp{sBH} that may form via direct collapse or a failed supernova may have zero or extremely low natal kicks \citep{2002ApJ...572..407B,2012ApJ...749...91F,2016ApJ...821...38S,2020A&A...639A..41B}. In the latter case, mass in the outer layers of an evolved star can fallback onto a newly formed \ac{sBH} which can result in reduced natal kicks and relatively small birth spin values \citep{2018ApJ...852L..19C,2021ApJ...914..140P}. In the latter case, the reduction in the natal kick depends on the mass fraction ($f_{\mathrm{fb}}$) of the stellar envelope that falls back on to the newly formed \ac{sBH}. For stellar/binary evolution population synthesis calculations that are employed within star cluster evolution codes, depending on the amount of fallback for a given SN mechanism, the 1D natal kick ($v_{\text {kick}}$) of an \ac{sBH} is often expressed as \citep{2012ApJ...749...91F}:

\begin{equation}
v_{\text {kick}}=\left(1-f_{\mathrm{fb}}\right) \sigma_{natal}
\label{eq:fallbackkick}
\end{equation}

where $\sigma_{natal}$ is drawn from a velocity distribution assuming that there is no fallback. The value of $f_{\mathrm{fb}}$ depends on the mass of the evolved core of the \ac{sBH} progenitor and the supernova prescription \citep[e.g., delayed explosion, rapid explosion][]{2012ApJ...749...91F,2019MNRAS.485..889S,2021A&A...645A...5S}. Many of these prescriptions predict that for evolved massive stars with carbon-oxygen core masses in excess of $11 \ \msun$, the fallback factor ($f_{\mathrm{fb}}$ ) value will be 1 \citep{2012ApJ...749...91F}, resulting in no natal kick (see Eq. \ref{eq:fallbackkick}). While the details of physical processes involved in different types of SNe are complicated \citep{2018SSRv..214...33B,2020ApJ...888...76M}., implified prescriptions are useful for carrying out rapid population synthesis studies and for evolving star cluster models.

If $v_{\text {kick}}$ is less than the escape speed ($v_{\text{esc}}$) of its birth star cluster at the time when the \ac{sBH} progenitor ended its life then the newly formed \ac{sBH} can be retained in the cluster. Therefore, the initial retention of an \ac{sBH} in a cluster depends on its natal kick ($v_{\text {kick}}$) and the central escape velocity ($v_{\text{esc}}$) of their birth environment which is determined by the central potential \citep{2013ApJ...763L..15M,2015ApJ...800....9M,2018MNRAS.479.4652A,2018A&A...617A..69P}. If all BHs get high natal kicks of the order of a few hundred $\rm km$ $\rm s^{-1}$ then they are extremely likely to escape the GC. Only dense GCs and NSCs with high escape velocities would be able to retain such sBHs. However, if BH natal kicks are low and of the order of few to several tens of $\rm km$ $\rm s^{-1}$ then a large number of BHs could be retained in a moderately dense GC following their formation from the evolution of massive stars. If sBHs form via a mechanism (e.g., direct collapse, failed supernova) that imparts very little or no natal kick on them then they can be retained even in loosely bound open clusters and stellar associations. The retention of at least a few sBHs in star clusters is particularly important for IMBH formation via repeated or hierarchical mergers of sBHs (see Section \ref{subsec:hierarchial-mergers}).

\begin{figure}[!htb]
\centering
\includegraphics[scale=0.35]{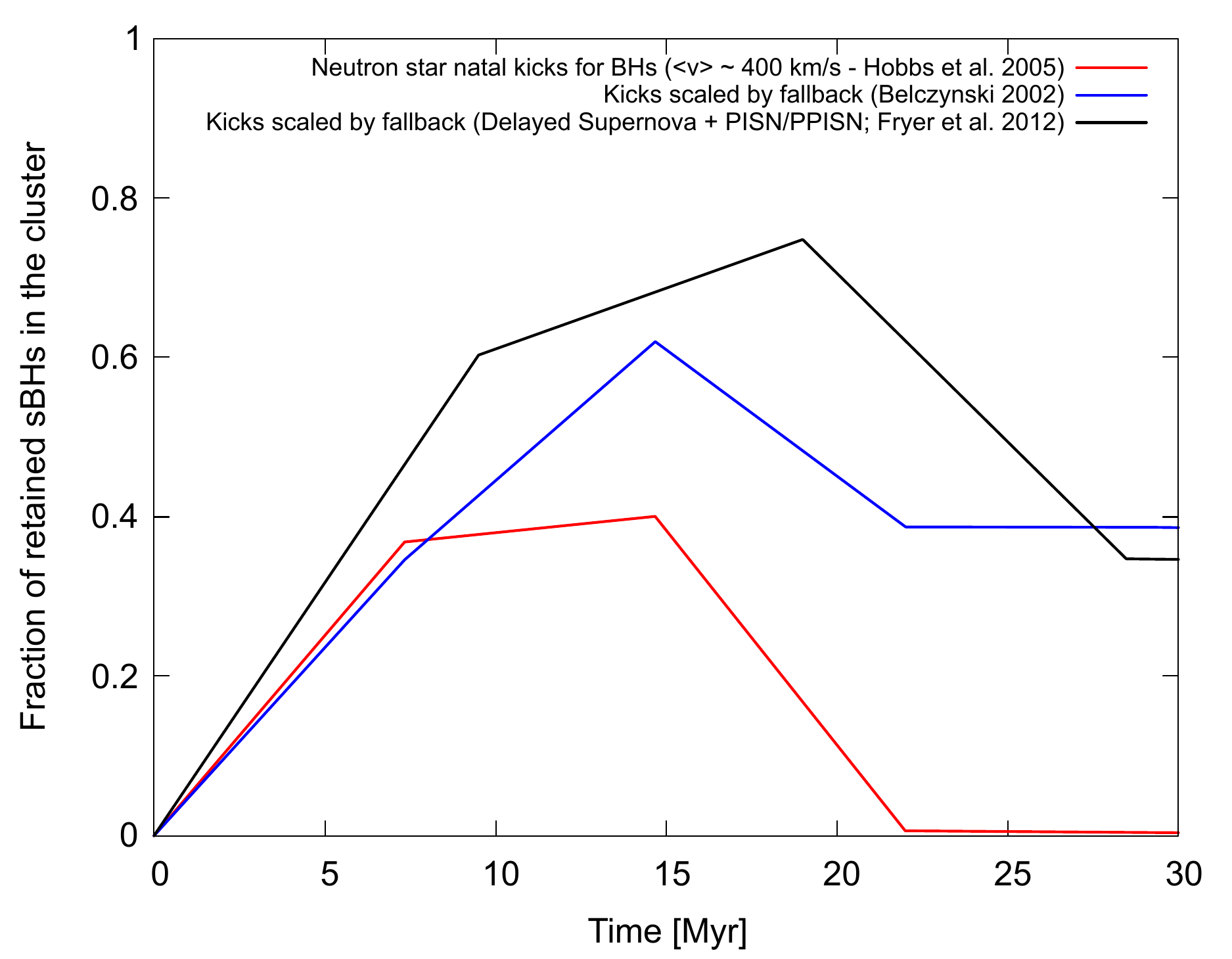}
\caption{Fraction of \acp{sBH} retained inside three star cluster models in the first 30 Myr of cluster evolution. The three clusters had the same initial number of objects ($N=7 \times 10^{5}$), binary fraction ($10\%$), half-mass (4.8 pc) and tidal (120 pc) radii, central density ($7.5 \times 10^{3} \ \msun \ \rm pc^{-3}$), escape velocity ($\sim 35 \ \kms$) and all stars had a metallicity ($\rm Z$) of $10^{-3}$. The main difference in the three simulations was connected to the way the natal kicks for \acp{sBH} were computed following the evolution of their progenitor star. For the model shown with the red line, \acp{sBH} were given natal kicks drawn from the same distribution as the one used for neutron stars \citep[derived from observed proper motions of pulsars;][]{2005MNRAS.360..974H}. For the models shown with blue and black lines, the natal kicks were modified according to the mass fallback (see Eq. \ref{eq:fallbackkick}) in which fallback factor was determined according to prescriptions by \citet{2002ApJ...572..407B,2012ApJ...749...91F}.}
\label{fig:bh-natal-kicks}
\end{figure}

The fraction of retained \acp{sBH} in the first 30 Myr of evolution for three star cluster models that were simulated using the MOCCA code (Monte Carlo \text{N}-body code) are shown in Fig. \ref{fig:bh-natal-kicks}. All three GC models had the same number of initial objects ($7 \times 10^{5}$), initial binary fraction ($10\%$), half-mass (4.8 pc), tidal radius (120 pc) and central density ($7.5 \times 10^{3} \ \msun \ \rm pc^{-3}$). The initial central escape speed from the cluster was about $35 \ \kms$. ZAMS masses for stars in all the three cluster models were sampled using the IMF from \citet{2001MNRAS.322..231K} with lower and upper mass limits of $0.08 \ \msun$ to $150 \ \msun$ and their metallicity was set to ($\rm Z=0.001$; $5\%$ of $\rm Z_{\odot}$). There were approximately 1800 \ac{sBH} progenitors in each of the three star cluster models. For each of these models, there were differences in the way \ac{sBH} natal kicks were computed. It can be seen from the Fig. \ref{fig:bh-natal-kicks} (see red line) that if the natal kicks for \ac{sBH} is drawn from the same distribution as that for neutron stars \citep{2005MNRAS.360..974H} then only a handful ($\sim 1\%$) of \acp{sBH} are retained in the cluster. In the two cluster models where the natal kicks were modified according to fallback prescriptions \citep[blue and black lines;][]{2002ApJ...572..407B,2012ApJ...749...91F} according to the formalism shown in Eq. \ref{eq:fallbackkick} then close to $42\%$ (around 750 BHs) of \acp{sBH} can be retained in the cluster after their formation. For star clusters, with larger initial escape velocities ($\sim 100 \ \kms$), the fraction of initially retained \acp{sBH} can be higher than $50\%$ \citep{2020IAUS..351..395A}. 

There has been considerable debate as to what would be the fate of \acp{sBH} if a significant fraction of them are retained in dense and massive star cluster (e.g, a typical GC). In the early nineties, theoretical studies \citep{1993Natur.364..421K,1993Natur.364..423S} had suggested that being the most massive objects in the GC, BHs would segregate to the center of the GC and form a subsystem that would dynamically decouple from the rest of the GC due to the Spitzer mass-segregation instability \citep{1969ApJ...158L.139S}. This instability occurs due to the inability to achieve energy equipartition when the total mass of the centrally segregated massive objects is comparable to the total mass of the lower mass surrounding objects. Due to this instability, the subsystem of sBHs would continue to contract and sBHs would eject each other due to strong dynamical interactions. \citet{1993Natur.364..421K,1993Natur.364..423S}  claimed that in a typical GC, such an sBH subsystem would deplete itself within a few hundred million years leaving behind at the most 1 or 2 stellar mass BHs in the GC. As the sBH subsystem depletes, a significant population of single BHs as well as sBHs in binary systems will be ejected from the GC. Many studies have postulated that these ejected binaries from GCs could be potential GW sources \citep[e.g.,][see {\color{blue} \textbf{Chapter 1.IV}} for details]{2000ApJ...528L..17P,2010MNRAS.407.1946D,2010MNRAS.402..371B,2016MNRAS.458.3075A}. 

\begin{figure}[!htb]
\centering
\includegraphics[scale=0.35]{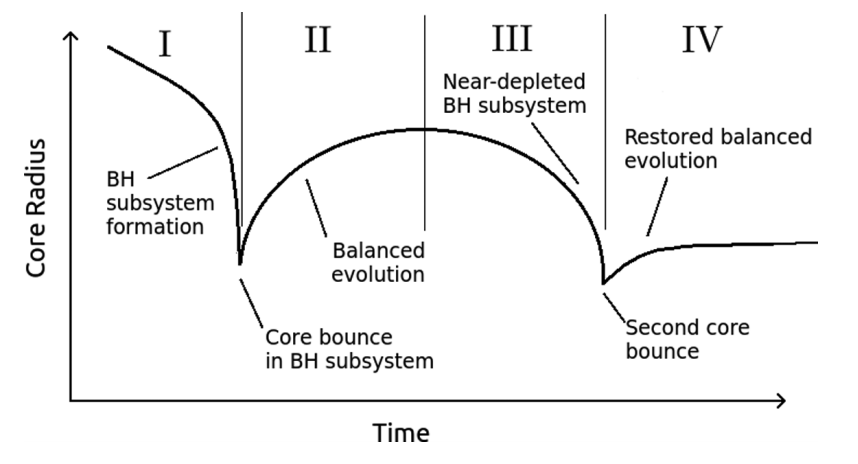}
\caption{This is a schematic illustration (taken from \citet{2013MNRAS.436..584B}) of the evolution of the core radius of a GC containing a subsystem of stellar mass BHs. The roman numerals mark 4 stages of the GC evolution. I: Initially the core of the cluster will evolve towards core collapse, as BHs segregate, they will form a subsystem that would arrest the core collapse. II: The subsystem comprising single and binary BHs will be an energy source for the GC and the core would start to expand. \citet{2013MNRAS.432.2779B,2013MNRAS.436..584B} showed that the duration of this stage would depend on the overall relaxation time of the whole GC. III: As the BH population depletes, the GC would start to collapse again. IV: The second core collapse will be halted by the formation of binaries of lower mass stars which would then supply energy to the GC.}
\label{fig:rc-bhs}
\end{figure}

In recent years, both numerical simulations and theoretical studies \citep[e.g.,][]{2007MNRAS.379L..40M,2008MNRAS.386...65M,2011ApJ...741L..12B,2013MNRAS.432.2779B,2013MNRAS.436..584B,2013ApJ...763L..15M,2013MNRAS.430L..30S,2015ApJ...800....9M}  have challenged the idea that BH populations deplete within a few hundred million years of GC evolution. These studies suggest that certain star clusters, depending on their initial conditions could sustain sizeable populations of BHs for several billions of years. These new results suggest the sub-cluster of BHs that would from the evolution of segregating BHs during the early evolution of the GC will not be completely decoupled from the entire cluster and that its evolution will be determined by the energy demands of the host GC \citep{2013MNRAS.432.2779B,2013MNRAS.436..584B}. The life span of this subsystem of BHs would thus depend on initial parameters of the GC. Fig. \ref{fig:rc-bhs} by \citet{2013MNRAS.436..584B} schematically shows the evolution of the core radius of a GC containing a subystem of sBHs. If the GC is initially not too dense, then its two-body relaxation time would be sufficiently long and this could allow a subsystem of BHs to sustain itself to up to a Hubble time or longer \citep[e.g.,][]{2016MNRAS.462.2333P,2018MNRAS.479.4652A,2018MNRAS.478.1844A,2019ApJ...871...38K,2020ApJ...898..162W}. These sBHs can interact with surrounding stars and hence provide an internal energy source in the star cluster that can halt and counteract core collapse \citep[see][and referennces therein]{2020IAUS..351..357K}. Direct \textit{N}-body simulations of massive star clusters initially containing up to a million stars \citep[(e.g.,][]{2016MNRAS.458.1450W} and having sufficiently long initial half-mass relaxation times ($\sim 7-8$ Gyr) have shown that few hundreds to up to a thousand BHs can survive in these GCs till up to 12 Gyr. 

Theoretical and numerical studies investigating evolution of BHs in GCs (cited in the previous paragraph) have been motivated by observational discoveries during the last ten years. Several multiwavelength observational studies have identified accreting BH candidates in Galactic and
extragalactic GCs \citep[e.g.,][]{2007Natur.445..183M,2011ApJ...734...79B,2012Natur.490...71S,2013ApJ...777...69C,2017MNRAS.467.2199B,2018ApJ...855...55S,2018ApJ...862..108D}. 
The most compelling evidence for the existence of sBHs in GCs comes from observations of the Galactic GC NGC 3201 by \citet{2018MNRAS.475L..15G,2019A&A...632A...3G}, who identified a couple of sBHs in detached binary systems with main sequence stars. Radial velocity variations of these two main sequence stars indicate that they are orbiting around sBH companions that have a minimum masss of $4.53 \pm 0.21$ and $7.68 \pm 0.50$ $\rm \ M_{\odot}$ \citep{2019A&A...632A...3G}. These observations point towards the possibility that certain GCs could contain a significant number of BHs. While most of the observed candidates are sBHs in binary systems, results from numerical simulations of GCs suggest that the presence of few sBHs in binary systems might be indicative of the presence of a sizeable number of single sBHs in dense star clusters \citep{2018ApJ...855L..15K,2018MNRAS.478.1844A,2019MNRAS.485.5345A, 2020ApJ...898..162W}.

\paragraph{Overview of IMBH formation pathways in star clusters}\label{subsubsec:overview}

\begin{figure}[t]
\centering
\includegraphics[width=\textwidth]{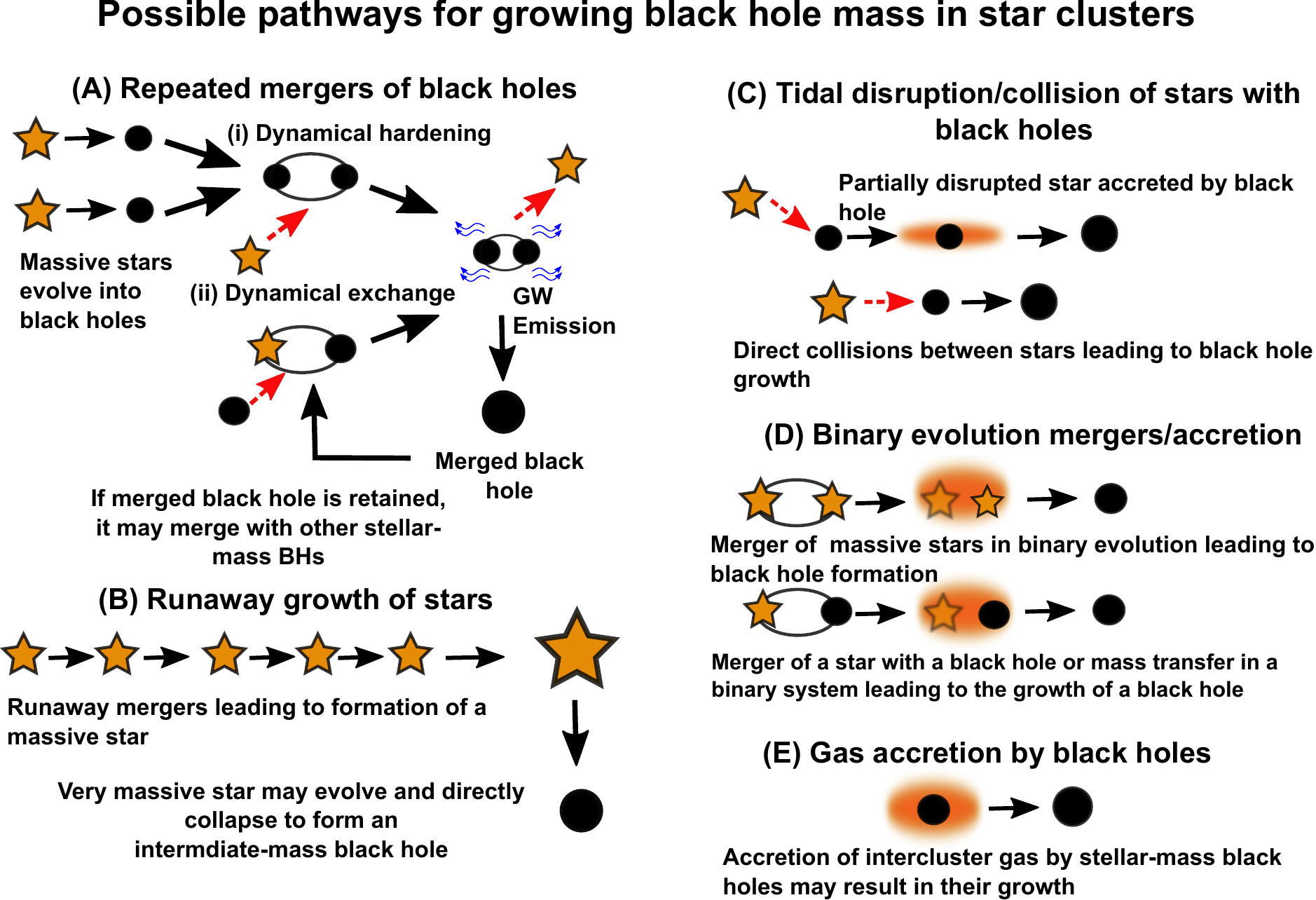}
\caption{
    Illustration of potential pathways by which IMBHs can grow in star clusters. \textit{(A)}: shows the merger of dynamically formed BBHs within the stellar cluster via GW emission. \textit{(B)}: For very dense ($\rm\gtrsim 10^{6} \ \msun \rm \ pc^{-3}$) star clusters, runaway mergers can occur between stars leading to the formation of a massive star (larger than a few 100 $\msun$). An IMBH of a few 100 $\msun$ could form via the collapse of such a star.\  \textit{(C)}: BHs can also grow by tidally disrupting or colliding with stars within the stellar cluster.\textit{(D)}: Binary evolution of two massive stars could form a massive single BH and accretion of stars by BHs in binaries can also result in their growth. \textit{(E)}: BHs may also accrete gas within star cluster }
\vspace*{-1mm}
\label{fig:formation-channnels}
\end{figure}

The idea that \acp{IMBH} could form and might exist in massive stellar clusters had been suggested by a number of papers in the 1970s \citep[e.g.,][]{1970ApJ...160..443W,1975IAUS...69....3S,1978MNRAS.185..847B}. One line of reasoning was that due to 2-body relaxation, the cluster core can evolve towards collapse leading to high central densities which  may lead to runaway collisions between stars \citep{1967ApJ...150..163C,1970ApJ...162..791S} and/or this process could drive mergers between sBHs leading to the formation of a massive BH \citep{1975IAUS...69....3S}. Several papers also investigated how the presence of a massive central BH would influence dynamics in a star cluster \citep[e.g.,][]{1976MNRAS.176..633F,1976ApJ...209..214B,1977ApJ...211..244L,1980ApJ...239..685M}. With the advent of X-ray astronomy in the 1970s, several X-ray sources associated with GCs had been detected. This led to studies suggesting that these sources may be powered by accreting massive BHs at the center of GCs \citep{1975Natur.256...23B}. Furthermore, extrapolation of the observed $M-\sigma$ relation between \acp{SMBH} and their host galaxies to dwarf galaxies and GCs also suggested that these systems could be harbouring \acp{IMBH} of $10^3$ to $10^{4} \ \msun$ \citep{2013A&A...555A..26L}.

As it was explained in Section \ref{subsec:model}, advancements in computational power and numerical methods over the past few decades have enabled the development of sophisticated simulation codes that take into account many important physical processes (e.g., close gravitational encounters, collisions, stellar/binary evolution, tidal field) that drive star cluster evolution. As a result, computational simulations of realistic star clusters have become an important tool for understanding how dynamics within dense environments can influence the formation of stellar exotica, \acp{sBH} and \acp{IMBH}.

In the subsequent sections, six pathways by which \acp{IMBH} could form and grow within dense stellar environments (YSCs, GCs and NSCs) will be highlighted. These are:

\begin{enumerate}
    \item \acp{sBH} that are initially retained in a star cluster will segregate to the core of the cluster. These \acp{sBH} can end up (or be born) in binary systems. If a binary \ac{sBH} merges inside the cluster due to GW radiation and the merged BH can be retained within the cluster then it may grow by merging with other BHs. This pathway for the \ac{IMBH} formation and growth is known as the `repeated mergers' or `hierarchical merger' channel (see Section \ref{subsec:hierarchial-mergers}). It is also illustrated in the top left panel (A) of Fig. \ref{fig:formation-channnels}.
    
    \item In initially dense star clusters, mergers between massive stars can lead to the runaway growth of a VMS which could potentially evolve to form an \ac{IMBH}. This scenario is known as the `fast runaway' scenario (see Section \ref{subsec:fast-runway}). It is also illustrated in the bottom left panel (B) of Fig. \ref{fig:formation-channnels}.

    \item An \ac{IMBH} may also form from the gradual growth of an \ac{sBH} by either accreting material from stars in the cluster during close encounters (collisions or tidal disruptions) or in very close binary systems. These processes are categorized within the 'slow runaway' formation channel (see Section \ref{subsec:slow-runaway}). These processes are illustrated in panels (C) and (D) on the right side of Fig. \ref{fig:formation-channnels}.
    
    \item Stellar mergers between an evolved star and a MS star during dynamical encounters or during binary stellar evolution may also rapidly form a low-mass \ac{IMBH} in a stellar cluster (see Section \ref{subsec:mergers-in-binaries}).

    \item Accreting \acp{sBH} in a gas rich dynamical environment may also grow to become \acp{IMBH} (see Section \ref{subsec:gas-accretion}). This process has been illustrated in panel (E) on the bottom right side of  Fig. \ref{fig:formation-channnels}. Furthermore, \acp{IMBH} may also grow through gas accretion in disks of AGNs (see Section \ref{subsec:agn-disks}).

\end{enumerate}

\subsubsection{Repeated or hierarchical mergers of stellar-mass BHs}\label{subsec:hierarchial-mergers}

The most straightforward way to grow an \ac{sBH} within a star cluster is to have it merge other \acp{BH} in the cluster \citep{2002MNRAS.330..232C,2002ApJ...566L..17M}.

\paragraph{Segregation of stellar-mass BHs due to dynamical friction}\label{subsubsubsec:hier-seg}

\acp{sBH} that form from the evolution of massive stars (also referred to as 1st generation or 1G \acp{BH}) have masses that can be 10 to 40 times more massive than the average mass of most of the stars in the cluster. These \acp{BH} can segregate to the dense central regions of the star cluster due to dynamical friction \citep{1943ApJ....97..255C,2008gady.book.....B}. A massive object (with mass $M$ moving in a sea of lighter object exchanges energy with the lighter stars and experiences a drag force that slows it down. As a result, the massive object sinks to the center of cluster. This process is a consequence of two-body relaxation and thus the timescale over which dynamical friction occurs is \citep{2018arXiv180909130M}:

\begin{equation}
t_{\mathrm{df}}(M) \simeq \frac{\langle m\rangle}{M} t_{\mathrm{rlx}}
\label{eq:timescale-df}
\end{equation}

where $\langle m \rangle$ is the average mass of the stars in the cluster and $M$ is the mass and $\rm t_{rlx}$ is the relaxation time defined in Equation \ref{eq:relaxation-time}. Using Equation \ref{eq:timescale-df}, we can estimate the dynamical friction timescale for a \ac{BH} of $20 \ \msun$ to sink to the center of a moderately dense star cluster ($\rho_{c} \sim 10^{5} \ \msun \ \mathrm{pc}^{-3} $) where the average mass of stars is $1 \ \msun$ and the relaxation time is a 500 Myr, to be $\rm t_{df} \sim 25 \ \rm Myr$. 

\paragraph{Dynamical BBH formation and hardening}\label{subsubsubsec:hier-dyn-form}

Due to this process, most sBHs that are retained in a star cluster  (see Section \ref{subsec:bh-formation-kicks-retention}) will sink to the high density central region of the cluster on timescales of about 10 to 100 Myr. In the center of a star cluster, these BHs can gravitationally interact with each other and form binary systems. Frequent three and few-body gravitational encounters can lead to the formation of BBHs through exchange interactions. These BBHs dynamically interact with other stars and this can tighten their orbit which will allow them to merge due to GW emission (see (A) in Fig.\,\ref{fig:formation-channnels}) (see also {\color{blue}\textbf{Chapter 1.IV}}). Furthermore, BH mergers can also occur due to GW capture during close single-single \citep{2018ApJ...860....5G,2020PhRvD.101l3010S,2021MNRAS.506.1665G}, binary-single \citep{2014ApJ...784...71S,2018PhRvL.120o1101R,2018PhRvD..97j3014S,2018ApJ...855..124S,2022MNRAS.514.5879M} and binary-binary \citep{2019ApJ...871...91Z,2021A&A...650A.189A,2022MNRAS.514.5879M} gravitational encounters in these dense environments. The latter may also lead to the formation of hierarchical triple systems \citep{2016ApJ...816...65A,2020ApJ...903...67M,2022MNRAS.511.1362T} in which inner binary sBHs can be driven towards GW merger through eccentricity oscillations due to the von Zeipel-Lidov–Kozai (ZLK) mechanism \citep{1910AN....183..345V,1962AJ.....67..591K}. 

\begin{figure}[t]
\centering
\includegraphics[scale=0.2]{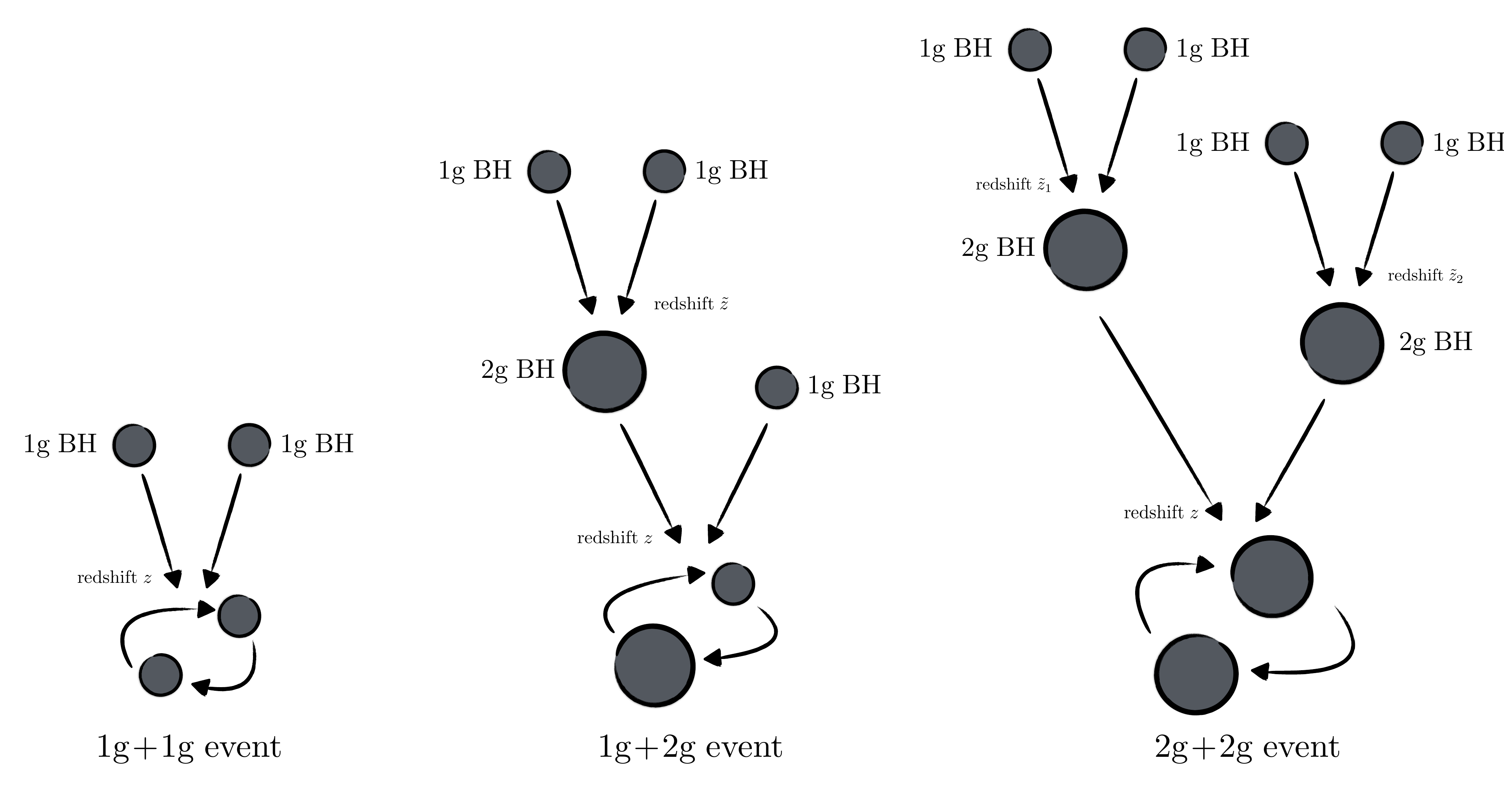}
\caption{
   Cartoon taken from \citet{2017PhRvD..95l4046G} illustrates how 1G sBHs can grow in dense environments through repeated mergers that result in the formation of 2g and Nth generation BHs.}
\vspace*{-1mm}
\label{fig:multiple-generation-bh}
\end{figure}

If the merged BH (also known as a 2nd generation BH) is retained in the star cluster, it can subsequently partner up and merge with another 1st or 2nd generation \acp{BH} which can form an Nth generation \acp{BH}. This pathway for BH growth and IMBH formation has been invoked in several papers to explain the high masses of the merging BHs in observed GW wave events like GW190521 \citep[e.g.,][]{2019PhRvD.100d3027R,2019MNRAS.486.5008A,2021ApJ...923..126S,2021arXiv210704639F,2020MNRAS.497.1043D,2021ApJ...920..128A,2023MNRAS.520.5259A,2023arXiv230704806A}. The cartoon shown in Fig. \ref{fig:multiple-generation-bh} (taken from \cite{2017PhRvD..95l4046G}) illustrates how hierarchical mergers of \acp{sBH} can form 2nd and Nth generation \acp{BH} in a dense stellar system.

\paragraph{BBHs from binary evolution and their hardening}\label{subsubsubsec:hier-prim-form-evol}

In addition to the dynamical formation of merging binary \acp{sBH}, many massive stars that are progenitors of \ac{sBH} are expected to be born in binary and multiple systems \citep{2012Sci...337..444S,2017ApJS..230...15M}. Depending on the birth binary and multiplicity fraction of a star cluster, there could be a sizeable number of initial binary systems containing two massive stars. These systems can potentially evolve into a tight binary \acp{sBH} within few to tens of Myr. The merger time due to GW emission ($t_{\rm gw}$) for these binaries will depend on the masses of the two \acp{sBH}, eccentricity and semi-major axis value of the binary systems:
\begin{equation}
    \label{peters-eq}
    t_{\rm gw} \simeq \ 10^{10} \ y r\left(\frac{a_{\rm bin}}{3.3 R_{\odot}}\right)^{4} \frac{1}{\left(m_{1}+m_{2}\right) m_{1} m_{2}} \cdot\left(1-e^{2}\right)^{7 / 2}
\end{equation}

where $a_{\rm bin}$ is the semi-major axis of the binary, $e$ is the eccentricity, and the masses of the two BHs are $m_{1}$ and $m_2$ in solar mass.

Within a dense dynamical environment (e.g., YMC or a GC), these BBHs would be considered hard binaries and they can further harden or become more eccentric due to frequent dynamical interactions (e.g, fly-by interactions with surrounding single and binary stars) and this may substantially decrease the merger time due to GW radiation (by increasing $e$ and reducing $a$ in Eq. \ref{peters-eq}). Therefore, dynamics can catalyze the merger of BBHs due to GW emission within a dense star cluster. However, if a BBH becomes sufficiently hard, binary-single and binary-binary scattering encounters could impart a dynamical recoil kick on the BBH due to momentum conservation. If the velocity of this dynamical recoil kick is larger than the escape speed of the cluster then the BBH will be expelled from the cluster. This is likely to occur if the timescale for a three-body encounter is shorter or comparable to the time needed for this hard binary to merge inside the cluster due to GW emission. While these dynamically ejected binaries could merge due to GW emission outside their host cluster, their ejection can inhibit the formation and growth of an IMBH seed.

The timescale over which a BBH will dynamically harden and eventually merge due to GW will be governed by the properties of the host cluster and the number of sBH it can retain. As it was discussed in Section \ref{subsec:bh-formation-kicks-retention}, the dynamical evolution of a subsystem of sBHs at the center of a GC is governed by the energy demands of the entire cluster \citep{2013MNRAS.432.2779B}. In an initial cluster that is not too dense and has a sufficiently long initial half-mass relaxation time, a long lasting sBH subsystem can generate energy which will result in cluster core expansion (see Fig. \ref{fig:rc-bhs}) and the BH subsystem depletion time will be longer for these clusters. In these clusters dynamically formed BBHs will be hard binaries at wider separations compared to denser clusters which will have a higher stellar velocity dispersion ($\sigma$) and thus a higher mean kinetic energy of stars. Therefore, it may take longer for dynamically formed BBH in low density cluster to merge as opposed to denser clusters with shorter initial half-mass relaxation time \citep{2019MNRAS.486.5008A}.




\paragraph{Gravitational wave recoil kick and merged BH retention}\label{subsec:gw-recoil-kicks}

While the pathway for building up BH mass through repeated or hierarchical mergers in star clusters is generally straightforward, it encounters a significant obstacle in the form of GW recoil kicks. As two BHs are in the process of merging due to the emission of GW radiation, the final inspiral occurs in less than one orbital period of the BBH. These GWs carry away linear momentum from the system and in order to compensate for this, the merged BH acquires a GW recoil kick  
\citep{1973ApJ...183..657B,1983MNRAS.203.1049F,1992PhRvD..46.1517W}.  The magnitude of this recoil kick depends strongly on the mass ratio of the merging BHs and their relative spin magnitudes and orientations \citep[see][]{2007ApJ...668.1140B,2008ApJ...682L..29B,2007PhRvL..98w1101G,2007ApJ...659L...5C,2012PhRvD..85h4015L}. For a perfectly symmetric binary system, comprising equal mass non-spinning BHs, the recoil kick will be zero; however, large spin values with asymmetric orientations can lead to recoil kicks that can be of order of a few thousand $\kms$ \citep{2008ApJ...686..829H,2016MNRAS.456..961B,2018ApJ...856...92F}. 

High velocity GW recoil kicks can eject the merged BH from its host cluster since the magnitude of this kick can be significantly higher than the typical escape speed of YMCs and present-day GCs. Several works have investigated the ejection of IMBH and their seeds from star clusters due to GW recoil kicks following the merger of two BHs \citep[e.g.,][]{2018MNRAS.481.2168M, 2019PhRvD.100d3027R,2021MNRAS.502.2682A, 2021ApJ...920..128A, 2021MNRAS.501.5257R}. The magnitude of these gravitational wave recoil kicks depends on the anisotropy in the masses and spins of the merging BHs \citep{2007PhRvL..98w1101G,2008ApJ...682L..29B}. If the kicks exceed the escape velocity of the cluster than this may result in the ejection of an IMBH seed and hence lower its retention probability \citep{2019PhRvD.100d1301G}. Such recoil kicks are more likely to eject potential IMBH seeds that are less massive than a $100 \ \msun$. Mergers with more extreme mass ratios result in lower recoil kicks, and the merged IMBH is more likely to be retained in dense environments with relatively large escape velocities \citep[e.g., NSC; see][]{2018MNRAS.481.2168M,2019MNRAS.486.5008A,2019PhRvD.100d3027R,2022ApJ...927..231F}. If a relatively massive seed can be built up through another mechanism (see subsection \ref{subsec:fast-runway} then it maybe possible to keep it in the cluster when it merges through GW emission with lower mass \acp{BH}. Therefore, if IMBH formation occurs from the mergers of stellar-mass BHs, then depending on the mass ratio, spin magnitudes and directions of the merging BHs, the GW recoil kick may eject the IMBH from the stellar cluster \citep{2004ApJ...616..221G,2006ApJ...637..937O,2008ApJ...686..829H,2021ApJ...915...56G}. \citet{2022MNRAS.514.5879M} showed that accounting for these GW recoil kicks may inhibit IMBH formation in \textsc{mocca} simulations in up to 70 per cent of the cases.

\begin{figure}[!htb]
\centering
\includegraphics[scale=0.35]{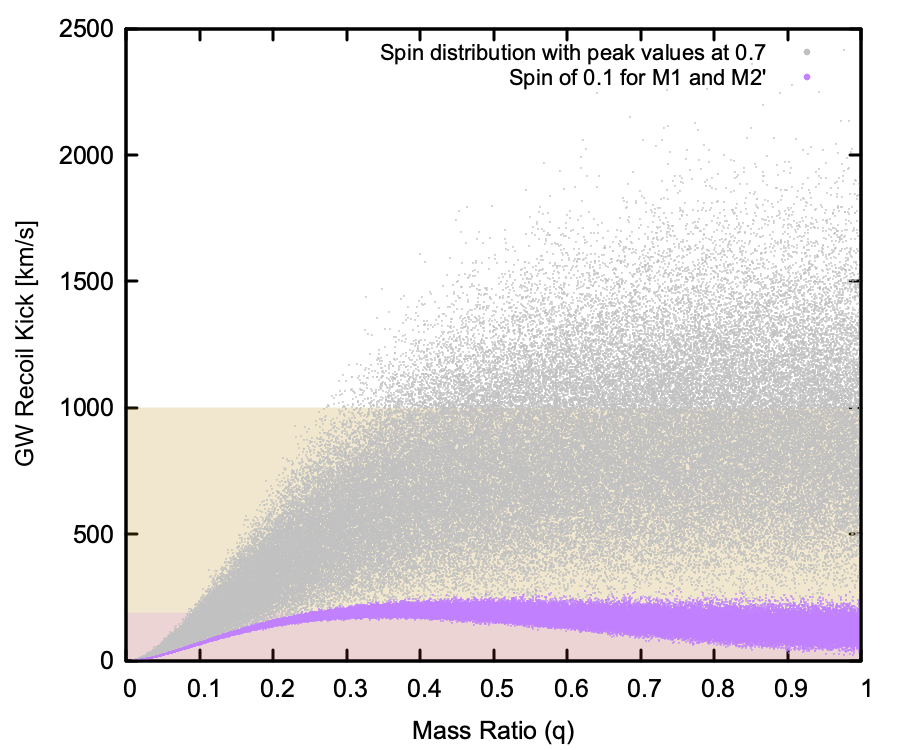}
\caption{GW wave recoil kick ($\rm v_{gr}$) as a function of the mass ratio (q) for $10^{5}$ merging BHs with a uniform mass ratio distribution. We consider two different cases for BH spin magnitudes: assuming high spins \citep[peak values close to 0.7;][shown as grey dots]{2012PhRvD..85h4015L} and low spins  \citep[values of 0.1;][shown as purple dots]{2019MNRAS.485.3661F,2020A&A...636A.104B}. The two shaded regions indicate escape speeds from dense environments like typical GCs to up to the most massive and dense NSCs}
\label{fig:gw-recoil-kick}
\end{figure}

\begin{figure}[!htb]
\centering
\includegraphics[scale=0.75]{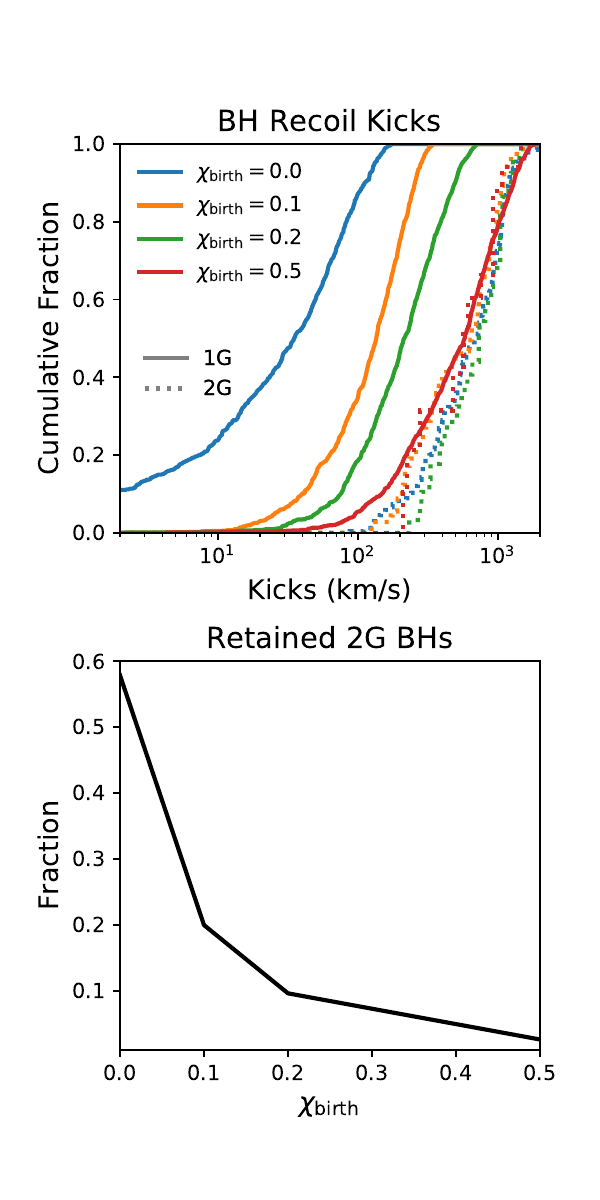}
\caption{The effect of initial BH spin on GW recoil kicks and
retention of BBH merger products in GC models (Fig.1 taken from \citet{2019PhRvD.100d3027R}) The top panel shows the cumulative distribution for kick velocities for different birth spins and the lower panel shows the fraction of retained 2nd generation BHs as a function of birth spins.}
\label{fig:kicks-rodri-1-gw-recoil-kick}
\end{figure}

Fig. \ref{fig:gw-recoil-kick} shows the dependence of the GW recoil kick on the mass ratio of merging BHs. This figure was produced by sampling a set of $10^{5}$ merging binary BHs with a uniform mass ratio distribution. Two cases for BH spins were considered. In the first case, it is assumed that the two merging BHs have high birth spins \citep[with peak values close to 0.7;][shown with grey dots]{2012PhRvD..85h4015L} and in the other the birth spins of the two merging BHs were set to 0.1 \citep[][shown with purple dots]{2019MNRAS.485.3661F,2022MNRAS.514.5879M}. An isotropic distribution of the spin orientation is assumed. For each binary system, the GW recoil kick is estimated using results from numerical relativity simulations provided by \citet{2008ApJ...682L..29B,2010ApJ...719.1427V}. It can be seen from the Fig. \ref{fig:gw-recoil-kick} that for the high spin values for BHs, the GW recoil kick magnitudes are on average larger than a few hundred $\kms$ for mass ratio values larger than about 0.15. For these spins and moderately high mass ratios, a merged BH would be difficult to retain in a GC However, for the low spin case, GW kick values can be less than a $100 \ \kms$ even for high mass ratios (see Fig. \ref{fig:kicks-rodri-1-gw-recoil-kick} taken from \citet{2019PhRvD.100d3027R}). Therefore, in the latter case, it might be possible to retain 2nd generation BHs within dense GCs and YMCs. However, the 2nd generation BH which forms from the merger of two low spin BHs might have a higher post-merger spin which could result in larger GW recoil kicks if it were to merge again \citep{2017PhRvD..95l4046G,2017ApJ...840L..24F}. Therefore, it could be difficult to retain multiple-generation within host stellar clusters with typical escape speeds $\lesssim 100 \kms$. 

Escape speeds of present-day MW GCs range from few to up to around a hundred $\kms$ for the densest and massive GCs \citep{2018MNRAS.478.1520B}. However, it is possible that the central escape speed from GCs was higher (by a factor of 2 to 3) during the very early stages of their evolution when they were more massive and potentially more concentrated \citep{2019MNRAS.486.5008A}. Many in-cluster BBH mergers are expected to take place during the early evolution of GCs \citep{2017MNRAS.464L..36A} and if they had higher central escape speeds (of the order of few hundred $\kms$) at that time) then they could potentially retain multiple generation IMBHs that form through hierarchical mergers.  Dense and massive \acp{NSC} found in the center of many galaxies can have escape speeds in excess of a few hundreds of $\kms$. These dense environments may be more suited to retaining multiple-generation BHs that form through hierarchical mergers \citep{2003ApJ...582..559V,2008ApJ...684..822B,2019MNRAS.486.5008A,2019PhRvD.100d1301G,2021ApJ...918L..31M,2021Symm...13.1678M,2022ApJ...933..170F,2022arXiv221206845K,2022ApJ...927..231F,2023MNRAS.523.4227A,2023MNRAS.526.4908C}.

In conclusion, stellar clusters that are initially massive (\rm $M \gtrsim 10^6 \mathrm{M}_{\odot}$), dense (\rm$\rho_{c}$ $\gtrsim 10^5 \mathrm{M}_{\odot} \ \mathrm{pc}^{-3}$) and have high escape speeds $\geqslant 300 \mathrm{~km} \mathrm{~s}^{-1}$ are likely to be the most efficient sites for producing IMBHs through hierarchical mergers of sBHs \citep{2019MNRAS.486.5008A,2022ApJ...927..231F}. Initially more massive clusters will contain more sBH progenitors which will lead to a larger reservoir of sBHs in the cluster. If this cluster is dense then two-body relaxation timescale will be short and this will result in the rapid segregation of sBHs to the cluster core\footnote{However, if the cluster is too dense then it can may lead to runaway collisions of massive stars in the cluster core (see Section \ref{subsec:fast-runway}).}. In this high density region, BH can interact with each other, forming binary systems that efficiently harden via dynamics and can merge due to GW radiation. A merged BH is likely to avoid ejection through GW recoil kicks if the escape speed from the cluster is of the order of few hundred $\kms$. With such high escape speeds, it would be possible to retain and grow an IMBH that formed through repeated mergers of sBHs within its host cluster.


\subsubsection{Fast runaway: stellar collisions resulting in IMBH formation}\label{subsec:fast-runway}

 \paragraph{Segregation of massive stars and collisional formation of a very masssive star}\label{subsubsubsec:coll-runawy}

Collisions between massive stars may occur in the cores of initially dense ($\rm \gtrsim 10^{6} \ \msun \ \rm  pc^{-3}$) star clusters during their early evolution (first 10 Myr). If these clusters have a sufficiently short initial half-mass relaxation time ($\sim 100$ Myr) then sBH progenitors (with initial masses ranging from $\sim 20 \ \msun$ to $\lesssim 150 \ \msun$) can segregate to the cluster center before evolving into BHs. From Eq. \ref{eq:timescale-df} it can be estimated that for stars with masses larger than about 50 times the average mass of stars in the cluster, the time needed to sink to the cluster center due to dynamical friction is less than $\sim 2$ Myr. This is shorter than the typical lifetime of these massive stars.  It is also possible that some clusters may already be strongly mass segregated at birth, in that case massive stars may preferentially form in the center of these clusters.

\paragraph{Evolution of a very massive star and IMBH formation}\label{subsubsubsec:vms-evolution}

As massive stars have large radii and therefore a larger collisional cross section, they can collide with other stars in the center of the cluster on timescales shorter than their lifetime. If these collisions between MS stars result in stellar mergers\footnote{Mergers in which a relatively small fraction of stellar mass is lost or ejected during the collision \citep{2010MNRAS.402..105G}.} then this can lead to the runaway growth of a \ac{VMS} ($\> 150 \ \msun$). In a high density cluster core, this \ac{VMS} will be the target for further collisions due to gravitational focusing and its large radius. This pathway for forming a \ac{VMS} was first demonstrated through \textit{N}-body simulations of the evolution of compact, initially dense ($\rm \sim 10^{6} \ \msun \ pc^{-3}$) low mass ($2 \times 10^{4} \ \msun) $ star clusters \citep{1999A&A...348..117P}. It was found in these simulations that massive stars quickly segregate to the cluster core where gravitational encounters can lead to the formation of MS stars more massive than $100 \ \msun$. Subsequent numerical works \citep{2002ApJ...576..899P,2004Natur.428..724P,2004ApJ...604..632G,2006MNRAS.368..141F,2012ApJ...752...43G} simulating dense star clusters with short initial half-mass relaxation time ($t \mathrm{rlx} \leqslant 25$ Myr) also found that the runaway growth of stellar collisions may form a \ac{VMS} that can have a mass which is $\lesssim 0.1 \%$ of the initial cluster mass. \citet{2004Natur.428..724P} found that in compact star cluster models with a high central concentration, massive stars rapidly sink to cluster center due to dynamical friction leading to runaway collisions that can form \ac{VMS} between $800$ to $\lesssim 3000 \ \msun$ within $\sim 5$ Myr. 

Whether these \acp{VMS} of $\sim 100 - 1000$ $\msun$ will end their lives as IMBHs depends on how they evolve \citep{2007ApJ...659.1576B}. One major hindrance to forming \acp{IMBH} from these \acp{VMS} are stellar winds. Mass loss from stellar winds can substantially reduce the mass of a post-merger object \citep{2006MNRAS.366.1424D,2009A&A...497..255G} and this could result in a lower BH mass that forms from the evolution of a \ac{VMS}. Furthermore evolved massive stars with helium core masses in the range of $135\gtrsim{}M_{\rm He}\gtrsim{}65$ M$_\odot$ can have extremely high core temperatures which leads to electron-positron pair production and the loss of photon pressure \citep{2016A&A...594A..97B,2021hgwa.bookE..16M}. Due to the loss of this pressure, the star contracts and this triggers explosive burning which can lead to the disruption of the entire star in what is known as pair instability supernova (PISN) \citep{1967ApJ...148..803R,1968Ap&SS...2...96F,2001ApJ...550..372F,2002RvMP...74.1015W}. For helium core masses in the range $65\gtrsim{}M_{\rm He}\gtrsim{}30$ M$_\odot$, pair production is not expected to disrupt the entire star but it can lead to a series of pulsations with enhanced mass loss (known as pulsational PISN \citep{2017ApJ...836..244W,2022MNRAS.512.4503R}) and significant reduction in the mass of the BH produced from the eventual collapse of this star \citep{2017MNRAS.472.1677S}.These two types of supernovae are responsible for generating the pair-instability mass gap which predicts that isolated stellar evolution of massive stars should not produce BHs in the mass range $\sim 60-120 \mathrm{M}_{\odot}$ \citep{2019ApJ...887...53F}. 

Despite these hindrances, in low metallicity ($\rm Z \lesssim 10^{-3}$) or metal-poor environments (like old GCs or clusters of pop III stars \citep{2018A&A...614A..14R,2020MNRAS.493.2352A,2022MNRAS.515.5106W}) where mass loss due to stellar winds is weak (also see Section \ref{sec:popIII}), \acp{VMS} might not lose a significant amount of their mass. In that case, they may evolve to form a relatively massive sBH of up to $\sim 60-70 \ \msun$ \citep{2016MNRAS.459.3432M,2018MNRAS.481..153O,2018MNRAS.478.2461G} or directly collapse to form an \ac{IMBH} (see (B) in Fig.\,\ref{fig:formation-channnels}) in the mass range of $100-1000 \ \msun$ \citep{2015MNRAS.454.3150G,2017MNRAS.470.4739S,2021MNRAS.507.5132D,2021MNRAS.505.2186D,2021MNRAS.505.2753S,2022MNRAS.512..884R}. The latter is most likely to occur for VMS that have evolved masses $\gtrsim 250 \ \msun$ \citep{2023arXiv230409350H} and metallicities which are $\lesssim 5\%$ of solar metallicity. Therefore, direct collapse of a VMS produced via runaway stellar collisions could be an effective pathway to rapidly form an \acp{IMBH} in dense, low metallicity environments. In massive clusters, Once a seed \acp{IMBH} forms it can gradually grow by merging with other stars and BHs (see Section \ref{subsec:slow-runaway}). In addition to YMCs and GCs, the fast scenario for IMBH formation through stellar collisions could also operate in NSCs and has also been invoked as a potential seeding mechanism for SMBHs \citep{2009ApJ...694..302D,2010A&ARv..18..279V,2018MNRAS.476..366B,2021MNRAS.503.1051D,2023MNRAS.tmp.1166V}. An important feature of this formation pathway for an IMBH is that unlike the repeated/hierarchical merger channel (see Section \ref{subsec:hierarchial-mergers}), it does necessitate that sBHs are retained in a star cluster (see Section \ref{subsec:bh-formation-kicks-retention}). Therefore, if sBHs receive substantial natal kicks at birth and are removed from their host cluster, this pathway may still lead to IMBH formation.

\subsubsection{Slow runaway: Gradual growth of a stellar-mass BH}\label{subsec:slow-runaway}

In contrast to the fast runaway scenario (see Section \ref{subsec:fast-runway}) which operates in a relatively dense stellar clusters and results in the rapid formation of an IMBH ($\rm t_{form} \lesssim$ 50 Myr), IMBHs may also form more gradually ($\rm t_{form} \gtrsim 50 -100$ Myr) in dense star clusters either through the growth of sBHs through a number of different pathways including merging and accretion of surrounding stars through dynamics or mass transfer in binary systems. 

\paragraph{Collision between BHs and stars}\label{collisions:bh-stars}

In the cores of dense star clusters, retained sBHs that form within a few tens of Myr of cluster evolution can undergo close encounters with surrounding stars. If the distance at closest approach during these encounters is comparable to the radius of the stars then this can lead to direct collisions between stars and sBHs. Results from both Monte Carlo \textit{N}-body  \citep{2015MNRAS.454.3150G,2021ApJ...911..104K} and direct \textit{N}-body simulations \citep{2021MNRAS.501.5257R,2021ApJ...920..128A,2022MNRAS.512..884R,2023arXiv230704806A} show that such encounters can occur in moderately dense ($\rho_{c} \gtrsim 10^{5} \ \msun \ \rm pc^{-3}$ GCs. The outcome of such a close encounter depends on the details of the interaction and the properties of the star. Results from hydrodynamical simulations show that such encounters between MS stars (between $0.5-20 \ \msun$) and sBHs ($10-30 \ \msun$) can result in the partial or complete disruption of the star \citep{2022ApJ...933..203K}. A fraction of the disrupted stellar material can become bound to the BH \citep{2022ApJ...933..203K}. Growth of an sBH through such collisions will depend on how much of this bound stellar material can actually be accreted by the sBH. Collisions between stars and sBHs has also been proposed as a potential pathway for forming IMBHs in galactic nuclei \citep{2022ApJ...929L..22R}

Within star cluster simulation codes, the end product of a star-BH collision is an sBH which absorbs a fraction of the stellar mass. This fraction is set by a collision factor ($f_{\mathrm{c}}$) which takes values between 0 (no mass is accreted by the BH) and 1 (all the stellar mass is absorbed by the BH). Therefore, the growth of an sBH and the formation of an IMBH through star-BH collisions strongly depends on the assumed value for the $f_{\mathrm{c}}$ factor \citep{2015MNRAS.454.3150G,2021MNRAS.501.5257R,2022MNRAS.512..884R}. If a VMS ($\gtrsim 150 \ \msun$) can form in the early evolution of a dense cluster through stellar mergers (see Section \ref{subsec:fast-runway} and this cluster also retains many sBHs then these can segregate to the cluster center due to dynamical friction. Therefore, there is a high probability for the VMS and an sBH to collide in the center of the cluster. A collision between a VMS and an sBH could result in significant growth of the sBH even if $f_{\mathrm{c}}$ values are between 0.1 to 0.25 \citep{2015MNRAS.454.3150G}. For instance, a collision between a VMS of $300 \ \msun$ and a 30 $\msun$ sBH could result in a final BH mass of $60 \ \msun$ (if $f_{\mathrm{c}}=0.1$) to $105 \ \msun$ (if $f_{\mathrm{c}}=0.25$). Several successive mergers between stars and a BH could potentially lead to the formation and growth of an IMBH. Once an IMBH forms it can grow further by either merging with other sBHs or by accreting mass from other stars \citep{2022MNRAS.514.5879M} in direct collisions and/or tidal disruption/capture events (see Section \ref{subsec:slow-tde-tdc}).

Monte Carlo \textit{N}-body simulations from \citet{2015MNRAS.454.3150G} found that initially massive and high central concentration star cluster models ($\rho_{c} \gtrsim 10^{7} \ \msun \ \rm pc^{-3}$ can rapidly form a VMS through runaway stellar collisions (see Section \ref{subsec:fast-runway}). In many of these models in which sBHs had low natal kicks, this VMS forms an IMBH following a collision with an sBH. These simulations either assumed that the sBH absorbed all the mass of the colliding VMS ($f_{\mathrm{c}}=1$) or absorbed only 25 per cent ($f_{\mathrm{c}}= 0.25$) of the stellar mass. Through this pathway, the densest star cluster models with initial masses of the order $10^{6} \ \rm \msun$ were able to form IMBHs that were able to grow to up to $10^{4} \  \rm \msun$ through mergers with other stars, BHs and mass transfer events in binary systems \citep{2006ApJ...642..427B}. 

\begin{figure}[htbp]
\centering
\begin{subfigure}[t]{.5\textwidth}
  \centering
  \includegraphics[width=\textwidth]{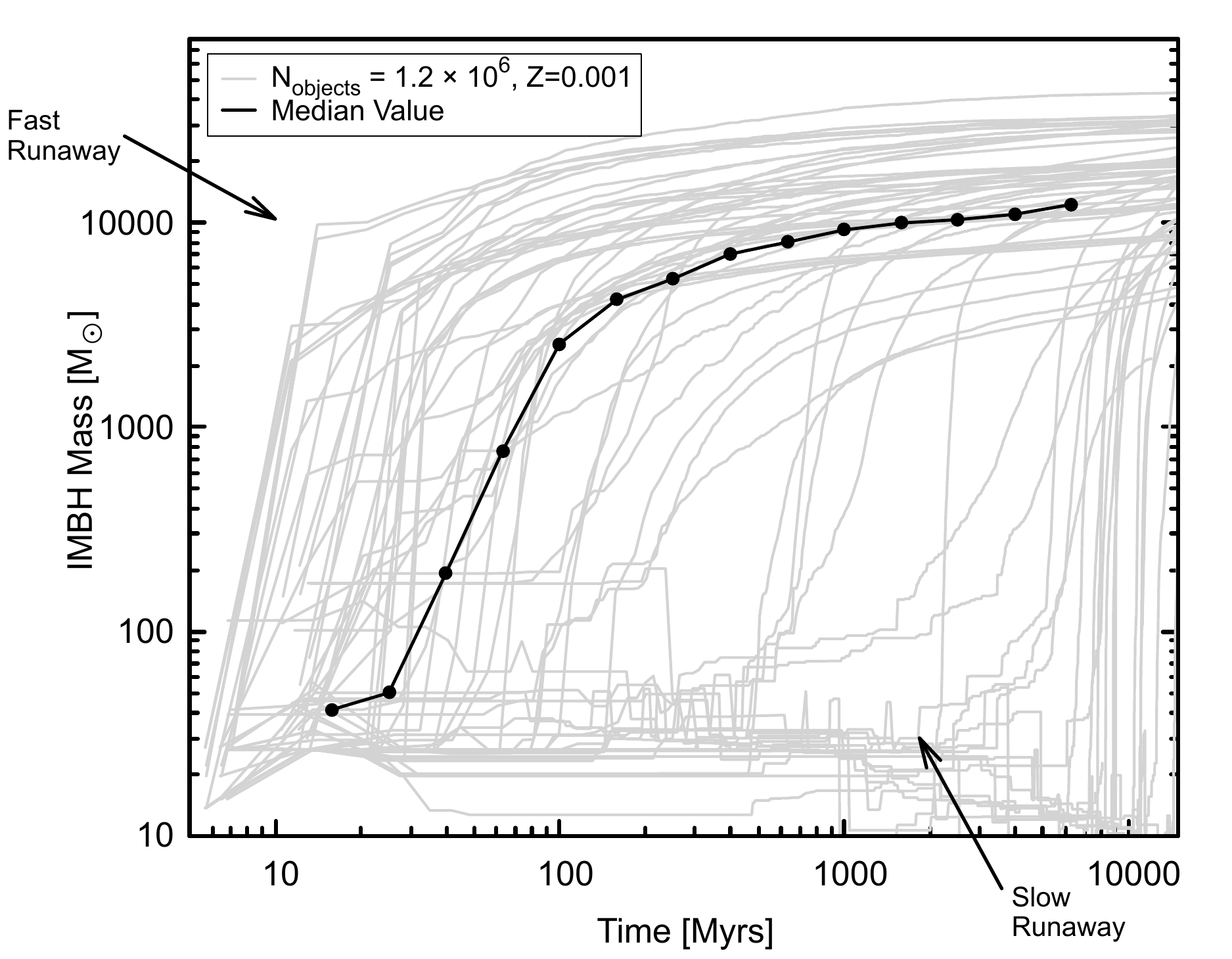}
  \caption{Time vs the mass of the most massive BH inside star cluster models simulated using the Monte-Carlo \textit{N}-body code  \textsc{MOCCA}. Each grey line represents a star cluster model which initially had $1.2 \times 10^6$ objects but varying combinations of the binary fraction, concentration, initial density and tidal radius. In all these models, a BH more massive than $100 \ \msun$ formed and was retained inside the star. These models show that both the fast (IMBH formation $\lesssim 100 \ \rm Myr$) and slow runaway scenario IMBH formation $\gtrsim 100 \ \rm Myr$) can lead to IMBH formation in these star cluster simulations. Combining data from all these models. the median growth rate of the BH is shown with the black lines.
 }
  \label{fig:sub1}
\end{subfigure}%
\begin{subfigure}[t]{.5\textwidth}
  \centering
  \includegraphics[width=\textwidth]{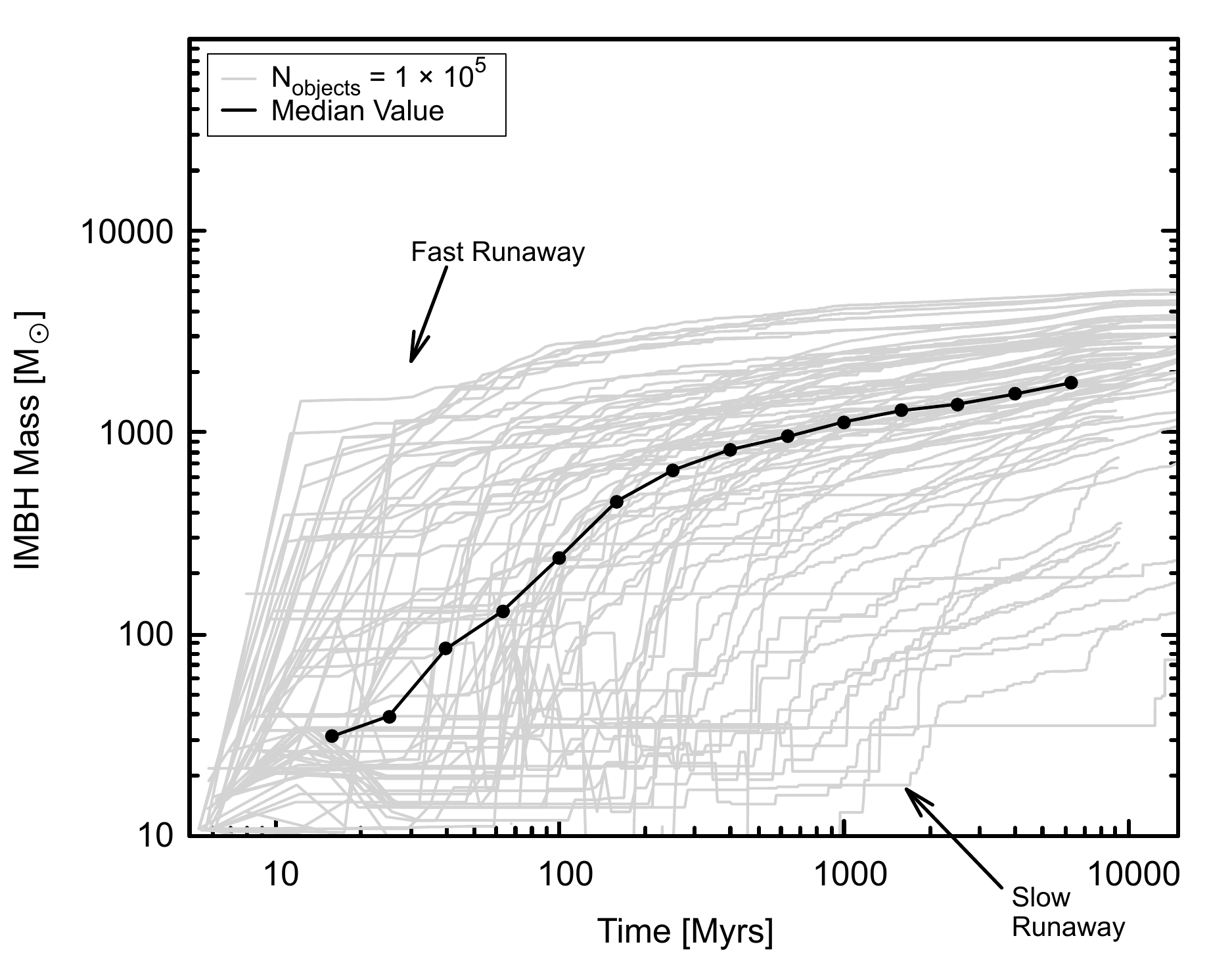}
  \caption{Similar to panel A, but in this case these are MOCCA models which initially had $1 \times 10^5$ objects. Both the fast and slow runaway scenarios can lead to the IMBH formation in these simulations. In contrast to the left panel for models that initially had $1.2 \times 10^{6}$, it can be seen that maximum IMBH mass in these lower N models is only about several $10^{3} \msun$ compared to few $10^{4} \ \msun$ in the left panel.}
  \label{fig:sub2}
\end{subfigure}
\caption{Figures show how \ac{IMBH} mass increases with time inside star cluster models simulated as part of the MOCCA-Survey Database I \citep{2017MNRAS.464L..36A}.}
\label{fig:imbh-growth-mocca}
\end{figure}

Based on results from hundreds of Monte Carlo \textit{N}-body simulations of stellar clusters with different initial parameters carried out using the \textsc{mocca} code \citep{2013MNRAS.429.1221H,2013MNRAS.431.2184G}. It can be seen in Fig.\,\ref{fig:imbh-growth-mocca} that in a significant fraction of these simulated stellar clusters, an IMBH forms within tens of Myr due to runaway mergers between massive main-sequence stars \citep{2002ApJ...576..899P,2015MNRAS.454.3150G,2021MNRAS.507.5132D} leading to the formation of a massive star which may evolve into an IMBH in low metallicity clusters. As discussed above, it is also possible for these stars to absorb a stellar-mass BH \citep{2021MNRAS.501.5257R,2022MNRAS.512..884R}.

In addition to the fast runaway scenario, it can also be seen that in a number of models, the IMBH formation and growth begins later in the cluster evolution and is more gradual compared to the fast runaway scenario. In these cluster models, there were only a handful of sBHs retained in the cluster. After a few Gyr of evolution, these models evolved towards core collapse which resulted in high core densities. This leads to an increased likelihood for an sBH to collide with a star during close binary-single and binary-binary scattering encounters. As these simulations assumed that $f_{\mathrm{c}}=1$, repeated mergers of the sBH with stars (in capture and accretion events) led to the growth of the sBH into an IMBH. \citet{2015MNRAS.454.3150G} found that this slow process for IMBH formation is more stochastic and produced low-mass IMBHs with a wider variety of masses (ranging between $10^{2}-10^{3} \ \msun$.

\begin{figure}[t]
\centering
\includegraphics[width = 0.7\textwidth]{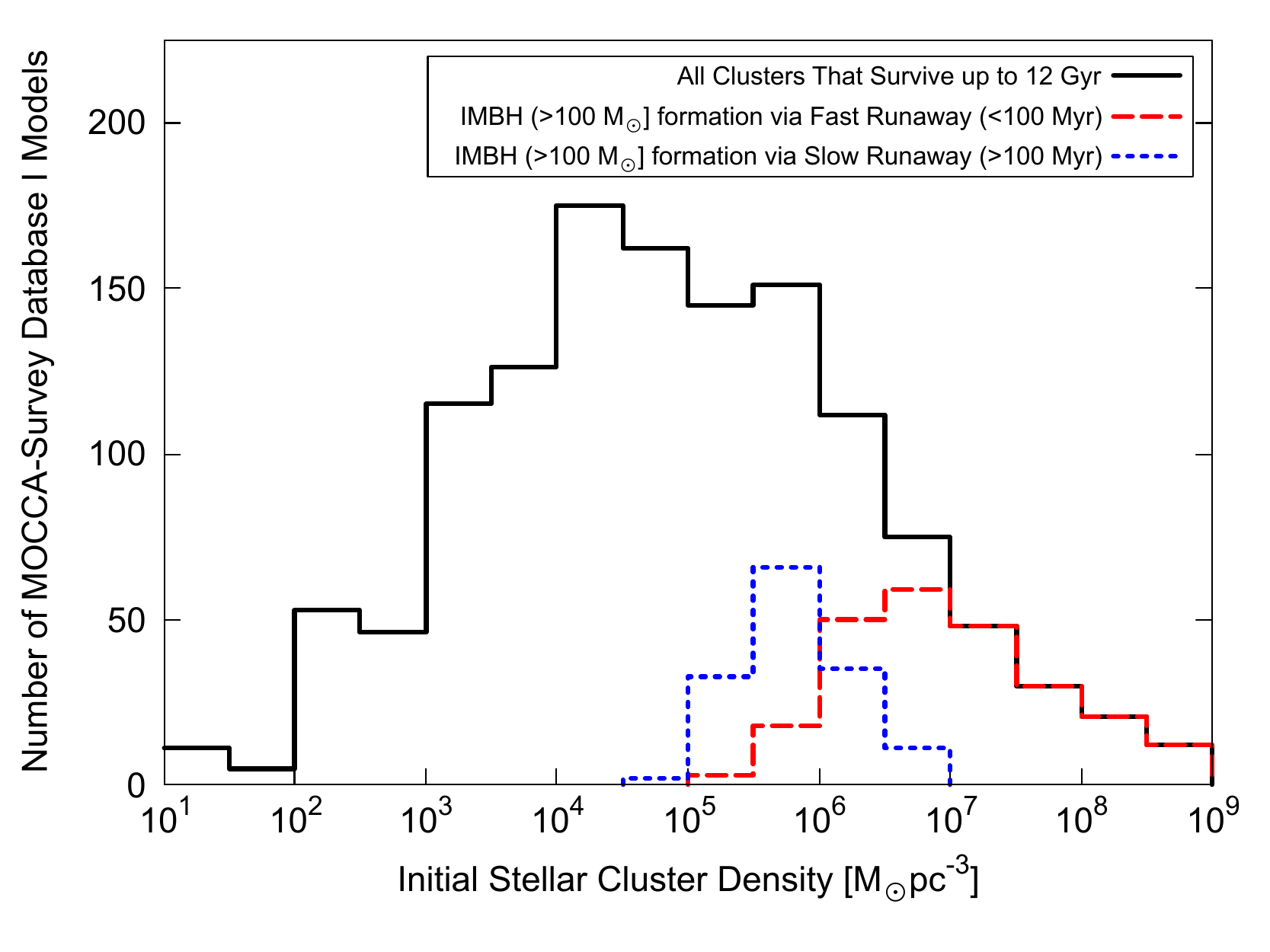}
\caption{
   Distribution of central density for all models in MOCCA-Survey Database I \citep{2017MNRAS.464L..36A} that survive up to 12 Gyr. It can be seen that these models span a wide range of initial central densities ranging from $10$ to up to $10^{8} \ \msun \ \rm pc^{-3}$. Initial densities of star cluster models that form an IMBH (which is at least more massive than $500 \ \msun$) are shown in red and blue. For the models with densities shown in red colours, the IMBH formation occurs via the fast runaway (within a 100 Myr) scenario while for the blue models it occurs through the slow runaway (IMBH forms after at least 100 Myr) scenario. A similar figure can also be found in \citep{2020MNRAS.498.4287H}}
\vspace*{-1mm}
\label{fig:density-distribution-mocca}
\end{figure}

Fig. \ref{fig:density-distribution-mocca} shows the distribution of the initial central density ($\rho_{c}$) of GC models in the MOCCA-Survey Database I \citep{2017MNRAS.464L..36A} that survived up to 12 Gyr. These GC models had different initial number of stars (with ZAMS masses sampled between $0.08 \ \msun$ to $100 \ \msun$  using a two component IMF\cite{2001MNRAS.322..231K}), metalliciy, initial binary fraction, half-mass and tidal radii. The histograms in red and blue indicate the distribution of the initial central density of models that formed an IMBH via the fast and slow runaway scenarios. It can be seen that in GC models with $\rho_{c}$ larger than  $10^{7} \ \msun \ \rm pc^{-3}$, an IMBH forms in all cluster models within few hundred Myr of cluster evolution. While the slow runaway channel can operate at lower $\rho_{c}$ values (ranging from $10^{5}-10^{6}\ \msun \ \rm pc^{-3}$. However, a smaller fraction of models with $\rho_{c}$ in this range acutally form an IMBH (about 30\%). For $\rho_{c} \gtrsim  10^{6}\ \msun \ \rm pc^{-3}$, the likelihood of IMBH formation via fast runaway is higher than the slow runaway channel. However, it needs to be cautioned that IMBH formation in these simulated GC models is optimistic due to $f_{\mathrm{c}}=1$ and the lack of inclusion of GW recoil kicks when two or more BHs merge \citep{2018MNRAS.481.2168M}. Furthermore, these simulations did not incorporate the latest stellar evolution prescriptions for BH progenitors (such as stellar winds and remnant masses).

\begin{figure}[t]
\centering
\includegraphics[width = 0.72\textwidth]{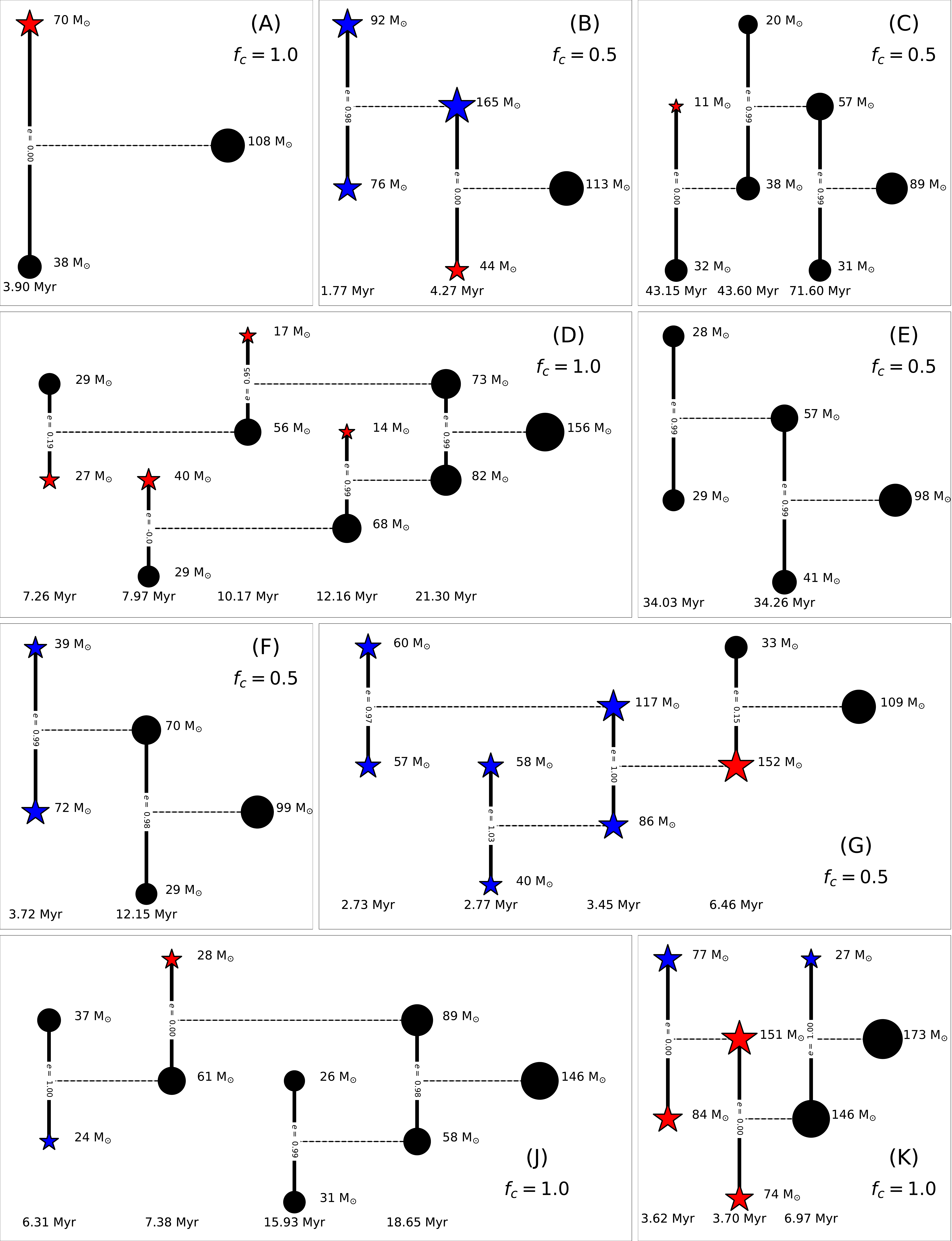}
\caption{
   Collision tree figure taken from \citet{2022MNRAS.512..884R} showing the formation of IMBHs star-BH collisions and BBH mergers in star cluster simulations carried out using the direct \textit{N}-body code \textsc{NBODY6++GPU} \citep{2015MNRAS.450.4070W}.}
\vspace*{-1mm}
\label{fig:rizzuto-collision-tree}
\end{figure}

Fig. \ref{fig:rizzuto-collision-tree} taken from \citet{2022MNRAS.512..884R} shows pathways for forming IMBHs in very metal-poor ($\rm Z=2 \times 10^{-4}$ star cluster models comprising $10^{5}$ stars that were simulated for 100 Myr with the direct \textit{N}-body code \textsc{NBODY6++GPU} \citep{2015MNRAS.450.4070W}. These simulations contained updated prescriptions for processes that govern stellar evolution of BH progenitors \citep[for details see][and references therein]{2020A&A...639A..41B,2022MNRAS.511.4060K} providing better treatment for stellar winds, supernovae and improved remnant masses. The initial models had a half-mass radius of 0.6 pc with $\rm \rho_{c} = 1.1 \times 10^{5} \ \msun \rm pc^{-3}$. \citet{2022MNRAS.512..884R} varied the $f_{\mathrm{c}}$ parameter value to 0.0, 0.5 and 1.0 and carried out simulations of 8 different realizations of their initial model for each $f_{\mathrm{c}}$ value. When $f_{\mathrm{c}}$ is set to 1, an IMBH forms in 6 out 8 simulations on timescales ranging from about 4 to 75 Myr. The top left panel in Fig. \ref{fig:rizzuto-collision-tree}, illustrates rapid IMBH formation of a $\sim 100 \ \msun$ IMBH through a star-BH merger. In other histories for IMBH growth a combination of a BBH mergers and star-BH mergers can contribute to IMBH Formation. For simulations with $f_{\mathrm{c}}=0.5$ and $f_{\mathrm{c}}=0$, 4 out of 8 and 2 out of 8 runs formed an IMBH. 

\paragraph{Tidal disruption and capture events}\label{subsec:slow-tde-tdc}

In addition to collisions with stars, BHs in dense star clusters can also grow by tidally disrupting, capturing \citep[e.g.,][]{2016ApJ...823..113P,2018ApJ...867..119F,2019ApJ...881...75K,2019MNRAS.484.1031P,2019ApJ...877...56L,2022MNRAS.515.4038T,2022MNRAS.516.2204R,2022ApJ...929L..22R} and subsequently accreting stars ((C) in Fig.\,\ref{fig:formation-channnels}). Once a seed IMBH has formed in a dense cluster, it can capture stars through tidal energy dissipation \citep{2006MNRAS.372..467b,2022MNRAS.517.1695H} and this capture can lead to merger and accretion events at pericenter passage which can gradually grow the IMBH mass. The growth of stellar-mass BHs and low-mass IMBH from the tidal capture and disruption of stars is most likely to occur in compact massive clusters (e.g., GCs and NSCs without an SMBH) with central densities larger than few $10^6 \ \msun \rm pc^{-3}$ \citep{2017MNRAS.467.4180S}. These extremely dense stellar systems will have central velocity dispersion values $\gtrsim 50 \ \kms$ and in these systems binary-single interactions will be unable to generate energy to heat the cluster and prevent core collapse \citep{2012ApJ...755...81M}. This is because at such high velocity dispersion, binaries are sufficiently tight and therefore binary-single encounters will result in stellar mergers. This will drive up the central density and can lead to the runaway growth of an sBH into an IMBH through tidal capture and disruption of surrounding stars \citep{2017MNRAS.467.4180S,2017NatAs...1E.147A}. In this scenario, tidal captures are expected to be more effective in initially growing the IMBH as they have a larger cross-section compared to tidal disruption events (TDEs).

In a recent work, \citet{2023MNRAS.521.2930R} carried out \textit{N}-body simulations of extremely dense (up to $10^{7} \ \msun \rm pc^{-3}$) idealized cluster models with a high initial central velocity dispersion ($\sim 50 \ \kms$) with the goal of investigating BH growth through runaway tidal captures proposed by \citet{2017MNRAS.467.4180S}.  They placed a single central BH in the cluster and found that stars on bound orbits around the BH get tidally disrupted and can grow a BH of a hundred solar masses to $2000-3000 \ \msun$ within a Gyr. In another work by \citet{2019MNRAS.484.4665S}, simulations of primordial star clusters in dark matter halos showed a central IMBH (of $\sim 10^{2}-10^{3} \ \msun$) could also grow through TDEs to masses of a few $10^{3} \ \msun$.

\subsubsection{Rapid stellar mergers leading to low-mass IMBH formation}\label{subsec:mergers-in-binaries}

In Section \ref{subsec:fast-runway}, it was discussed that PISNe can prevent the formation of BHs in the mass range of $\sim 60-120 \mathrm{M}_{\odot}$ through the isolated evolution of an initially massive star. However, several recent works \citep{2019MNRAS.485..889S,2020ApJ...903...45K,2020MNRAS.497.1043D,2021ApJ...908L..29G,2021MNRAS.507.5132D,2022MNRAS.516.1072C,2023MNRAS.519.5191B} suggest that low mass IMBHs ($\sim 100 \ \msun$) can form from the evolution of the merger between a MS star and an evolved star. These mergers could occur either through collisions during close encounters \citep{2020ApJ...903...45K,2023MNRAS.519.5191B,2023arXiv230704806A} in YSCs or during the close binary evolution (as illustrated in 1(D) in Fig.\,\ref{fig:formation-channnels}) of massive stars \citep{2020MNRAS.497.1043D,2021ApJ...908L..29G,2022MNRAS.517.2953T}. Since the conditions to trigger PISN depends on the helium core mass of an evolved star, a merger between a star that does not have a helium core massive enough to undergo PISN and a star on the MS could produce an evolved star with a large hydrogen envelope. Such a merger product could potentially circumvent PISN and result in the formation of a massive sBH or a low-mass IMBH \citep{2019MNRAS.485..889S,2019MNRAS.487.2947D,2022MNRAS.516.1072C,2022Galax..10...76S} in the pair-instability gap. Monte Carlo and direct \textit{N}-body simulations from \citet{2019MNRAS.487.2947D,2020ApJ...903...45K,2020MNRAS.497.1043D,2021ApJ...908L..29G} find that an IMBH seed can form through this pathway in the first 5 Myr of cluster evolution. In the dense environment of a star cluster, this BH could be an IMBH seed which subsequently grows further either through mergers with other BHs (see Section \ref{subsec:hierarchial-mergers}) or grow gradually through mergers, disruptions and accretion of surrounding stars (see Section \ref{subsec:slow-runaway}). 

\begin{figure}[t]
\centering
\includegraphics[width = 0.70\textwidth]{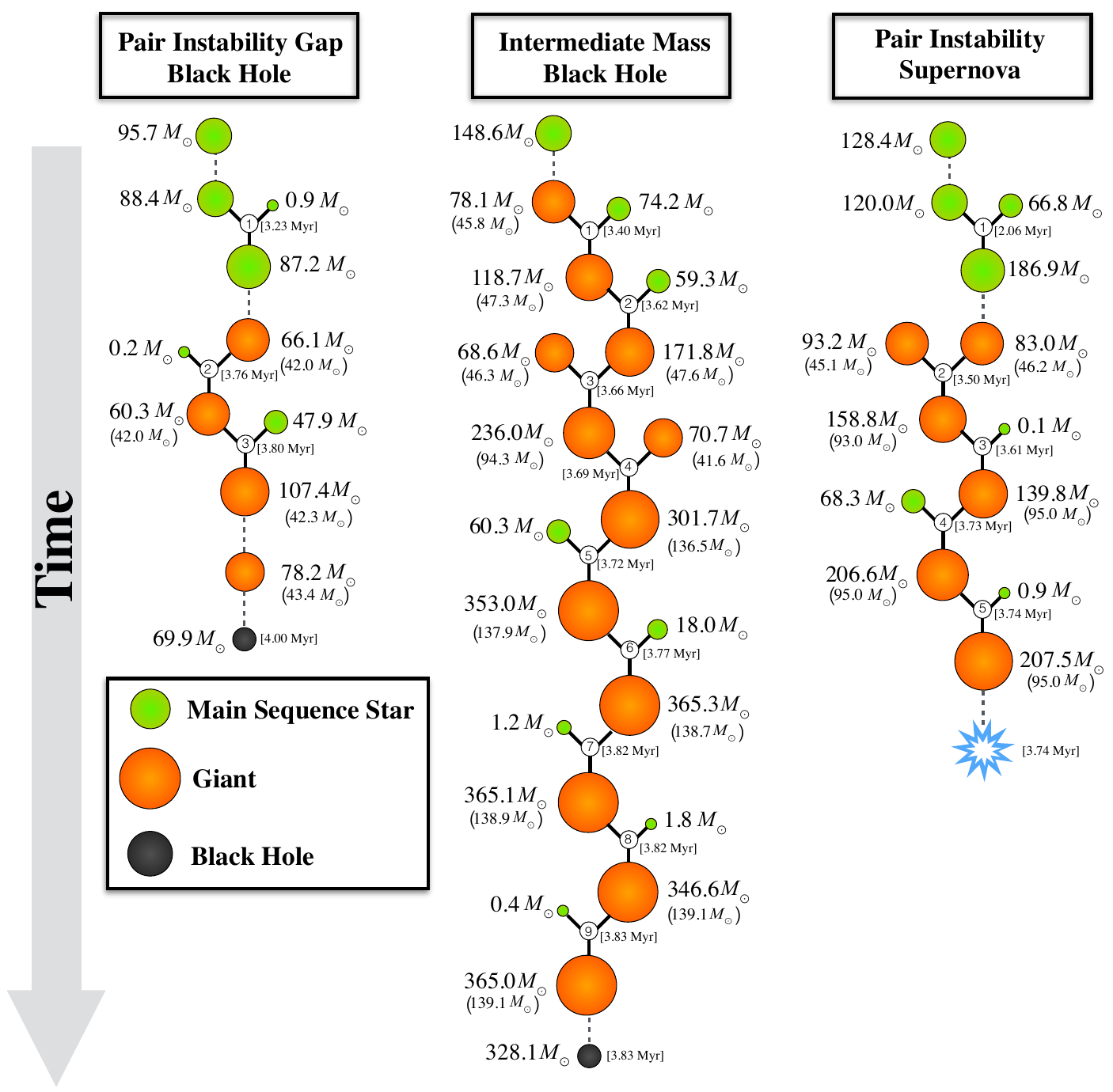}
\caption{
   Illustration taken from \citet{2020ApJ...903...45K} showing the formation of an IMBH (middle panel) through collisions in a star cluster simulated using the CMC Monte Carlo \textit{N}-body code \citep{2022ApJS..258...22R}.}
\vspace*{-1mm}
\label{fig:kremer-spera-2020-cartoon}
\end{figure}

\begin{figure}[hbt!]
\centering
\includegraphics[width = 0.8\textwidth]{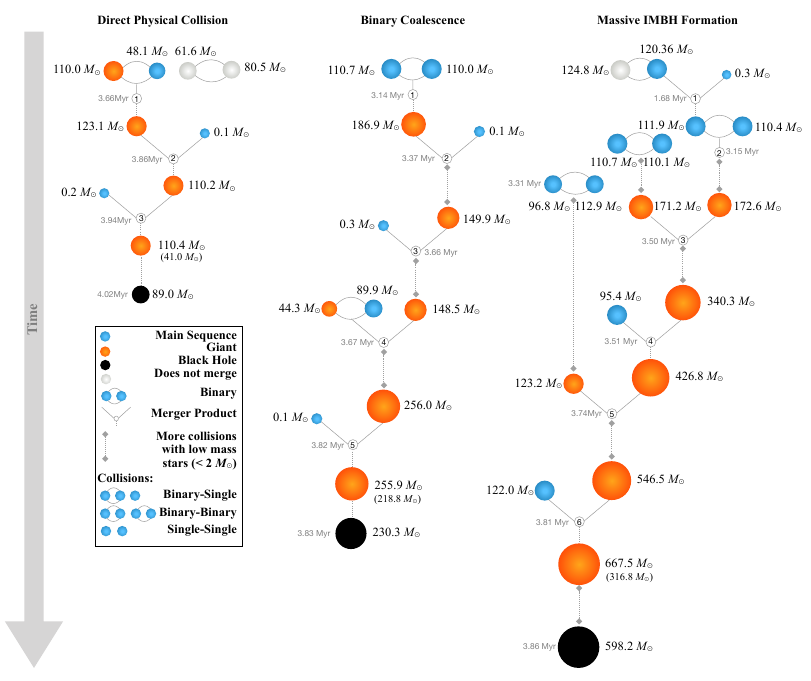}
\caption{
   Cartoon from \citet{2021ApJ...908L..29G} illustrating the formation of an IMBH through mergers during binary evolution in a star cluster simulated using the CMC Monte Carlo \textit{N}-body code \citep{2022ApJS..258...22R}.}
\vspace*{-1mm}
\label{fig:gonzalez-et-al-2021-tree}
\end{figure}

Fig. \ref{fig:kremer-spera-2020-cartoon} taken from \citet{2020ApJ...903...45K} illustrates (in the middle panel) how runaway stellar collisions with a massive evolved giant star in a low metallicity ($Z=0.002\left(0.1 Z_{\odot}\right)$) YMC can form an IMBH of up to 328 $\msun$ by forming a star which can avoid PISN and directly collapse into a BH. The evolved giant undergoes a series of collisions. Additional simulations from \citet{2021ApJ...908L..29G} also found that mergers during binary evolution can also facilitate the formation of an IMBH (as illustrated in Fig. \ref{fig:gonzalez-et-al-2021-tree}) between $\sim 80$ to $600 \ \msun$. This shows that this pathway could be effective in making multiple massive sBH in stellar clusters in which the initial binary fraction of massive stars is significantly high.

Low-mass IMBHs expected to form through this channel could have high birth spins \citep{2020ApJ...888...76M}, therefore if they were to pair up with another BH forming a BBH that would merge due to GW emission then the resulting GW recoil kick will be higher (see Section \ref{subsec:gw-recoil-kicks}). This could potentially eject the seed IMBH from the YMC. Furthermore, this pathway could be an effective way of forming relatively massive sBHs that have been observed with LVK. However, it is sensitive to the treatment of post-stellar merger stars and the prescriptions for their evolution.

\subsubsection{Gas accretion by stellar-mass BHs}\label{subsec:gas-accretion}

Another potential pathway for forming and growing IMBHs in dense environments is through gas accretion onto sBHs. \citet{2005ApJ...628..721K} suggested that radiation drag exerted on interstellar gas from bright stars in young GCs could potentially result in the gas losing angular momentum. This would allow young GCs to accrete the interstellar gas on timescales of 10-100 Myr. This gas would be channelled into the cluster center where it could be accreted by sBHs. This process would only be effective in massive GCs or nuclei of dwarf galaxies more massive than $\sim 5 \times 10^{6} \ \msun$ and can lead to the formation of an IMBH of around $260 \ \msun$.

A similar scenario for forming IMBHs in young GCs was also proposed by \citet{2010ApJ...713L..41V}. In that scenario, stellar material ejected due to wind mass loss from massive giant stars can become centrally concentrated in the core of a cluster. This can increase the central gas density and lead to substantial accretion of gas onto existing sBH resulting in the formation and growth of an IMBH. This gas could also form polluted (second generation) stars in GCs. It has been suggested that accretion of interstellar gas onto an sBH in a massive star cluster can significantly grow a massive sBHs by a few orders of magnitude in only a few tens of Myr \citep{2013MNRAS.429.2997L} and that his accretion could also be responsible for depleting primordial gas in young GCs. 

It has also been suggested that the inflow of a significant amount of gas (where gas mass is comparable to cluster mass) in a GC or a NSC can result in cluster contraction, driving up central density and velocity dispersion such that stellar binaries can no longer effectively heat the cluster. This will lead to core collapse and the possible formation of a $10^{5} \ \msun$ SMBH \citep{2011ApJ...740L..42D,2015ApJ...812...72A,2016MNRAS.461.3620G}

One recently proposed formation mechanism for IMBHs invokes build-up of BH mass in gas-rich NSCs throughout cosmic time \citep{2021MNRAS.501.1413N} . In this scenario, a seed sBH wanders \citep{2023A&A...673A...8D} in the cluster center due to interactions with surrounding stars. This wandering allows the seed to initially grow rapidly through wind-fed accretion. Depending on the amount of gas supply in these NSCs, an IMBH of $10^{2}-10^{5} \ \msun$ could form through this channel.

\subsubsection{IMBH formation in AGN disks}\label{subsec:agn-disks}

It has also been proposed \citep{2011PhRvD..84b4032K,2012MNRAS.425..460M,2014MNRAS.441..900M} that IMBHs can grow efficiently in the gas disk around an SMBH in the nuclei of active galaxies (AGNs) through accretion. Stars \citep{2003MNRAS.339..937G,2004ApJ...608..108G}, compact objects and binary systems trapped in the accretion disk can be driven towards mergers and collisions within this environment. This is because damping by the disk gas can reduce relative velocities between the objects trapped in the disk and this reduction in relative velocity increases the collision rate between objects in the disk.
A seed IMBH can thus efficiently grow by collisions with other objects in the disk and it can also accrete surrounding gas. Since the IMBH can migrate in the disk, it can keep expanding its feeding zone which can permit it to grow quite rapidly through accretion. Migrating objects in these disks can also be trapped at specific parts in the disk due to torques exerted by density perturbations \citep{2016ApJ...819L..17B}.
This could also lead to accumulation and collision between objects trapped in the disk. Furthermore, BBH mergers due to GW emission in AGN disks \citep{2017ApJ...835..165B,2020A&A...638A.119G,2022Natur.603..237S,2023Univ....9..138A} could also lead to hierarchical mergers due to the high escape speeds from these environments \citep{2019PhRvL.123r1101Y,2021ApJ...908..194T}.

\section{Gravitational waves from IMBH mergers with other BHs}\label{sec:gws}

This discovery of the \ac{GW} merger event GW190521 by the ground-based LVK detectors marks the first confirmed discovery of what can be regarded as a low-mass IMBH (as discussed in Section \ref{sec:1.1}). Despite the fact that current ground-based GW detectors are not ideally suited for detecting IMBHs due to their sensitivity to high frequency GW mergers, a handful of BBH mergers leading to the formation of low-mass IMBHs have been detected since GW190521 \citep{2021arXiv211103606T}. More such events could possibly be detected with future observational runs of ground-based LVK detectors \citep{2022A&A...659A..84A,2022MNRAS.516.5309V}. However, better insight into how IMBHs form and what role they play in seeding SMBHs will be provided by  GW observatories\footnote{Sensitive to decihertz \citep{2013CQGra..30p5017B,2019arXiv190811410B,2020CQGra..37u5011A,2021hgwa.bookE..50I,2023arXiv230716628T} and lower frequency GWs} and instruments that are currently being developed and will be deployed over the course of the next few decades\footnote{e.g.,proposed ground and space-based Atom Interferometer (AI) GW detectors \citep{2020CQGra..37v5017C,2020EPJQT...7....6E,2023AVSQS...5d5002T}}. In this section, we discuss formation channels of astrophysical sources of GWs which could potentially be generated by IMBHs and how they may be detectable with current and future GW observatories.


\subsection{Light IMRIs: Mergers between IMBH and sBHs}\label{subsec:light-imri}

If a relatively heavy IMBH ($\gtrsim 10^{3} \ \msun$) can form and be retained in the center of a dense star cluster through one of the many channels highlighted in Section \ref{sec:Channels}) then it may grow by merging with surrounding compact objects and sBHs. These mergers can be sources of GWs that could be detectable with future space-based GW detectors like LISA \citep{2017arXiv170200786A,2023LRR....26....2A}. Such mergers are classified as light intermediate-mass ratio inspirals (IMRI) and their detection would provide immense insights into whether IMBHs can form and grow within our Universe. A few studies \citep{2020ApJ...904L..26F,2021ApJ...907L...9N} that re-examined the LVK detection of GW190521 have suggested that depending on the priors for the binary masses, the merger signal could be interpreted as a light IMRI between a $\sim \rm 170 \ M_{\odot}$ IMBH and a $\sim 16 \ \msun$ sBH \citep{2021ApJ...907L...9N}.  

Light IMRIs could originate in dense ($\rho_{c} \gtrsim 10^{5} \ \msun \ pc^{-3}$) dynamical environments where IMBH formation is likely to occur. If these environments also contain a sizeable population of sBHs then it is likely that these sBHs will segregate in the vicinity of the IMBH due to dynamical friction. Close gravitational encounters between sBHs and an IMBH can potentially lead to formation of an IMBH-sBH binary \citep{1994A&A...283..301P,2004ApJ...616..221G,2004ApJ...613.1143B,2007MNRAS.379...93H,2016ApJ...819...70M,2020ApJ...897...46F} that may harden and merge due to GW emission. If the magnitude of the GW recoil kick from these mergers is less than the escape speed of the cluster then the merged IMBH can be ejected out of the host cluster (as discussed in Section \ref{subsec:gw-recoil-kicks}). However, if it is retained in the environment it may merge again \citep{2020ApJ...897...46F}. 

\citet{2013A&A...557A.135K} carried out \textit{N}-body simulations of GC models with an IMBH ($500-1000 \msun$) at their center and with the inclusion of relativistic effects including energy dissipation due to GWs and recoil kicks. In one of their simulations they were able to obtain an IMRI which resulted in a GW recoil kick which ejected the IMBH from the cluster. \citet{2013A&A...557A.135K} found that GWs emitted by this merger would be detectable with space based detectors like LISA \citep{2007CQGra..24R.113A,2018LRR....21....4A,2020ApJ...897...46F,2023LRR....26....2A} and Tian-Qin \citep{2016CQGra..33c5010L,2023arXiv230716628T}. It could be possible to retain the merged IMBH following an IMRI in a NSCs and in dwarf galaxies. In such environments, the IMBH could continue to dynamically interact with surrounding BHs and may form binary systems that merge \citep{2014MNRAS.444...29L,2015MNRAS.454.3150G,2022MNRAS.515.5106W}. \citet{2016ApJ...832..192H} also carried out \textit{N}-body simulations of star clusters containing stars, sBHs and a central IMBH of up to $200 \ \msun$. They found that a tight IMBH-sBH binary can form and be driven to high eccentricities through dynamical encounters with surrounding stars. Additionally, the IMBH-sBH binary at the center of the cluster can also eject stars and BHs from the clusters in few-body encounters. \citet{2022MNRAS.514.5879M,2022ApJ...940..131G} found that in Monte Carlo \textit{N}-body simulations of dense star clusters that can form IMBHs and contain many sBH, hard IMBH-sBH binaries (separations less than few $\rm R_{\odot}$) can be ejected out of the cluster in a series of strong binary-single and binary-binary encounters. Some of these IMBH-sBH can merge outside their birth cluster due to GW emission. These binaries could also be potential sources of GW radiation that could be detected with ground-based and planned space-based GW detectors (like LISA). IMBH-sBH mergers may also occur during close passages in three (binary-single) or few-body (binary-binary) encounters that could frequently take place in the core of a dense stellar cluster \citep{2015MNRAS.454.3150G,2020ApJ...897...46F,2021A&A...650A.189A,2023arXiv230307421D}.

\begin{figure}[t]
\centering
\includegraphics[width = 0.85\textwidth]{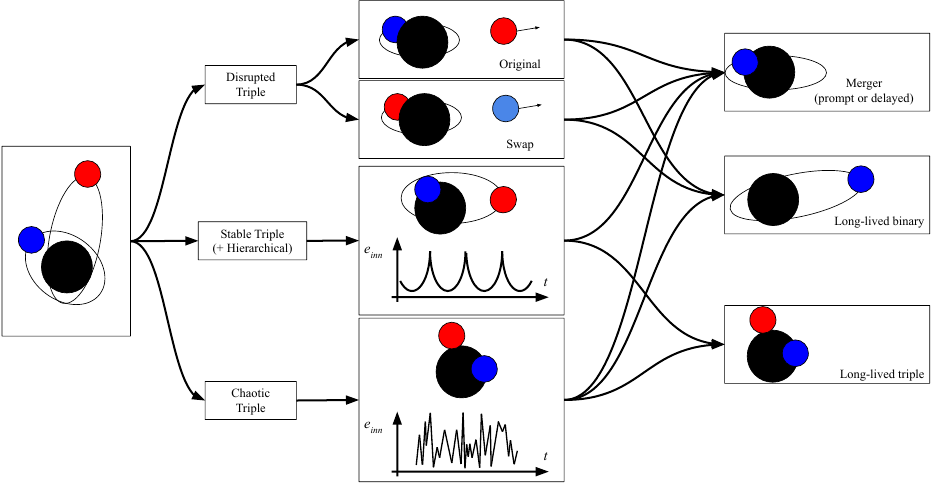}
\caption{
    Schematic taken from \citet{2021A&A...652A..54A} (Fig. 2) showing possible outcomes of IMBH-sBH-sBH triple evolution in a dense environment that could lead to light IMRI formation. Mergers between an IMBH and an sBH through this channel will be detectable with future GW observatories.}
\vspace*{-1mm}
\label{fig:imbh-bh-bh-triple}
\end{figure}

The presence of a central IMBH in a dense star cluster could also lead to the formation of triple systems with binary sBH (see Fig.\ref{fig:imbh-bh-bh-triple} that shows possible outcome of triple evolution). In that case, the binary sBH may be driven to high eccentricties via the ZLK mechanism which could result in eccentric mergers \citep{2019MNRAS.488.4370F,2020ApJ...903...67M,2021A&A...652A..54A}. A chaotic triple interaction may also result in an IMBH-sBH binary merger \citep{2021A&A...652A..54A}. Results from numerical simulations suggest that the presence of an IMBH at a center of a cluster is likely to lead to the formation of an IMBH-sBH binary and other binary sBHs may be dynamically ejected or disrupted in such clusters \citep{2007MNRAS.374..857T,2014MNRAS.444...29L,2020MNRAS.498.4287H,2020MNRAS.499.4646A,2021MNRAS.508.4385A}. Furthermore, IMBH-sBH binary may also be driven to high eccentricity in an NSC which contains an SMBH \citep{2020ApJ...901..125D}.

Depending on how effective IMBH formation is in dense environments, it has been predicted that LISA might be able to detect 0.01-60 light IMRIs from stellar clusters per year (out to a redshift of $z \sim 2$ \citep{2020CQGra..37u5011A,2021A&A...652A..54A,2023Univ....9..138A}. The proposed space-based DECi-hertz Interferometer GW Observatory (DECIGO) could detect 6-3000 IMRIs \citep{2021A&A...652A..54A} per year and with planned third-generation, underground GW detectors like the Einstein Telescope \citep[ET,]{2011GReGr..43..485G,2021arXiv211204058S}, up to 1-600 IMRIs from GCs could be detected \citep{2021A&A...652A..54A}. \citet{2020ApJ...897...46F} predict that eccenntric IMBH-sBH mergers occurring in galactic nuclei in which the IMBH has a mass ranging between $5000-20000 \ \msun$ will have peak GW frequencies between $\sim 0.05-0.2 \ \rm Hz$ and willl be detectable with DECIGO and ET.

Multiple channels for forming an IMBH (described in Section \ref{sec:Channels}) can operate in the dense dynamical environments of NSCs found at the center of most galaxies \citep{2020A&ARv..28....4N,2022ApJ...929...84B}. These environments with their high stellar density, high escape speeds and presence of gas could be perfect nurseries for forming/and or growing IMBHs that can seed the formation of SMBHs (see also {\color{blue}\textbf{Chapter 3.IV}}). If an IMBH (or a seed SMBH) forms in-situ or is delivered to an NSC then this seed can grow through repeated mergers with stellar BHs \citep{2022ApJ...933..170F}. Depending on initial seed BH mass, the birth spins of sBHs and the formation history of the NSC, \citet{2022ApJ...933..170F} predict that the maximum number of IMBH-sBH binary mergers could range from $\sim 10^{-3}$ to $\sim 100$ mergers per year for LISA, $\sim 10^{-2}$ to $\sim 10^{4}$ mergers per year for DECIGO, $\sim 10^{-2}$ to $\sim 10^{4}$ for ET. 

IMBH could also form in massive star-forming clumps of gas\footnote{These massive and dense clumps are likely to be proto star clusters.} that are observed in a large fraction of massive galaxies at redshifts between $z \sim 1-3$ \citep{2016ApJ...821...72S,2021MNRAS.500.4628P}. IMBHs can grow within these massive stellar clumps through GW mergers with sBHs and other compact objects. \citet{2021MNRAS.500.4628P} use Fokker-Planck simulations to estimate that $10^{-7} \mathrm{yr}^{-1}$ IMBH-sBH mergers could occur in these massive clumps and GWs from such mergers could be detected with LISA. 

In addition to IMRIs forming dynamically in star clusters, it is also possible to get light IMRIs in AGN disks \citep{2023LRR....26....2A}. If IMBHs can form and grow through the channel described in Section \ref{subsec:agn-disks} then it could possibly merge with an sBH \citep{2020MNRAS.498.4088M}. Particularly, if this IMBH sits in a migration trap within the AGN disk \citep{2016ApJ...819L..17B}. It has also been suggested that IMBH-sBH mergers in these environments could have high eccentricity \citep{2021ApJ...907L..20T,2022Natur.603..237S,2022arXiv220406002M}.

\subsection{Binary IMBHs}\label{subsec:binary-imbh}

If two or more light IMBHs ($\gtrsim 100 \ \msun$) can form (via the channels described in Section \ref{sec:Channels}) and end up in a dense stellar environment then there is a possibility that the IMBHs can pair up and merge due to emission of GWs \citep{2020CQGra..37u5011A}. These binary IMBH mergers could potentially play a role in seeding the formation of SMBHs \citet{2004ApJ...614..864M,2019BAAS...51c.432C}. For instance, some GCs might be initially dense enough to have multiple VMS form through collisional runaways \citep{2006ApJ...640L..39G} (or binary mergers) and this could lead to the formation of a binary IMBH inside the cluster \citep{2005MNRAS.364.1315M,2006ApJ...646L.135F,2020ApJ...899..149R}. Such binaries may also form dynamically at high redshifts within pop III and metal-poor star clusters which may contain multiple IMBH progenitors \citep{2019BAAS...51c.175B,2020ARA&A..58..257G,2021MNRAS.505L..69H}. Binary hardening in a dense environment due to interactions with surrounding stars could drive the IMBH binary to merger due to GW radiation. While the formation of a binary IMBH may be rare in GCs, \citet{2020ApJ...899..149R} estimate that one inspiralling binary IMBH from a GC could be detected by LISA during its planned 4 year mission. With planned improvements in sensitivity, ground-based LVK detectors are expected to better observe binary mergers in which one (in the case of light IMRIs) or both of the BHs have masses in the range of $\sim 100-500 \ \msun$ \citep{2014PhRvD..90f3002M,2018PhRvD..97l3003K,2022A&A...659A..84A,2022ApJ...924...39M} out to cosmological distances.

\begin{figure}[t]
\centering
\includegraphics[width = 0.6\textwidth]{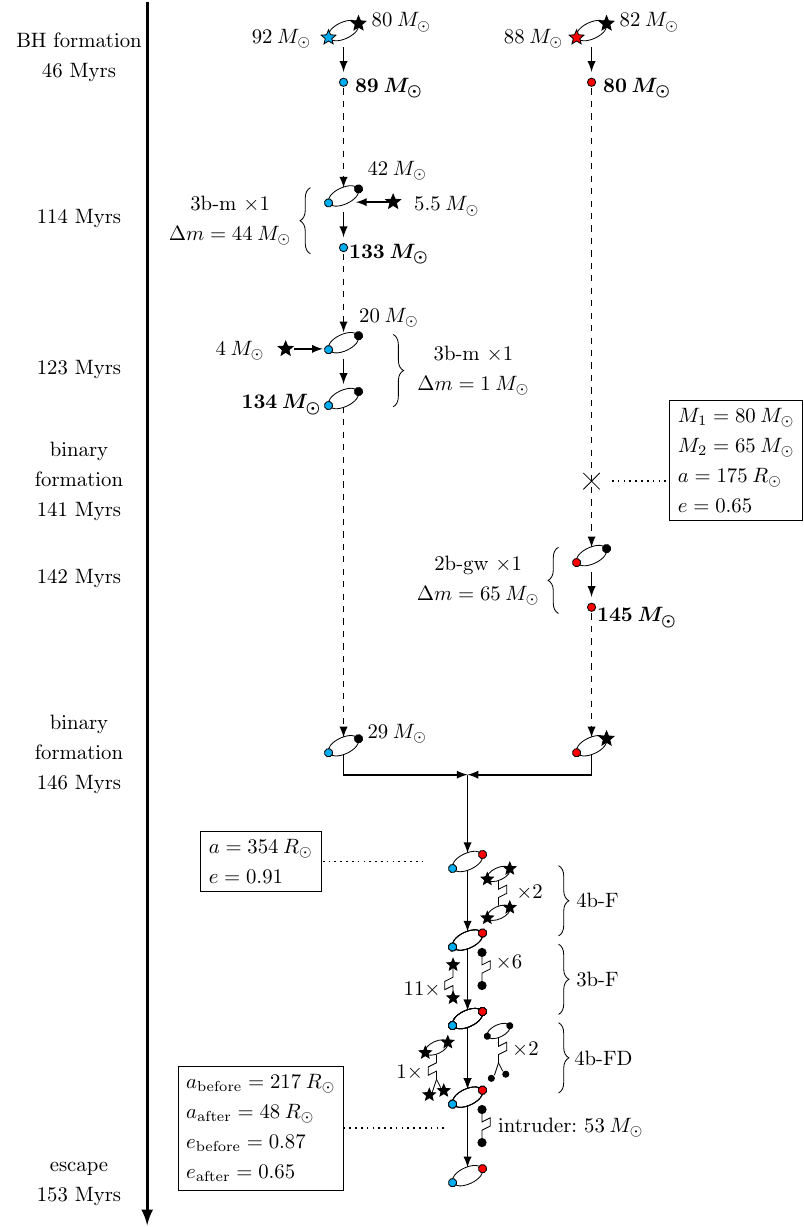}
\caption{
    Illustration taken from \citet{2022MNRAS.514.5879M} showing the formation history of an escaping binary IMBH in a star cluster simulated using the \textsc{MOCCA} Monte Carlo \textit{N}-body code. Abbreviations used in the description of events: 3b-m -- merger due to three-body interaction, 2b-gw -- BBH merger due to the emission of GW, 4b-F -- four-body flyby, 4b-FD -- four-body flyby resulting in binary disruption, 3b-F -- three-body flyby.}
\vspace*{-1mm}
\label{fig:maliszewski-escaping-imbh-history}
\end{figure}

\citet{2022MNRAS.514.5879M} also found that a binary IMBH with component masses of $145 \ \msun$ and $134 \ \msun$ forms due to exchange encounters in a Monte Carlo \textit{N}-body simulations of a GC model which had an initial density of $2.5 \times 10^{7} \ \msun \rm \ pc^{-3}$ with more than a million objects. This IMBH binary is then ejected from the cluster in a binary-single scattering encounter with a $53 \ \msun$ sBH. The formation history of this binary is illustrated in Fig. \ref{fig:maliszewski-escaping-imbh-history}.

\begin{figure}[t]
\centering
\includegraphics[width = 0.8\textwidth]{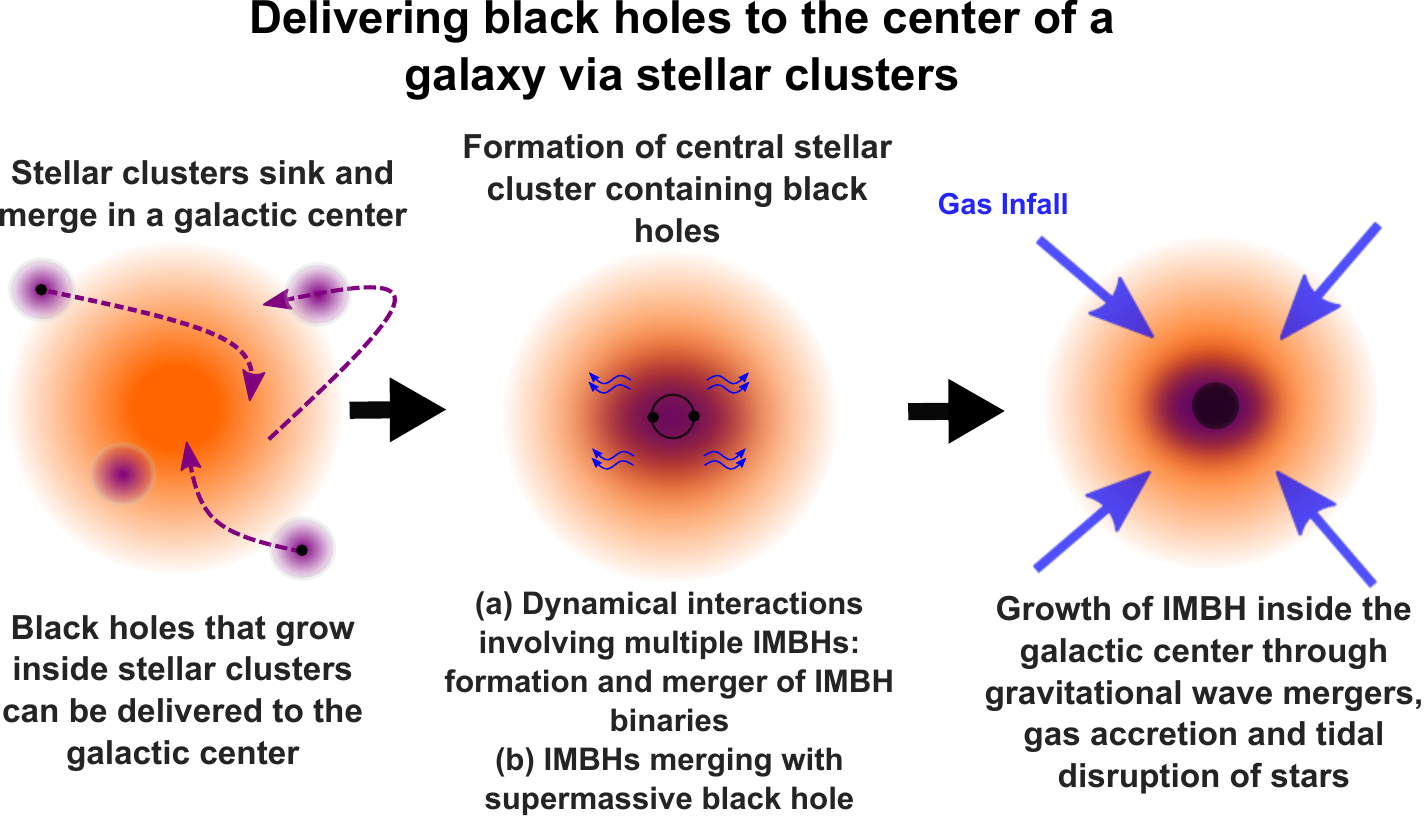}
\caption{
    Illustration showing how star clusters can sink to form an \ac{NSC} in a galactic nucleus. These clusters can deliver \acp{IMBH} to the nucleus of the Galaxy. These IMBHs can form binaries and merge with each other or they may merge with the \ac{SMBH} resulting in \acp{IMRI}.}
\vspace*{-1mm}
\label{fig:formation-nsc-imbh-cartoon}
\end{figure}

It has also been shown that a binary IMBH could form via the merger of two or more stellar clusters that contain an IMBH. Using \textit{N}-body simulations, \citet{2019arXiv190605864A} found that the merger of two GCs hosting IMBHs results in the formation of an IMBH binary that can efficiently harden and become eccentric through interactions with surrounding stars leading to GW mergers. Dense and massive clusters that form IMBHs can also sink and merge in the nucleus of their host galaxy due to dynamical friction. Thereby, they can effectively deliver multiple IMBHs to a galactic nucleus \citep{2001ApJ...562L..19E,2006ApJ...641..319P,2017MNRAS.464.3060A}. \citet{2021MNRAS.502.2682A} simulated the merger of three stellar clusters in the nucleus of a galaxy with no pre-existing SMBH. They found that in the runs where two of three merging clusters hosted an IMBH in their center, the IMBHs segregated to the center of the merged cluster on timescale of few tens of Myr and formed a binary sytem which also gradually hardened and became more eccentric. These binaries (with component mass ratios ranging between 0.1 to 0.5) can merge efficiently due to GW radiation on the timescale of a few hundred Myr. If the merged IMBH is retained in the NSC, then it could pair up with other IMBHs that were subsequently delivered to the NSC through dense infalling star clusters \citep{2022MNRAS.511.2631A}. Growth of an IMBH through these repeated mergers with other IMBHs could be a potential pathway for seeding SMBHs \citep{2021ApJ...923..146G}. Additionally, this seed IMBH could also grow further through gas accretion (see Section \ref{subsec:gas-accretion}) or tidal disruptions/capture of stars (see Section \ref{subsec:slow-tde-tdc}) in the NSC \citep{2023ApJ...944..109L}. This process for forming binary IMBHs and potentially seeding an SMBH is illustrated in Fig. \ref{fig:formation-nsc-imbh-cartoon}. The formation of merging binary IMBHs through this pathway will depend on the growth history of an NSC and the likelihood of IMBH formation within stellar clusters that merge with the NSC \citet{2022MNRAS.511.2631A}. GW recoil kicks (see Section \ref{subsec:gw-recoil-kicks}) from binary IMBH mergers could also eject IMBHs from galactic nuclei that have low escape speeds. If SMBHs are seeded from binary IMBH mergers then these GW recoil kicks can potentially decrease SMBH occupation fraction in low mass and density NSCs. These GW recoil kicks could also produce IMBHs that could be wandering inside a galaxy \citep{2022MNRAS.515.2110S,2022MNRAS.511.2229W,2022arXiv221014960D}. 

Results from direct \textit{N}-body simulation of an extremely dense ($\rho_c \sim 4.4 \times 10^{15} \ \mathrm{M}_{\odot} \mathrm{pc}^{-3}$) cluster of IMBHs\footnote{400 IMBHs each with a mass of $10^{4} \ \msun$} found that runaway GW mergers between IMBHs can grow and form a single massive BH which makes up for about 25 per cent of total cluster mass \citep{2021MNRAS.504.3909C}. The build-up of an IMBH through these runaway mergers could be observed with LISA and that may help constraining the viability of this seeding mechanism.

Direct \textit{N}-body simulations of the evolution of a binary IMBH in the center of nucleated dwarf galaxy by \citet{2021MNRAS.508.1174K} also showed that the binary can efficiently harden due to dynamical encounters and can merge due to GW radiation. Therefore, mergers between dwarf galaxies that could each be harbouring an IMBH could lead to the both IMBHs ending up in the nucleus of the merged galaxy. These IMBHs can form a binary system that could merge due to GW emission which could be observed with LISA. Results from \textit{N}-body simulations \citep{2020MNRAS.496..921W} have also demonstrated that IMBHs can sink to the center of ultracompact dwarf galaxies (UCDs) through dynamical friction. The timescale for IMBHs to sink to the center of an UCD depends on the mass ratio between an IMBH and its host UCD. IMBHs in the center of UCDs are likely to end up in binary systems and GWs fom such binaries could potentially be observed with LISA \citep{2020MNRAS.496..921W}. \citet{2018ApJ...864L..19T} investigated binary IMBH formation and evolution in dwarf galaxy mergers with collisionless \textit{N}-body simulations and found that dwarf galaxies with cuspy dark matter profiles are more favourable for forming hard IMBH binaries that will merge due to GW emission. 

It has also been suggested the presence of an SMBH could also lead to the formation of a hierarchical triple system in which the inner binary comprises two IMBHs and the outer perturber is an SMBH. Such binaries may also be driven towards merger through eccentricity oscillations due to the ZKL mechanism \citep{2011GReGr..43..485G,2019MNRAS.483..152A}.

If there is a gas disk around an accreting IMBH then other IMBHs may form or grow in that gas disk. This could result in the formation of a binary IMBH that is embedded in a gas disk \citep{2022MNRAS.517.1339G}. The presence of gas could influence the orbital evolution of this binary as it evolves due to GW emission and this may produce detectable signatures in GW waveform \citep{2022MNRAS.517.1339G}.

\subsection{Heavy IMRIs: Mergers between IMBH and SMBHs}\label{subsec:heavy-imri}

Another class of IMBH mergers are heavy IMRIs. These are IMRIs between an SMBH and an IMBH. Similar to IMBH-sBH binary mergers, the mass ratio of the merging components for heavy IMRIs is also between $10^{-5}-10^{-2}$ \citep{2010ApJ...713.1016H,2023LRR....26....2A}. SMBH-IMBH binary mergers are expected to occur in the nuclei of galaxies where SMBHs are typically located. Both galaxy mergers and the assembly of a galactic nucleus can contribute to the formation of heavy IMRIs. Future GW instruments and observatories that will be sensitive at low-frequencies are expected to detect SMBH-IMBH mergers. This will enable constraints to be placed on how SMBHs grow and whether they merge with IMBHs.

As previously discussed, masses of SMBH correlate with the velocity dispersion and mass of their host galaxy. Therefore, the nuclei of dwarf and low-mass galaxies (with stellar masses between $\left(\sim 10^7-10^9 \mathrm{M}_{\odot}\right)$) could harbour IMBHs in their nuclei \citep{2018MNRAS.478.2576M,2018ApJ...863....1C,2021ApJ...918...18W}. Relatively massive galaxies evolve and build up through mergers with dwarf galaxies. These mergers can lead to IMBHs ending up in the center of the nuclei of the more massive galaxy through dynamical friction. This IMBH can then pair up with the SMBH existing in the larger galaxy and this process can lead to the formation of heavy IMRIs \citep[see][and references therein]{2020MNRAS.498.2219V,2023LRR....26....2A}. In such hierarchical galaxy mergers, the timescales for the actual merger of the SMBH-IMBH binary due to GW emission could be substantially long as it would depend on the time it takes for the galaxy to merge, the orbit of the IMBH and how long it takes for it to sink to the center due to dynamical friction \citep{2019MNRAS.482.2913B,2020MNRAS.498.2219V}.

Similarly, dense and massive clusters that could be potential sites for forming IMBHs can also merge with the nucleus of a galaxy through dynamical friction. Thereby, they can deliver IMBHs to the nucleus of galaxy that may already have an SMBH at its center. The timescale for a cluster to merge with the center of its host galaxy will depend on its birth position. Clusters that are born close or end up close to the center of their host galaxy will infall due to dynamical friction and merge with the galactic nucleus on a shorter timescale \cite{2008gady.book.....B}. This process can lead to the formation and/or growth of a NSC and can also form heavy IMRIs (as illustrated in Fig. \ref{fig:formation-nsc-imbh-cartoon}). Several studies \citep[e.g.,][]{2005ApJ...635..341L,2014ApJ...796...40M,2017MNRAS.467.3775P,2018MNRAS.477.4423A,2019MNRAS.483..152A,2022ApJ...939...97F} have investigated the scenario whereby IMBHs can be delivered to an NSC which contains an SMBH. Using direct \textit{N}-body simulations, \citet{2018MNRAS.477.4423A} found that dense star clusters can efficiently deliver IMBHs to an NSC where they can merge with the SMBH due to GW radiation on timescales of 3-4 Gyr and the expected merger rate for such events is $\sim 3 \times 10^{-3} \ \mathrm{yr}^{-1} \mathrm{Gpc}^{-3}$. \citet{2022ApJ...939...97F} developed a semi-analytical framework to model SMBH-IMBH binary formation and found the comoving merger rate to be $\sim 10^{-4} \ \mathrm{yr}^{-1}\mathrm{Gpc}^{-3} $. With LISA, these merger events could be observed up to high redshifts and \citet{2022ApJ...939...97F} predict that more than 90 per cent of such mergers will have signal-to-noise ratio larger than 10. 

Heavy IMRIs may also form in AGN disks. If an IMBH forms within AGN disks (as described in Section \ref{subsec:agn-disks}) then this IMBH will form a binary system with the SMBH. The SMBH-IMBH binaries formed through this channel can merge due to GW emission on timescales of a few hundred Myr \citep{2016ApJ...819L..17B,2023LRR....26....2A}. Results from \textit{N}-body simulations investigating the dynamical evolution of an IMBH inside an NSC that contains an SMBH and a nuclear stellar disk find that an IMBH can warp the disk \citep{2021ApJ...919..140S}. This process can accelerate the capture of an IMBHs in an AGN disk.  

\section{Summary} \label{sec:summary}

IMBHs with masses between $10^{2}-10^{5} \ \msun$ are crucial to our understanding of BH demographics in our Universe. They are the link between two well-observed categories of BHs: sBHs ($< 100 \ \msun$) and SMBHs ($\gtrsim 10^{6} \ \msun$). In this part of the chapter, we highlighted various mechanisms for IMBH formation. We focused on explaining the proposed pathways by which IMBHs can form and grow within dense star clusters. Several theoretical and computational studies have shown how the interplay between gravitational dynamics and stellar/binary evolution can facilitate IMBH formation within dense stellar environments. Additionally, the presence of gas in star clusters and in galactic nuclei can also assist in the formation and growth of IMBHs. Below, we summarize IMBH formation and growth channels that were explained in this chapter. 

 \begin{itemize}

\item \textbf{Evolution of Pop III and metal-poor stars:} The stellar evolution of massive metal-free (pop III stars) or extremely metal-poor stars formed in the early Universe could lead to the formation of an IMBH in the mass range of $\sim 100-1000 \ \msun$. The formation of an IMBH through this pathway depends on a number of uncertain initial parameters and physical processes involved in the evolution of these stars. Among others, this includes limits on their initial masses, their exact mass function and multiplicity, the mass they lose through stellar winds. IMBHs that potentially form through this pathway could have grown further by accreting surrounding gas.

\item \textbf{Repeated/hierarchical BH mergers:} \acp{sBH} that are retained in a star cluster will segregate to the dense core of the cluster due to dynamical friction. Frequent gravitational encounters between BHs in the core of a star cluster can lead to BBH mergers through GW emission. If the merged BH is retained in the cluster it is likely to dynamically pair up and merge within another BH. This process of repeated mergers can result in the growth of an sBH into an IMBH. The efficacy of this IMBH formation channel depends on a number of uncertainties. These include how many sBHs can be initially retained in a given star cluster which depends on the uncertain natal kicks which sBHs receive at birth. Whether the merged BH can be retained in the cluster after the GW recoil kick it will receive also affects the efficacy of this scenario in building up IMBH mass. GW recoil kick velocity (can be up to thousands of $\kms$) depends on the mass ratio of the merging BHs and their uncertain spin magnitudes and orientation. If this velocity exceeds the escape speed of the host environment then the merged BH will be ejected. Therefore, the repeated/hierarchical BH mergers pathway for forming and growing IMBHs is most likely to be effective in dense GCs and NSCs that have high escape speeds. 

\item \textbf{Runaway stellar collisions:}  In dense star clusters with short initial half-mass relaxation times, massive stars can rapidly ($\lesssim 5$ Myr) segregate to the cluster center where they can undergo collisions and build up a very massive star ($\gtrsim 250 \ \msun$). Depending on how this VMS evolves, it could form an IMBH. This pathway for IMBH formation will be most effective in dense low metallicity ($\rm \lesssim 0.05 Z_{\odot}$) star clusters. Mass loss through stellar winds is expected to be weaker for low metallicity stars and a VMS which forms in this environment is likely to directly collapse to form an IMBH. This formation scenario does not require the retention of sBHs.

\item \textbf{Close encounters between sBHs and stars:} An IMBH may also form through the collision between an sBH and a massive star. Such collisions are likely to occur in the core of a dense cluster that retains sBHs. IMBH formation and growth through this channel depends on the uncertain amount of mass absorbed by the sBH in such a collision event. Collisions between an sBH and a VMS in which the sBH accretes even a small fraction ($\lesssim 0.2$) of the VMS can result in the formation of an IMBH. However, if the accreted mass is significantly lower then this can inhibit IMBH formation and growth through this pathway.

Tidal disruption and/or capture of stars by BHs during close encounters may also lead to gradual growth of an sBH through accretion of the disrupted star. This could potentially result in the slow formation ($\gtrsim 100$ Myr) of an IMBH.

\item \textbf{Binary and stellar mergers:} Stellar mergers between an evolved massive star and a massive MS star during dynamical encounters or during binary stellar evolution could also result in the rapid formation of a low-mass \ac{IMBH}. These merger products could potentially avoid pair and pulsational pair instability supernovae and could directly collapse to form IMBHs. The efficacy of this pathway depends on the uncertain properties, structure and subsequent evolution of the merged star.

\item \textbf{Accreting sBHs in gas rich environments:}  sBH embedded in gas rich environments can potentially grow to become IMBHs. This pathway depends on the availability of gas and the efficiency of its accretion. sBH trapped in disks of active galactic nuclei (AGNs) can also significantly grow through gas accretion in order to produce IMBHs.

\end{itemize}

If IMBHs can form via one or more of the processes described above then they could end up forming GW sources. 
The expected peak frequencies of these GW sources are lower than the frequencies to which current ground-based LVK detectors are sensitive. Therefore, GW sources comprising IMBHs are expected to be detected with planned and proposed next-generation of space and ground-based GW detectors that include LISA, ALIA, DECIGO, TianQin/Taiji, Einstein Telescope and Cosmic Explorer. The detection of these GW sources will enable us to better understand IMBH demographics and constrain their formation pathways. Furthermore, the detection of GW sources from IMBHs will also shed light on their role in potentially seeding and contributing to the growth of SMBHs. Below, we summarize the different type of GW sources in which an IMBH may merge with an sBH, another IMBH or an SMBH and how they may form.

\begin{itemize}
    \item \textbf{IMBH-sBH mergers,} also known as light intermediate-mass ratio inspirals (IMRIs). These GW sources are expected to be produced in dense environments like star clusters and galactic nuclei where an IMBH could merge with an sBH through various dynamical pathways. The mass ratio of the merging BHs for light IMRIs is $10^{-5}-10^{-2}$. Their expected merger rate depends on the efficacy of IMBH formation pathways. Depending on their properties, these binaries and their mergers are expected to generate GWs peaking in the decihertz range. They will be detectable through space-based detectors ALIA, DECIGO, LISA and TianQin and could also be detected with the next generation ground-based GW detectors and even potentially with LVK\footnote{at design sensitivity and for relatively low-mass IMBHs}.

    \item \textbf{Binary IMBH mergers:} If an IMBH is in a binary system with another IMBH then the evolution of this binary and its merger through GW emission will be detectable with observatories sensitive to decihertz frequencies. Such binary systems are likely to dynamically form in dense environments (e.g., NSCs, cores of merging dwarf galaxies, dense and massive GCs) where dynamical encounters can lead to their merger. Their formation and subsequent evolution could be closely linked with how galaxies form and grow their nuclei and SMBHs.

    \item \textbf{SMBH-IMBH mergers,} also known as heavy IMRIs are expected to occur in the nuclei of galaxies where an SMBH could pair up with an IMBH. The IMBH could either form in-situ (e.g., in an AGN disk around the SMBH) or be delivered to the nucleus by either merging dense and massive star clusters or dwarf galaxies. These mergers could be a key process in enabling the growth of SMBHs in the early Universe. The expected merger rate for SMBH-IMBH binaries depends on the efficacy of IMBH formation pathways and the assembly history of a galaxy and its nucleus. These binaries and their mergers are expected to generate low-frequency GWs that will be detected with spaced-based detectors like LISA.

\end{itemize}

\part{Intermediate-Mass Black Holes in Star Clusters and Dwarf Galaxies: Observations 
\\ {\large Authors: Vivienne F. Baldassare and Mar Mezcua}}

\section{Introduction}
\label{sec:Introduction}

The population of BHs in globular clusters (GCs) and dwarf galaxies can provide constraints for theories of BH formation \cite{2020ARA&A..58..257G}. The SMBHs that reside in the centers of all massive galaxies, with typical masses larger than several $10^5\Ms$ and up to $10^{10}\Ms$, are thought to form at high-redshift, though the mechanism by which these ``BH seeds" form remains unclear. However, different SMBH seed formation theories make different predictions for the population of SMBHs in dwarf galaxies. For example, if high-redshift SMBH seeds form through the deaths of Population III stars, we expect most dwarf galaxies today to contain an SMBH (i.e., SMBH occupation fraction $\sim$100\%). This also predicts a continuum of masses from the initial seed mass (which could be as low as $100\;M_{\odot}$) up through SMBHs with masses of $10^{10}\;M_{\odot}$. On the other hand, SMBH seeds which form through direct collapse could start at $10^{5}\;M_{\odot}$ and are expected to have been less common, yielding a MBH occupation fraction in today's dwarf galaxies of $\sim$50\%. This would also predict a dearth of SMBHs in galaxy centres with masses less than $10^{5}\;\rm{M_{\odot}}$. These different formation scenarios also make disparate predictions for SMBH-galaxy scaling relations, such as the $M_{\rm BH}-\sigma_{\ast}$ relation between SMBH mass and host galaxy stellar velocity dispersion \cite{2000ApJ...539L..13G, 2000ApJ...539L...9F, 2020ApJ...898L...3B}.
There is also a third possible SMBH formation pathway that does not rely only on conditions found at high redshift. Some theoretical works predict that IMBHs could form in dense stellar environments (see Section \ref{Sec21}), including GCs and the dense nuclear star clusters (NSCs) that reside at the centers of most dwarf galaxies. 

\begin{figure}
    \centering
    \includegraphics[width=0.7\textwidth]{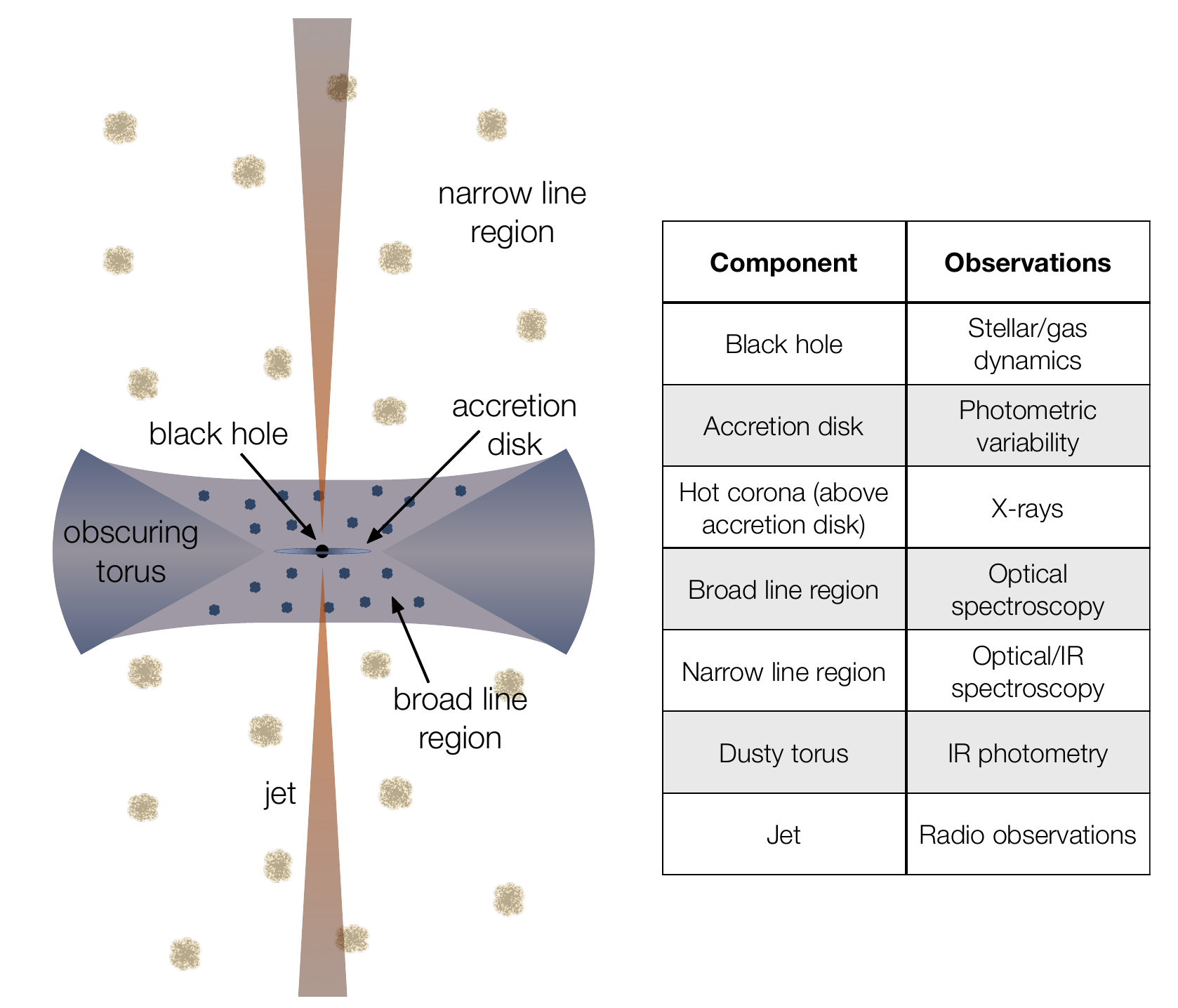}
    \caption{Simplified AGN schematic adapted from \cite{2017PhDT.......196B}. We list out techniques that are commonly used to probe each component of the AGN.}
    \label{fig:schematic}
\end{figure}

In this chapter, we review multi-wavelength evidence for MBHs in GCs and dwarf galaxies\footnote{Owing to the blurred transition region from IMBHs to SMBHs, with massive black holes (MBHs) we hereby refer to all objects falling in the mass range between a few $10^{2}\Ms$ and a few $10^{6}\Ms$.}. We also discuss future prospects for detection of IMBHs in these environments using gravitational wave observatories. Detecting IMBHs can be challenging; in the IMBH regime, it is much easier for host galaxy light to outshine the central MBH, or for stellar processes to produce observational signatures that mimic those of MBHs. Thus, many different selection techniques are used to identify MBHs in GCs and dwarf galaxies, each with their own advantages and limitations. At the end of this introduction, we provide a summary table of the different methods discussed throughout this chapter (Table~\ref{tab:det_summary}) and the challenges associated with each. In Figure~\ref{fig:schematic}, we show a simplified schematic of an active galactic nucleus (AGN; figure adapted from \cite{2017PhDT.......196B}) and list the techniques that can probe each component.

\begin{table}[h]
    \centering 
    \begin{tabular}{| p{1.5in} | p{2.3in} | p{2.3in} |}
        \hline
        \textbf{Selection technique} & \textbf{Description} & \textbf{Limitations}  \\
        \hline
        Dynamical detection & Presence of BH inferred from stellar or gas dynamics & Requires high spatial resolution, only feasible for nearby ($\lesssim5\; \rm{Mpc}$) objects \\ 
        \hline
        Optical spectroscopy (broad Balmer lines) & Presence of BH inferred by Doppler broadened emission lines originating near the BH & Confusion with lines broadened by supernovae and massive stars \\ 
        \hline
        Optical spectroscopy (narrow lines) & Presence of relatively strong high-ionization lines from excitation by optical/UV accretion disk photons & Signature can be diluted by strong star formation, diagnostics not calibrated for low-metallicity galaxies. \\
        \hline
        Photometric variability & Detection of stochastic variability from an AGN accretion disk & Requires repeat imaging observations over months-to-years, variability can be hard to detect for IMBHs \\ 
        \hline
        Infrared color diagnostics & IR colors indicative of dust heated by the AGN & Extreme star formation in low-mass galaxies can replicate AGN IR colors \\
        \hline 
        Infrared coronal lines & Presence of high-ionization coronal lines photoionized by UV/X-ray accretion disk photons & Observationally expensive, potential confusion with lines generated by supernovae  \\
        \hline
        X-ray detection & Detection of X-rays from the hot X-ray emitting corona of an AGN & At IMBH masses, there is potential confusion with X-rays emitted by X-ray binaries \\ 
        \hline 
        Radio detection & Detection of compact radio continuum or extended jet emission & Potential confusion with star-formation processes (e.g., supernovae, supernova remnants)   \\ 
        \hline
             
    \end{tabular}
    \caption{Summary of BH detection techniques and limitations.}
    \label{tab:det_summary}
\end{table}

\section{Globular clusters}
\label{sec:GCs}

\subsection{Dynamical BH mass measurements}
\label{sec:GCdynamical}

Globular clusters (GCs) have long been suggested as possible hosts for IMBHs with masses in the range $10^2-10^5\Ms$ \cite{2002MNRAS.330..232C}. Though there have been many studies finding some evidence for IMBHs in globular clusters, there has been no definitive detection. Some of the strongest candidates so far are the Milky Way GC $\omega$ Cen and the M31 GC G1. Several papers have claimed a $10^{4-5}\;M_{\odot}$ IMBH in $\omega$ Cen based on kinematics \cite{2008ApJ...676.1008N, 2010A&A...514A..52M, 2010ApJ...719L..60N, 2012A&A...538A..19J}. However, other works have failed to replicate those results, finding that models with a flat core do not require a central massive object \cite{2010ApJ...710.1063V} or that the kinematics can be explained with a dense cluster of stellar-mass BHs \cite{2019MNRAS.488.5340B}. In the case of G1, there are similar controversies, with some studies finding evidence for a $\sim10^{4}\;M_{\odot}$ IMBH \cite{2002ApJ...578L..41G, 2005ApJ...634.1093G} and others finding that the data is well-described without invoking a IMBH \cite{2003ApJ...589L..25B}. Recently, high-resolution integral field spectroscopy was used to discover a $10^{5}\;M_{\odot}$ IMBH in B023-G078, the most massive GC of M31 \cite{2022ApJ...924...48P}. However, this object raises the interesting issue of distinguishing between GCs and stripped galaxy nuclei, as they present compelling evidence that B023-G078 is a stripped nucleus. Going forward, integral field spectroscopy observations with JWST may help better constrain the dynamics at the centres of nearby GCs and yield definitive results. 

\subsection{Signatures of BH accretion}
\label{sec:GCaccretion}
GCs have little gas and dust for the putative IMBH to be actively accreting. However, very low accretion rates (sub-Eddington rates of $<$ 2\%) are expected from the gas supplied by the evolving stars in the GC (\cite{Strader2012}), which should produce a synchrotron-emitting jet detectable in deep radio observations (e.g., \cite{Maccarone2004, Maccarone2005}). Most efforts to detect such radio signatures have however yielded no compelling evidence for the presence of an IMBH of more than 1000 \Ms~ in GGs (\cite{Maccarone2008}; \cite{Strader2012}; \cite{Wrobel2015}; \cite{Tremou2018}; \cite{Wrobel2020}). 

No definitive X-ray detections consistent with an IMBH have been found either in deep \textit{Chandra} observations (e.g., \cite{Haggard2013}; \cite{Lin2015}). \cite{Lin2015} find 0.3-8 keV luminosities of L$_\mathrm{X} \sim$ 10$^{36}$-10$^{39}$ erg s$^{-1}$ in 49 GCs in the nearby lenticular galaxy NGC 3115, but conclude these are consistent with low-mass X-ray binaries. There have also been ultraluminous X-ray sources detected in many GCs (e.g., \cite{2021MNRAS.504.1545D, 2023MNRAS.518.3386T, 2023MNRAS.518..855A}), the origin of which remains unclear.

The cluster G1 in M31 is the only GC for which the detection of both X-ray (L$_\mathrm{X} \sim$ 2 $\times$ 10$^{36}$ erg s$^{-1}$; \cite{Pooley2006}) and radio emission (\cite{Ulvestad2007}) have been reported. However, later deep high-angular resolution radio observations by \cite{Miller-Jones2012} ruled out the detection of radio emission, concluding that the X-ray emission originated most likely from a low-mass X-ray binary.

The next-generation Very Large Array (ngVLA) offers a promising prospect for detecting IMBHs in GCs in the radio regime (\cite{Wrobel2021}; \cite{2022MNRAS.515.2110S}).

\section{Dwarf galaxies}
\label{sec:dwarves}

\subsection{Dynamical BH masses}
\label{sec:dwarvesdynamical}

Dynamical mass measurements of MBHs involve studying the motion of stars or gas in the immediate vicinity of the MBH. This is one of the only ways to measure the mass of a quiescent MBH. Thus, this technique is particularly important for measuring the MBH occupation fraction for dwarf galaxies. 
Specifically, dynamical BH measurements rely on being able to resolve the MBH gravitational sphere of influence. The sphere of influence of a MBH is $GM_{\rm BH}/\sigma^{2}$, where $G$ is the gravitational constant, $M_{\rm BH}$ the BH mass, and $\sigma$ the stellar velocity dispersion. For a typical MBH in a dwarf galaxy with $M_{\rm BH} = 10^{5}\;M_{\odot}$ and $\sigma=30\;\rm{km\;s^{-1}}$, this sphere of influence is $\sim0.5$ pc. At a distance of 10 Mpc, this corresponds to roughly 0.01$''$. Thus, dynamical BH detection is currently only feasible for very nearby dwarf galaxies. 

Despite challenges imposed by the high angular resolution needed for these measurements, dynamical BH masses have been estimated for a handful of nearby, early-type galaxies. In a set of papers, \cite{Nguyen2017, Nguyen2018} and \cite{Nguyen2019} measured MBH masses for five low-mass early-type galaxies within 3.5 Mpc. These systems are M32, NGC 205, NGC 404, NGC 5102, and NGC 5206. Though this sample is small, it does point towards a relatively high BH occupation fraction for local early-type low-mass galaxies. Interestingly, many of these systems also host a dense central NSC. NSCs are the densest known stellar environments, and tend to reside in galaxies with stellar masses of $10^{8-10}\;M_{\odot}$ \cite{2020A&ARv..28....4N}. The co-existence of NSCs and MBHs in these low-mass systems raises many interesting questions about MBH formation and growth inside these dense environments.

In a handful of cases, tidal disruption events (TDEs) have been used to estimate BH masses in dwarf galaxies. Specifically, MBH masses can be estimated by modeling the light curve from the TDE \cite{2019ApJ...872..151M}. Recently, \cite{2022arXiv220900018A} analyzed the TDE 2020neh, which took place in a dwarf galaxy. They measure a MBH mass of $\sim10^{5}\;M_{\odot}$ and find that it is consistent with the extrapolation of the relationship between MBH mass and host stellar velocity dispersion. BH mass measurements from TDEs are useful in that they allow us to measure MBH masses in quiescent galaxies beyond the limit imposed by resolving the sphere of influence. 

\subsection{Optical spectroscopy}
\label{sec:dwarvesopticalspec}
\begin{figure}
\centering
\includegraphics[width=\textwidth]{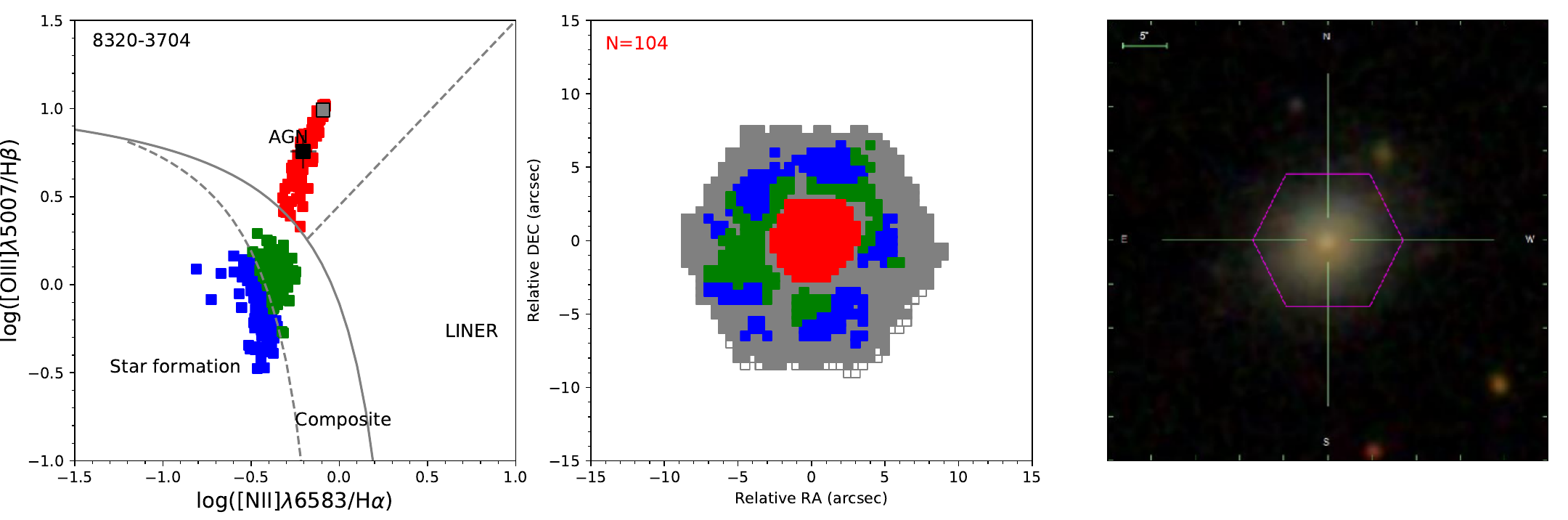}
\caption{\textbf{Left}: Spatially resolved [OIII]$\lambda$5007/H$_{\beta}$ versus [NII]$\lambda$6583/H$_{\alpha}$ emission line diagnostic diagram for the dwarf galaxy 8320-3704 derived using MaNGA (Mapping Nearby Galaxies at Apache Point Observatory; \citet{Bundy2015}) integral field spectroscopy. The location of each MaNGA spaxel on the diagram is used to distinguish between ionisation by AGN/LINER (red spaxels), star-formation (blue spaxels), or composite (green spaxels). The black square marks the median location of those spaxels classified as AGN/LINER, the grey square the Sloan Digital Sky Survey (SDSS, single-fiber spectroscopy) location. \textbf{Center}: Spatial distribution of the emission line diagnostic diagram-classified spaxels (color-coded as in the left panel). Empty squares mark the coverage of MaNGA's integral-field unit, grey squares those spaxels with continuum signal-to-noise ratio $>$ 1. The N shows the total number of AGN/LINER spaxels. \textbf{Right}: SDSS composite image. The pink hexagon shows the integral-field unit coverage. Figure and caption from \cite{Mezcua2020}.}
\label{BPT}
\end{figure}

Given the scarceness of dynamical BH mass measurements, most searches for MBHs have made use of optical spectroscopy to identify AGN in dwarf galaxies. On the one hand, the use of optical emission line diagnostics diagrams such as [OIII]$\lambda$5007/H$_{\beta}$ versus [NII]$\lambda$6583/H$_{\alpha}$,[OIII]$\lambda$5007/H$_{\beta}$ versus [SII]$\lambda$6717,6731/H$_{\alpha}$, or [OIII]$\lambda$5007/H$_{\beta}$ versus [OI]$\lambda$6300, allows distinguishing between gas ionisation produced by AGN, star formation, and LINERs\footnote{Low Ionization Emission Line Region (LINER).} (e.g., \cite{Baldwin1981}; \cite{Kewley2001}; \cite{Kewley2006}; \cite{Kauffmann2003}; see Fig. \ref{BPT}). On the other hand, the detection of broad Balmer lines (e.g., H$\alpha$, H$\beta$) can be used to estimate the BH mass under the assumption that the gas is moving in Keplerian orbits around the MBH. The combination of both methods has been commonly used to infer the presence of low-mass AGN (i.e. with $M_{\rm BH} \lesssim 10^{6}\;M_{\odot}$) in dwarf galaxies, yielding hundreds of candidates \citep[e.g.,][]{Kunth1987,Filippenko1989,Reines2013,Moran2014,2015ApJ...809L..14B,Baldassare2016,Baldassare2017a,Sartori2015,Marleau2017,Chilingarian2018,Mezcua2020,Molina2021,Polimera2022,Salehirad2022}. 

Additional diagnostics sometimes combined with the above include the WHAN diagram \citep{CidFernandes2010} to differentiate between post-AGN stars (which can produce emission line ratios as high as those of LINERs but have typically an H$\alpha$ equivalent width (EW) below 3$\mathring{{\rm A}}$) and true AGN (which have EW(H$\alpha >$) 3$\mathring{{\rm A}}$; e.g. \cite{Mezcua2020}), or the optical emission line diagnostic based on the He II$\lambda$4686 line \citep[e.g.,][]{Sartori2015}.

In the last couple of years, the detection of optical coronal lines (i.e. with a ionisation potential $\geq$ 100 eV that is more likely to be produced by AGN than stellar processes) such as [NeV] $\lambda$3347, $\lambda$3427, [FeVII] $\lambda$3586, $\lambda$3760, $\lambda$6086, or [FeX] $\lambda$6374 has been also used to identify possible AGN in dwarf galaxies \citep[e.g.][]{Molina2021,Negus2021,Salehirad2022}.

\subsection{Optical variability}
\label{sec:dwarvesopticalvariability}

In the last several years, analysing optical photometric variability has become an important tool for identifying AGN in dwarf galaxies. AGN tend to vary across the electromagnetic spectrum on timescales ranging from days to years due to instabilities and turbulence in the accretion disk. It is now feasible to search for AGN variability in large samples of dwarf galaxies thanks to an abundance of publicly available data from repeat imaging surveys such as Stripe 82 \citep{Aihara2011}, Palomar Transient Factory (PTF) \citep{Law2009, Rau2009}, the Zwicky Transient Facility (ZTF) \citep{Graham2019}, and Pan-STARRS \citep{Chambers2016}. Crucially, we are nearing the start of operations for the Vera Rubin Observatory (VRO; \cite{Ivezic2008}). The VRO's Legacy Survey of Space and Time (LSST) will improve on depth and resolution compared to existing surveys. This will greatly improve our ability to detect active MBHs in dwarf galaxies using this technique. 

There have now been several systematic searches for AGN in dwarf galaxies via optical photometric variability. \cite{Baldassare2018} used repeat imaging observations from SDSS Stripe 82 to study the variability properties of $\sim25,000$  nearby galaxies from the NASA-Sloan Atlas. They used difference imaging to construct light curves and assessed whether any detected variability was AGN-like by evaluating the goodness-of-fit to a damped random walk model. A major takeaway of this study was that many of the low-mass galaxies with AGN-like variability did not have optical emission line ratios indicative of an AGN. These systems would have been missed by studies that select AGN using the BPT diagram. This result was confirmed by \cite{Baldassare2020a} with a larger sample of galaxies using data from the PTF survey and by \cite{2022ApJ...936..104W} with data from ZTF. In another set of papers, \cite{2022MNRAS.516.2736B, 2022MNRAS.tmp.2505B} show how variability studies can be used to constrain the overall BH occupation fraction. This technique is desirable as variability selection doesn't depend on galaxy properties as strongly as i.e., the BPT diagram or mid-IR colors. 

Variability may also offer a way to estimate BH masses. \cite{2021Sci...373..789B} finds a correlation between the damping timescale and BH mass such that lower-mass MBHs have a shorter damping timescale. This correlation spans at least six orders of magnitude and is a promising method for measuring BH masses in dwarf galaxies provided sufficiently high-cadence light curves.

Variability searches for AGN in dwarf galaxies are also moving beyond optical wavelengths. \cite{2020ApJ...900...56S} used multi-epoch catalogues from the Wide-Field Survey Explorer to search for variability in a sample of $\sim2000$ dwarf galaxies. They find only two variable dwarf galaxies, possibly due to the low cadence of the observations. The low spatial resolution of the observations also makes it impossible to separate the nucleus from the underlying galaxy. Searching the JWST continuous viewing zones may yield a higher fraction of IR-variable AGN in dwarf galaxies. \cite{2022ApJ...933...37W} used UV observations from the GALEX Time Domain Survey to search for AGN. Out of $\sim500$ dwarf galaxies, they find six new variable low-mass AGN. Most would not be selected using the BPT diagram.

\subsection{Infrared spectroscopy}
\label{sec:dwarvesIRspec}
In the infrared (IR), AGN can be identified via the detection of high-ionisation IR coronal lines (with ionisation potential $>$ 100 eV) such as [Si VI]$\lambda$1.96$\mu$m, [Si VII]$\lambda$2.48$\mu$m, or [Ca VIII]$\lambda$2.46$\mu$m (e.g., \cite{Reunanen2003}; \cite{Prieto2005}). These coronal lines are less affected by dust extinction and star formation dilution than the lines used in BPT diagnostics (which have ionisation potentials $<$ 35 eV). Massive stars cannot produce the high radiations needed to ionise coronal lines and the infrared coronal lines produced by type II supernovae is weak and short-lived (e.g., \cite{Satyapal2021}). Hence infrared coronal lines have been used to detect AGN in bulgeless galaxies (e.g., \cite{Satyapal2007}; \cite{Satyapal2008}; \cite{Satyapal2009}) and low-metallicity dwarf galaxies (e.g., \cite{Cann2020}; \cite{Cann2021}) in addition to massive galaxies (e.g., \cite{Riffel2006}; \cite{MullerSanchez2018}).

\subsection{Mid-infrared colours}
\label{sec:dwarvesMIR}

Mid-IR colour selection is frequently used for AGN identification \citep{2005ApJ...631..163S, 2012ApJ...753...30S, 2009ApJ...696..891H, 2015ApJS..221...12S, 2021ApJ...908..185C}. This technique is particularly useful for identifying obscured AGN as the mid-IR is less impacted by dust extinction compared to optical and ultraviolet observations. 

Several studies have applied this selection to search for AGN in dwarf galaxies. \cite{2014ApJ...784..113S} searched for AGN using mid-IR colors in a sample of $\sim14000$ bulgeless galaxies. Bulgeless galaxies also tend to be low-mass; NGC 4395 is a canonical example. Depending on the particular criterion, they identify between 30 and 300 AGN in bulgeless galaxies. Most of the systems they identify are not classified as AGN on the BPT diagram. \cite{Sartori2015} carried out a similar exercise, looking at the mid-IR colors of almost $50000$ dwarf galaxies. They identify $\sim200$ mid-IR selected AGN, again with little overlap with those selected by optical spectroscopic diagnostics. 

Mid-IR colors make for an attractive AGN diagnostic as the criteria are straightforward and there is excellent sky coverage by the Wide-Field Infrared Survey Explorer. However, there has been some debate over whether this technique is broadly applicable to dwarf galaxies. \cite{2016ApJ...832..119H} find that the majority of dwarf galaxies selected by mid-IR AGN criteria are actually young, compact, blue star-forming galaxies. Extreme star formation is capable of heating dust enough to produce AGN-like mid-IR colors. \cite{2016ApJ...832..119H} urges caution in selecting AGN in dwarf galaxies via this technique since samples can be contaminated by star-forming dwarf galaxies. A follow-up X-ray study of WISE mid-IR selected dwarf galaxies found no X-ray evidence for AGN in the IR-selected systems that were star-forming on the BPT diagram \citep{2021ApJ...914..133L}. These results indicate that caution should be taken when selecting AGN in dwarf galaxies using mid-IR color criteria. 

\subsection{Radio observations}
\label{sec:dwarvesradio}
Radio observations offer also a dust-unbiased way to identify AGN either via the detection of extended, jet radio emission or of pc-scale compact radio emission with brightness temperature $>10^5$ K (indicative of non-thermal emission). To distinguish between AGN emission and star formation processes one can derive the maximum radio luminosity expected from a supernova remnant, from a population of supernova remnants and supernovae (e.g., (\citep{Chomiuk2009}), and from the galaxy-wide SFR (e.g. \citep{Filho2019}) and compare these to the measured radio luminosity. The finding of a radio excess has been used to identify tens of AGN in dwarf galaxies (e.g., \citep{Mezcua2019}; \citep{Reines2020}; \citep{Davis2022}). For these sources, the finding of jet powers as high as in massive galaxies additionally suggests that (mechanical) AGN feedback could be significant in the low-mass regime (e.g., \citep{Mezcua2019}; \citep{Davis2022}), contrarily to what had been so-far assumed in simulations. 

\subsection{X-ray observations}
\label{sec:dwarvesXray}

Bright X-ray emission (i.e., $L_{X}\gtrsim10^{42}\ \rm{erg\;s^{-1}}$) is usually interpreted as unambiguous evidence for AGN activity. There are challenges associated with using X-rays to identify AGN in dwarf galaxies; a $10^{5}\;M_{\odot}$ MBH accreting at its Eddington luminosity will have an X-ray luminosity of $\sim10^{42} \ \rm{erg\;s^{-1}}$. Thus, most X-ray detections of AGN in dwarf galaxies are below this canonical threshold and furthermore overlap in luminosity with ultraluminous X-ray sources, X-ray binaries, and sometimes supernovae. Nevertheless, X-ray observations are still extremely useful in searching for or confirming the presence of AGN in dwarf galaxies, provided those potential sources of contamination are accounted for. 

One of the best studied dwarf AGN in X-rays is the canonical NGC 4395, which shows bright and highly variable X-ray emission \cite{2005AJ....129.2108M, 2005MNRAS.356..524V}. X-ray emission has been found from other broad-line AGN in dwarf galaxies; \cite{2017ApJ...836...20B} analyzed Chandra X-ray Observatory imaging of 10 broad-line, BPT AGN in dwarf galaxies and found that 100\% had X-ray emission. The inferred Eddington fractions for this sample ranged from $0.1-50$\%. 

Large X-ray data sets have also been used to search for AGN in samples of thousands of dwarf galaxies. \cite{2020MNRAS.492.2268B} searched for AGN in a sample of over 4000 dwarf galaxies with XMM-Newton coverage. They found 61 dwarf galaxies with nuclear X-ray emission consistent with an AGN. Interestingly, most did not have optical spectroscopic AGN signatures. Similar studies have been carried out by \cite{2021ApJ...922L..40L} using early results from eROSITA and \cite{2015ApJ...805...12L} using the \textit{Chandra} Source Catalog. 

X-ray observations are also useful for probing the population of AGN in dwarf galaxies at higher redshift. \citealt{2016ApJ...817...20M} found that a population of MBHs with X-ray luminosities as low as $\sim10^{39-40}\rm{erg\;s^{-1}}$ must exist in dwarf galaxies at least out to z=1.5, which was later supported by the finding of sample of dwarf galaxies with AGN of L$_{X}\sim10^{39-44}\rm{erg\;s^{-1}}$ out to photometric z$\sim$2.4 and spectroscopic z=0.505 (\citealt{2018MNRAS.478.2576M}). \cite{2019ApJ...885L...3H} reported the discovery of a $10^{6}\;M_{\odot}$ MBH in a LMC-mass dwarf galaxy at a spectroscopic redshift z=0.56, and \cite{2019ApJ...882..181B} released a catalog of \textit{Chandra} X-ray sources likely to be MBHs out to z=0.9. More recently, \cite{Mezcua2023} reported the finding of a sample of seven broad-line AGN in dwarf galaxies at spectroscopic z$\sim$0.9, six of which have X-ray luminosities $>10^{43}\rm{erg\;s^{-1}}$. Interestingly, these seven sources are found to have MBH masses $>10^{7}$ M$_{\odot}$ and thus to be more massive than expected from MBH-galaxy scaling relations. Another overmassive BH in a dwarf galaxy was also found by \cite{Ubler2023} at z=5.55 using James Webb Space Telescope NIRSpec\footnote{Near Infrared Spectrograph} observations, which constitutes the redshift record-holder for an AGN in a dwarf galaxy.

The use of X-ray observations, in combination with large surveys, additionally allow us to derive an AGN fraction (proxy for the MBH occupation fraction) corrected for completeness. 
Using the \textit{Chandra} COSMOS-Legacy survey, \cite{2018MNRAS.478.2576M} found an AGN fraction of $<$1\%, in agreement with that from \cite{2020MNRAS.492.2268B, Birchall2022} found with the XMM-Newton Serendipitous Sky Survey, and that the AGN fraction decreases with stellar mass. The latter is predicted by cosmological simulations of dwarf galaxies hosting direct collapse BHs, suggesting that the early Universe seed MBHs predominantly formed via direct collapse.

\section{Gravitational waves}
\label{sec:GWs}

\subsection{LIGO detections}
\label{sec:GWsdetections}
The detection of gravitational waves resulting from a binary black hole merger, such as that occurring when a stellar-mass one is captured by an MBH (intermediate-mass ratio inspirals -- IMRIs) in a globular cluster or a dwarf galaxy, offers a new avenue for the detection of MBHs (e.g., \cite{Amaro2020}). IMRIs are not yet detectable by current interferometers; however, the current LIGO and Virgo interformeters should be able to detect the merger of two black holes of up to 100 M$_{\odot}$. One such case already exists, GW190521, in which the mergers of two black holes of 85 $M_{\odot}$ and 66 $M_{\odot}$ resulted in the formation of an MBH of 142 $M_{\odot}$. The transformation of 9 $M_{\odot}$ into energy in the form of gravitational waves was detected by the LIGO and Virgo interformeters \citep{2020ApJ...900L..13A}.

\subsection{LIGO-LISA predictions}
\label{sec:GWspredictions}

The Laser Interferometer Space Antenna (LISA; \cite{2022arXiv220306016A}) will be revolutionary for constraining the population of MBHs. Estimating the overall population of MBHs in dwarf galaxies is fraught with observational challenges related to the low luminosity and relatively small sphere of influence for a $\sim10^{5}\;M_{\odot}$ MBH. However, this mass regime is optimal for LISA. \cite{2019MNRAS.482.2913B} uses high-resolution cosmological zoom-in simulations to study MBHs in dwarf galaxies. They find that mergers between MBHs in dwarf galaxies occur at all redshifts and their GWs will be detectable by LISA. This is key as they also find MBHs in dwarf galaxies tend to accrete very little gas; GW detections may be the only way of detecting this population at high redshift.
\cite{2021MNRAS.508.1174K} find that MBH mergers in dwarf galaxies are extremely efficient due to the high central stellar densities of NSCs that are typically present. Equal-mass mergers are particularly efficient and are expected to be promising LISA GW sources. \cite{2021arXiv211115035D} estimate that as many as 80\% of close MBH pairs in dwarf galaxies will merge in less than a Hubble time and be detectable by LISA. 
\cite{2022arXiv221014960D} find that mergers of MBHs will produce the peak of GW events at redshift z=2, almost all of which will be detectable by LISA. MBH mergers in dwarf galaxies  
LISA may even yield observations of MBHs in nearby GCs. \cite{2022PhRvD.105l4048S} predict that a few galactic GCs could produce at least one GW event that is detectable by LISA.


\section{Acknowledgements}
\vs{-3mm}
Abbas Askar (AA), Vivienne Baldassare (VB) and Mar Mezcua (MM) thank the editors of the book for giving them the opportunity to contribute to this chapter. AA acknowledges support for this work from project No. 2021/43/P/ST9/03167 co-funded by the Polish National Science Centre (NCN) and the European Union Framework Programme for Research and Innovation Horizon 2020 under the Marie Skłodowska-Curie grant agreement No. 945339. For the purpose of Open Access, the authors have applied for a CC-BY public copyright licence to any Author Accepted Manuscript (AAM) version arising from this chapter. AA also acknowledges support from the Swedish Research Council through the grant 2017-04217 and from the Polish NCN through the grant UMO-2021/41/B/ST9/01191. MM acknowledges support from the Spanish Ministry of Science and Innovation through the project PID2021-124243NB-C22. This project was also partially supported by the program Unidad de Excelencia María de Maeztu CEX2020-001058-M. AA would also like to thank Fatimah Ali and Huzaima Bukhari for proofreading part I of the manuscript and Swayamtrupta Panda for his helpful suggestions and references.


\interlinepenalty=10000 
\section{Bibliography}
\bibliography{Biblio,Biblio2,Biblio_Ch2}

\end{document}